\documentclass[letterpaper,11pt]{article}
\usepackage{jheppub}
\usepackage{xcolor}
\usepackage{braket}
\usepackage{graphicx}
\usepackage{soul}
\usepackage{amssymb,amsmath}
\DeclareMathOperator{\tr}{tr}
\usepackage{verbatim}
\usepackage{anyfontsize}
\usepackage{dsfont}
\usepackage{tikz}
\usepackage{sidecap}
\usepackage{subfig}
\sidecaptionvpos{figure}{t}
\usepackage[english]{babel}
\usepackage{enumitem}  
\usepackage{setspace}
\usepackage[permil]{overpic}
\usepackage{graphicx}
\usepackage{mathtools}
\usepackage{wasysym}
\usepackage{graphicx}
\usepackage{placeins}
\usepackage{bm}
\newcommand{\be}{\begin{equation}}
\newcommand{\ee}{\end{equation}}
\newcommand{\GW}{\textcolor{magenta}}
\usepackage{amsmath}
\usepackage{amsfonts}
\usepackage{amssymb}
\usepackage{graphicx}
\usepackage{wrapfig}
\usepackage{braket}
\usepackage{hyperref}
\usepackage{amsthm,verbatim,cancel}
\usepackage{color}
\usepackage{caption}
\usetikzlibrary{decorations.markings, arrows.meta}

\newcommand{\pd}{\partial}
\newcommand{\nn}{\nonumber\\}
\newcommand{\bea}{\begin{eqnarray}}
\newcommand{\eea}{\end{eqnarray}}

\usepackage{tikz}
\usetikzlibrary{positioning,decorations.pathmorphing}
\usepackage{hyperref}
\usepackage{amssymb}
\usepackage{amsmath}
\usepackage{color}
\usepackage{tikz}

\usepackage{marginnote}

\newcommand{\mathtikz}[2][]{\quad\vcenter{\hbox{\tikz[#1]{#2}}}\quad}

\newcommand{\vhalf}[1]{%
  \begin{scope}[yscale=.5, transform shape]
    #1
  \end{scope}%
}

\newcommand{\vthird}[1]{%
  \begin{scope}[yscale=.3, transform shape]
    #1
  \end{scope}%
}

\newcommand\pairAT[2]{ 
\begin{scope}[xshift=#1,yshift=#2]
\draw (-0.75,0) -- (-0.25,0) to [out=-90,in=180] (0,-0.33) to [in=-90,out=0] (0.25,0) 
      -- (0.75,0) node[right, blue] {$T$}
      to [out=-90,in=0] (0,-0.83) to [out=180,in=-90] (-0.75,0);
\draw[blue] (0.25,0) -- (0.75,0);
\end{scope}
}

\newcommand\stripRR[2]{ 
    \begin{scope}[xshift=#1,yshift=#2]
        \draw[
            line width=1.5pt,
            decoration={
                markings,
                mark=at position 0.55 with {\arrow[line width=1.5pt, scale=1]{Stealth}}
            },
            postaction={decorate}
        ] (-0.5,0) -- (-0.5,-2);
        \draw[
            line width=1.5pt,
            decoration={
                markings,
                mark=at position 0.55 with {\arrow[line width=1.5pt, scale=1]{Stealth}}
            },
            postaction={decorate}
        ] (0.5,-2) -- (0.5,0);
        \draw[dashed, line width=1pt] (-0.5,-1) -- (0.5,-1);
        \node[above, font=\small] at (-0.5,0) {$R^*$};
        \node[above, font=\small] at (0.5,0) {$R $};
    \end{scope}
}

\newcommand{\CthreeVertex}[1][1]{%
  \begin{scope}[
      scale=#1,
      edge/.style    = {thick, ->, black!70},
      framing/.style = {->, black!55, thick},
      nodelbl/.style = {font=\small},
      framlbl/.style = {font=\footnotesize, black!60},
  ]
    \fill[black!70] (0,0) circle (2.5pt);

    \draw[edge] (0,0) -- ({-2*0.707},{-2*0.707})
        node[nodelbl, below left] {$v_1$};
    \draw[edge] (0,0) -- (0,2)
        node[nodelbl, above right] {$v_2$};
    \draw[edge] (0,0) -- (2,0)
        node[nodelbl, right] {$v_3$};

    \draw[framing] ({-0.707},{-0.707}) -- ++(0,0.7)
        node[framlbl, left] {$f_1$};
    \draw[framing] (0,1) -- ++(0.7,0)
        node[framlbl, above] {$f_2$};
    \draw[framing] (1,0) -- ++({-0.7*0.707},{-0.7*0.707})
        node[framlbl, below right] {$f_3$};
  \end{scope}%
}

\newcommand\nablaAflip[2]{ 
\begin{scope}[xshift=#1,yshift=#2,yscale=-1]
\draw (-0.25,-0.5) -- (0.25,0) -- (0.75,0) -- (0.25,-0.5) --(0.25,-1) -- (-0.25,-1) -- (-0.25,-0.5) -- (-0.75,0) -- (-0.25,0) -- (0,-0.25);
\draw[dashed] (-0.25,-0.5) -- (0.25,-0.5);
\end{scope}
}

\newcommand\idC[2] { 
\begin{scope}[xshift=#1,yshift=#2]
\filldraw[left color=lightgray, right color=white] (-0.25,0) -- (0.25,0) -- (0.25,-1) to [in=-90,out=-90] (-0.25,-1) -- (-0.25,0);
\filldraw[left color=white,right color=lightgray] (0,0) ellipse (0.25 and 0.1);
\draw[dotted] (0.25,-1) arc (0:180:0.25 and 0.1);
\end{scope}
}

\newcommand\HidC[2] { 
\begin{scope}[xshift=#1,yshift=#2]
\filldraw[top color=lightgray, bottom color=white]
  (0,-0.25) -- (0,0.25) -- (1,0.25)
  to [out=0,in=0] (1,-0.25)
  -- (0,-0.25);

\filldraw[top color=white,bottom color=lightgray]
  (0,0) ellipse (0.1 and 0.25);

\draw[dotted] (1,0.25) arc (90:270:0.1 and 0.25);

\node[left] at (-0.15,0) {$\bra{\tilde{e}}$};
\node[right] at (1.15,0) {$\ket{e}$};
\end{scope}
}

\newcommand\idA[2] { 
\begin{scope}[xshift=#1,yshift=#2]
\filldraw[fill=white,draw=black] (-0.25,0) rectangle (0.25,-1);
\end{scope}
}

\newcommand\idAT[2]{
\begin{scope}[xshift=#1,yshift=#2]

\filldraw[fill=white,draw=black] (-0.25,0) rectangle (0.25,-1);

\filldraw[fill=white,draw=black] (-0.25,-1) rectangle (0.25,-2);

\draw[blue,thick] (-0.25,-1) -- (0.25,-1);

\node[blue,right] at (0.25,-1) {$T$};

\end{scope}
}

\newcommand\muC[2]{ 
\begin{scope}[xshift=#1,yshift=#2]
\filldraw[left color=lightgray, right color=white] (-0.25,0) to [out=-90,in=180] (0,-0.33) to [in=-90,out=0] (0.25,0) to  (0.75,0) to [in=90,out=-90] (0.25,-1) to [out=-90,in=-90] (-0.25,-1) to [in=-90,out=90] (-0.75,0);
\filldraw[left color=white,right color=lightgray] (-0.5,0) ellipse (0.25 and 0.1);
\filldraw[left color=white,right color=lightgray] (0.5,0) ellipse (0.25 and 0.1);
\draw[dotted] (0.25,-1) arc (0:180:0.25 and 0.1);
\end{scope}
}

\newcommand\pairC[2]{ 
\begin{scope}[xshift=#1,yshift=#2]
\filldraw[left color=lightgray, right color=white] (-0.25,0) to [out=-90,in=180] (0,-0.33) to [in=-90,out=0] (0.25,0) to  (0.75,0) to [out=-90,in=0] (0, -0.83) to [out=180,in=-90]
(-0.75,0);
\filldraw[left color=white,right color=lightgray] (-0.5,0) ellipse (0.25 and 0.1);
\filldraw[left color=white,right color=lightgray] (0.5,0) ellipse (0.25 and 0.1);
\end{scope}
}

\newcommand\copairC[2]{ 
\begin{scope}[xshift=#1,yshift=#2]
\filldraw[left color=lightgray, right color=white] (-0.25,0) to [out=90,in=180] (0,0.33) to [in=90,out=0] (0.25,0) to [out=-90,in=-90] (0.75,0) to [out=90,in=0] (0, 0.83) to [out=180,in=90]
(-0.75,0) to [out=-90,in=-90] (-0.25,0);
\draw[dotted] (-0.25,0) arc (0:180:0.25 and 0.1);
\draw[dotted] (0.75,0) arc (0:180:0.25 and 0.1);
\end{scope}
}

\newcommand\deltaC[2]{
\begin{scope}[xshift=#1,yshift=#2]
\filldraw[left color=lightgray, right color=white] (-0.25,-1) to [out=90,in=180] (0,-0.66) to [in=90,out=0] (0.25,-1) to [out=-90,in=-90] (0.75,-1) to [in=-90,out=90] (0.25,0) to (-0.25,0) to [in=90,out=-90] (-0.75,-1) to [out=-90,in=-90] (-0.25,-1);
\filldraw[left color=white,right color=lightgray] (0,0) ellipse (0.25 and 0.1);
\draw[dotted] (-0.25,-1) arc (0:180:0.25 and 0.1);
\draw[dotted] (0.75,-1) arc (0:180:0.25 and 0.1);
\end{scope}
}

\newcommand\muA[2]{ 
\begin{scope}[xshift=#1,yshift=#2]
\draw (-0.75,0) -- (-0.25,0) to [out=-90,in=180] (0,-0.33) to [in=-90,out=0] (0.25,0) -- (0.75,0) to [in=90,out=-90] (0.25,-1);
\draw (-0.25,-1) -- (0.25,-1);
\draw (-0.75,0) to [in=90,out=-90] (-0.25,-1);
\end{scope}
}

\newcommand\pairA[2]{ 
\begin{scope}[xshift=#1,yshift=#2]
\draw (-0.75,0) -- (-0.25,0) to [out=-90,in=180] (0,-0.33) to [in=-90,out=0] (0.25,0) -- (0.75,0) to [out=-90,in=0] (0,-0.83) to [out=180,in=-90] (-0.75,0);
\end{scope}
}

\newcommand\copairA[2]{ 
\begin{scope}[xshift=#1,yshift=#2]
\draw (-0.75,0) -- (-0.25,0) to [out=90,in=180] (0,0.33) to [in=90,out=0] (0.25,0) -- (0.75,0) to [out=90,in=0] (0,0.83) to [out=180,in=90] (-0.75,0);
\end{scope}
}

\newcommand\copairAT[2]{ 
\begin{scope}[xshift=#1,yshift=#2]
\draw (-0.75,0) -- (-0.25,0) to [out=90,in=180] (0,0.33) to [in=90,out=0] (0.25,0);
\draw[blue] (0.25,0) -- (0.75,0);
\draw (0.75,0) to [out=90,in=0] (0,0.83) to [out=180,in=90] (-0.75,0);
\end{scope}
}

\newcommand\deltaA[2]{ 
\begin{scope}[xshift=#1,yshift=#2]
\draw (-0.75,-1) -- (-0.25,-1) to [out=90,in=180] (0,-0.66) to [in=90,out=0] (0.25,-1) -- (0.75,-1) to [in=-90,out=90] (0.25,0) -- (-0.25,0) to [in=90,out=-90] (-0.75,-1);
\end{scope}
}

\newcommand\semiwidemuA[2]{ 
\begin{scope}[xshift=#1,yshift=#2,yscale=-1]
\draw (-1,-1) -- (-0.5,-1) to [out=90,in=180] (0,-0.66) to [in=90,out=0]
(0.5,-1) -- (1,-1) to [in=-90,out=90] (0.25,0) -- (-0.25,0) to [in=90,out=-90] (-1,-1);
\end{scope}
}

\newcommand\zipper[2]{ 
\begin{scope}[xshift=#1,yshift=#2]
\draw (-0.25,-1) -- (0.25,-1);
\filldraw[right color=white,left color=lightgray] (-0.25,0) to (-0.25,-1) to [out=90,in=225] (0,-0.5) to [out=-45,in=90] (0.25,-1) to (0.25,0);
\filldraw[left color=white,right color=lightgray] (0,0) ellipse (0.25 and 0.1);
\end{scope}
}

\newcommand\cozipper[2]{ 
\begin{scope}[xshift=#1,yshift=#2]
\draw (-0.25,0) -- (0.25,-0);
\filldraw[right color=white,left color=lightgray] (-0.25,-1) to (-0.25,0) to [out=-90,in=135] (0,-0.5) to [out=45,in=-90] (0.25,0) to (0.25,-1) to [in=-90,out=-90] (-0.25,-1);
\draw[dotted] (0.25,-1) arc (0:180:0.25 and 0.1);
\end{scope}
}

\newcommand\epsilonC[2]{ 
\begin{scope}[xshift=#1,yshift=#2]
\filldraw[right color=white,left color=lightgray] (-0.25,0) to [out=-90,in=180] (0,-0.33) to [in=-90,out=0] (0.25,0);
\filldraw[left color=white,right color=lightgray] (0,0) ellipse (0.25 and 0.1);
\end{scope}
}

\newcommand\etaC[2] { 
\begin{scope}[xshift=#1,yshift=#2]
\filldraw[right color=white,left color=lightgray] (-0.25,0) to [out=90,in=180] (0,0.33) to [in=90,out=0] (0.25,0) to [in=-90,out=-90] (-0.25,0);
\draw[dotted] (0.25,0) arc (0:180:0.25 and 0.1);
\end{scope}
}

\newcommand\epsilonA[2] {
\begin{scope}[xshift=#1,yshift=#2]
\draw (-0.25,0) -- (0.25,0);
\draw (-0.25,0) to [out=-90,in=180] (0,-0.33) to [in=-90,out=0] (0.25,0);
\end{scope}
}

\newcommand\etaA[2] {
\begin{scope}[xshift=#1,yshift=#2]
\draw (-0.25,0) -- (0.25,0);
\draw (-0.25,0) to [out=90,in=180] (0,0.33) to [in=90,out=0] (0.25,0);
\end{scope}
}

\newcommand\rightbraidA[2]{ 
  \begin{scope}[xshift=#1,yshift=#2]
    \draw (-0.75,0) -- (-0.25,0) 
      to [out=-90,in=90,looseness=0.5] (0.75,-1) -- (0.25,-1) 
      to [out=90,in=-90,looseness=0.5] (-0.75,0);

    \filldraw[fill=white,draw=black] (0.75,0) -- (0.25,0) 
      to [out=-90,in=90,looseness=0.5] (-0.75,-1) -- (-0.25,-1) 
      to [out=90,in=-90,looseness=0.5] (0.75,0);
  \end{scope}
}

\newcommand\leftbraidA[2]{ 
  \begin{scope}[xshift=#1,yshift=#2]
    \draw (0.75,0) -- (0.25,0) 
      to [out=-90,in=90,looseness=0.5] (-0.75,-1) -- (-0.25,-1) 
      to [out=90,in=-90,looseness=0.5] (0.75,0);

    \filldraw[fill=white,draw=black] (-0.75,0) -- (-0.25,0) 
      to [out=-90,in=90,looseness=0.5] (0.75,-1) -- (0.25,-1) 
      to [out=90,in=-90,looseness=0.5] (-0.75,0);
  \end{scope}
}

\begin{document}
\author{Gabriel Wong}
\emailAdd{gabrielwon@gmail.com}
\affiliation{Mathematical Institute, University of Oxford, Andrew Wiles Building, Radcliffe Observatory Quarter, Woodstock Road, Oxford, OX2 6GG, U.K.}

\title{Entanglement and geometric transitions
in topological string theory }
\abstract{ How do we define a bulk subsystem in  quantum gravity?  In \cite{Wong:2025kpz}, we argued that such a subsystem must support local holographic degrees of freedom. These are gravitational edge modes, whose entanglement creates a backreaction that fuses together subregions of spacetime.  In this work we give a realization of these ideas in topological string theory, building upon \cite{Donnelly:2020teo,Jiang:2020cqo}.   In this theory, a subsystem for closed strings consists of open strings ending on entanglement branes, which play the role of a dynamical entangling surface.  Local holography is implemented by the geometric transition of these branes.  We define a subregion open string algebra and develop a diagrammatics for open string modular flow for arbitrary states and subregion.  We check that the entanglement entropy of these open strings reproduces the gravitational entropy of the associated closed string background.   Finally, we relate these local transitions to defect holography.
}  
\maketitle

\section{Introduction}
While the backreaction of spacetime to stress-energy is a well known gravitational phenomenon, the idea that spacetime backreacts to entanglement is a newer and more radical concept.  A striking example appears in the analysis of black hole evaporation,  where the growing entanglement between Hawking pairs shifts the saddle of the gravitational path integral, leading to a topology change that restores unitarity. \cite{Penington2020EntanglementWedgeReconstructionInformationParadox,Almheiri2020PageCurveHawkingRadiationSemiclassicalGeometry,Almheiri2020ReplicaWormholesEntropyHawkingRadiation} 
\begin{figure}
    \centering
\includegraphics[width=1\linewidth]{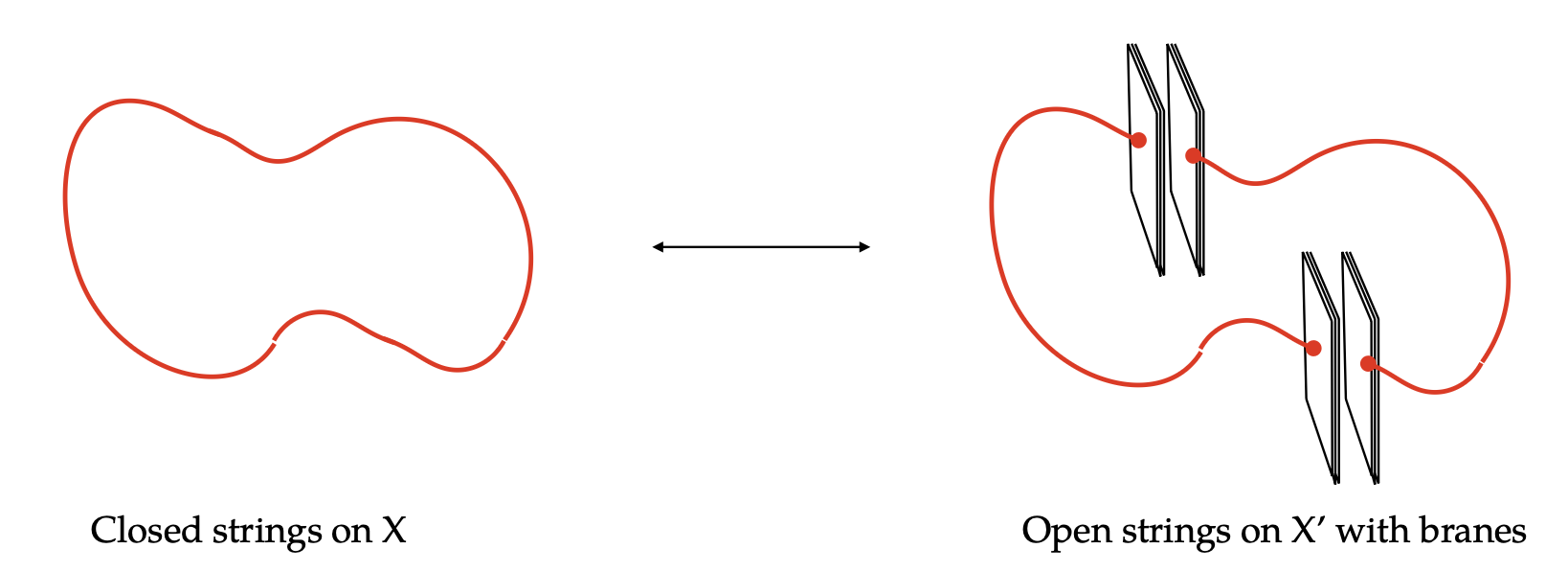}
    \caption{ The entanglement of  subregion open strings produce a backreaction leading to closed strings on a new background}
\label{fig:entangledstrings}
\end{figure}
A deeper manifestation of gravitational backreaction to entanglement arises in the context of the ER=EPR correspondence, where entanglement between microscopic degrees of freedom  produces a connected geometry \cite{Maldacena:2013xja, Maldacena:2001kr,VanRaamsdonk:2009ar,VanRaamsdonk:2010pw}.

Even though ER=EPR is well understood from the boundary perspective, a bulk formulation remains elusive.  This is because the gravitational constraints obstruct an obvious factorization into subsystems\footnote{ Indeed, it remains an unsolved challenge to even define a  subsystem in gravity, since a rigid partition of spatial degrees of freedom would be inconsistent with diffeomorphism invariance.  One proposal for a classical description of such a subsystem involves dynamical gravitational edge modes \cite{Donnelly:2016auv}: our work is in the same spirit as these ideas.   }, so it is unclear what plays the role of bulk qubits. 
In low energy effective field theory, this obstruction can be resolved by introducing edge modes localized to the entangling surface that lifts the constraints. In \cite{Wong:2025kpz}, we argued that in quantum gravity, the edge mode paradigm is elevated to a type of local holography: the bulk entangling surface is a gravitational object, and supports holographic degrees of freedom that backreacts on the spacetime. 

In this work, we provide a systematic realization of local holography in the A model topological string theory, building upon 
\cite{Donnelly:2020teo, Jiang:2020cqo,Wong:2025asd}.
In this context, local holography  is implemented by local geometric transitions in which entanglement  branes are replaced by their backreacted geometry. These branes are dynamical edge modes that play the role of the entangling surface in string theory.  The backreaction of branes as a mechanism for holography is well known in open-closed string dualities of the A model  \cite{Gopakumar:1998ki}, \cite{Okuda:2007ai, Gomis:2006mv, Gomis:2007kz}, as well in physical string dualities like AdS/CFT \cite{Maldacena:1997re}.   However, our work provides a new setting in which these transitions arise from entanglement of subregion open strings ending on  entanglement branes. In \cite{Donnelly:2020teo, Jiang:2020cqo,Wong:2025asd}, we provided a consistency check showing how the entanglement entropy  of open strings reproduced the gravitational entropy of a closed string background. This gave a precise illustration of how a closed string background can emerge from the entanglement of open strings.  
 
In this work, we develop this picture further.  We provide a geometric description of the entanglement brane and show that it can be resolved into a complete set of stringy objects. The transition of the entanglement branes  is then implemented by summing over vacuum loops of these objects, and their pair creation is identified as the source of bulk entanglement.   

We will give a new quantization of  subregion open strings and derive their open string algebra, using the recent proposal for topological subregions in \cite{Wong:2025asd}. 
These are quantum group algebras because the effective open string degrees of freedom
become anyonic due to their interactions with the entanglement branes.
We define an associated quantum trace, whose large $N$ limit provides the algebraic
description for the transition of entanglement branes. 

Using anyon diagrammatics for the subregion algebra, we give a complete, algebraic description of open string modular flow.  We generalize the replica trick defined in \cite{Jiang:2020cqo,Donnelly:2020teo}  and checked that the resulting gravitational entropy agrees with the anyonic entanglement entropy of open strings, defined with respsect to the large N quantum trace $\widetilde{\tr}$:
\begin{align}\label{openclosedentropy}
S_{\rm replica}
&=-\widetilde{\tr}(\rho\log\rho)
\end{align}
While we focus on computation of entropies, our exact description of the open string reduced density matrix enables one to compute general anyonic entanglement measures using standard techniques in anyon diagrammatics.

\paragraph{Summary and outline}
To complete our introduction, we flesh out some details of our results and give an outline of the paper.  

We begin by placing our proposal for local holography in a more familiar context.
A standard example of a holographic correspondence is the relation between a Euclidean hyperbolic disk and the trace of a boundary quantum system: 
\begin{align}\label{globalholography}
    \includegraphics[scale=.2]{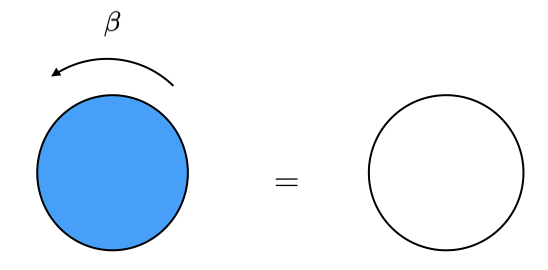}
\end{align}
In the bulk, the temperature circle shrinks and obstructs a bulk trace interpretation: this is another version of the  Hilbert space factorization problem. 

The idea behind local holography is to overcome this obstruction via a spacetime surgery in which the insertion of local holographic degrees of freedom removes part of the bulk so that the temperature circle becomes non-contractible:
\begin{align}
\includegraphics[scale=.2]{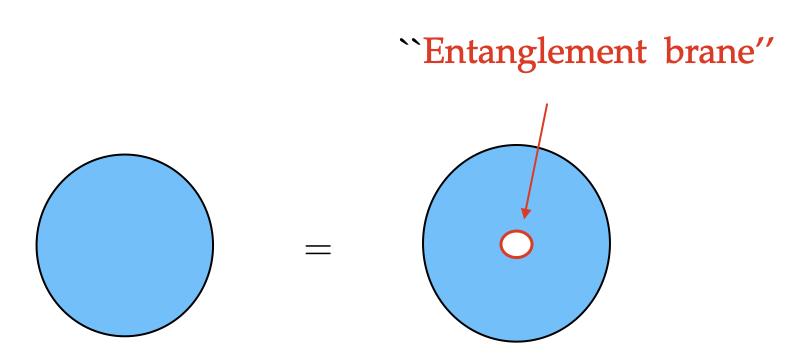}
\end{align}
Note that the red circle that is homologous to the bulk temperature circle need not be a physical boundary in the  higher dimensional geometry on which the full quantum gravity theory is defined.  For example, it could represent the end of space due to cycles contracting in the internal dimensions.  In the context of the FZZ duality, it represents a winding condensate.    We refer to the degrees of freedom needed to shrink the bulk temperature circle generically  as ``entanglement branes".

In Euclidean de Sitter space, local holography becomes indispensible, as there is no spatial infinity at which to anchor a holographic screen. In EdS$_2$ such a duality would take the form 
\begin{align}\label{dS2}
\includegraphics[scale=.3]{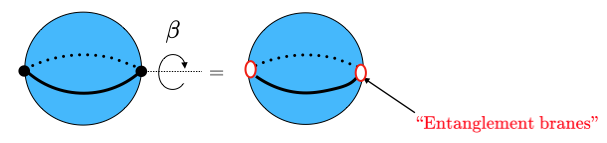}
\end{align}

In the A model topological string theory, this EdS2 geometry appears as the base space of the resolved conifold geometry, which is a non-compact Calabi Yau.  More generally, we will consider non compact Calabi Yau's that are fiber bundles over a Riemann surface. In this context, we show there is an exact realization of the surgery operation in \eqref{dS2}, which is implemented by the transition of entanglement branes. 
This can be interpreted as a local version of the usual spacetime surgery which describes EAdS/CFT: in this case, a \emph{relative}
homology cycle ending on branes closes up into an absolute homology cycle when these branes
dissolve. These transitions are a stringy realization of the shrinkable boundary condition in QFT, which is a boundary condition at the entangling surface that satisfies the  ``hole-closing" condition \eqref{dS2} \cite{Jafferis:2019wkd,Donnelly:2018ppr, Hung:2019bnq, Wong:2025asd}.   
They enable us to give a canonical open string interpretation of gravitational entropy by reformulating closed string amplitudes in terms of open string traces.  

\begin{figure}[h]
    \centering
    \includegraphics[width=0.75\linewidth]{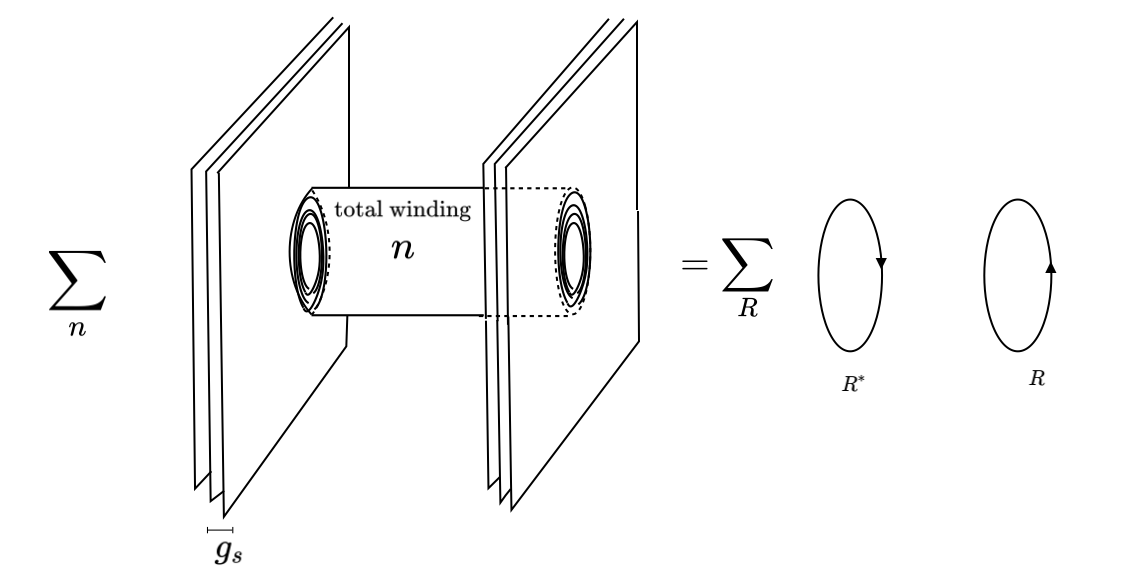}
    \caption{The LHS of the diagram shows open strings ending on stacks of entanglement branes.  Each element of the stack is separated from the next by $g_{s}$, with a total thickness of $g_{s}N$.  The sum over string worldsheets with arbitrary winding $n$ produces a trace which implements the geometric transition of these branes.   On the RHS, we show and equivalent interpretation of this trace as a  sum over vacuum loops of stringy objects labelled by an irreducible representation $R$ of the quantum group $U(\infty)_{q}$ }
    \label{fig:Ebranes}
\end{figure}

To make this paper self contained, we will review some ideas proposed in \cite{Donnelly:2020teo, Jiang:2020cqo}.  For the reader's convenience, we highlight the new contributions:

\begin{enumerate}

\item \textbf{Geometric interpretation of the entanglement brane.}  In \cite{Donnelly:2020teo, Jiang:2020cqo}, we
showed the entanglement brane in the A model is a large N stack of non-compact Lagrangian branes with a distinguished value of the worldvolume holonomy.  Here we give a geometric interpretation of this worldvolume holonomy as the position moduli for the non-compact branes.  Due to a special feature of A model, in the large $N$ t'Hooft limit, turning on this worldvolume holonomy thickens the stack into an equally spaced configuration of D branes, with separation distance $g_{s}$ and total thickness $g_{s}N$. See figure \ref{fig:Ebranes}

\item \textbf{Backreaction and negative branes.}
We clarify the role of backreaction in shifting the global, closed string background relative to the subregion, open string background (figure \ref{fig:entangledstrings}). 
This is an essential feature of quantum gravity which distinguishes it from QFT. 
We also clarify the role of negative branes and their annihilation with ordinary branes in the entangling process. 

\item \textbf{Diffeomorphism-invariant quantization of the subregion.}
In  \cite{Donnelly:2020teo, Jiang:2020cqo}, we applied a standard 
gapless QFT boundary conditions to quantize the subregion theory.  This breaks the the diffeomorphism invariance of the theory, leading to  UV-divergent contributions to the entanglement entropies that needs to be subtracted. In this work, we provide an alternative quantization \cite{Mertens:2025ydx, Alekseev:1994au,Alekseev:1994pa,Wong:2025asd}which preserves topological invariance and gives finite entropies.  

\item \textbf{Framing}
We include the effects of framing in the topological string.
Certain choices of cobordism generators in \cite{Donnelly:2020teo, Jiang:2020cqo}
excluded these. 

\item \textbf{General backgrounds, and disconnected subregions}
We generalize computations of the gravitational replica trick, open string modular flow, and open string anyonic entanglement entropy to an infinite class of Calabi Yau backgrounds which are rank 2 bundles over a Riemann surface. We also compute the modular flow for multiple disconnected regions, and interpret the  modular evolution in terms of  pair creation and annihilation of stringy objects.
\item \textbf{Gravitational entropy of Bubbling Calabi Yau's}
We incorporate D branes into the preparation of closed string states.  Via a known large N transition, these D branes backreact to produces bubbling Calabi Yau geometries, whose gravitational entropies we compute. 
\end{enumerate}

The paper is organized as follows.
Section \ref{sec:closedstrings} defines the A model TQFT, which provides an analogue of a spacetime path integral for the topological string.   We define and compute a backreacted, gravitational
replica entropy.
Section \ref{sec:extendedTQFT} explains how local holography is implemented via the open-closed extension of the A model TQFT, and gives a formal definition of the entanglement brane as a boundary state in spacetime.  
Section \ref{sec:Dbranewfn} reviews the formulation of closed string wavefunctions in the A model  as D brane amplitudes, and clarifies the role of framing and negative branes.  

Section \ref{section:transition} applies the closed string technology developed in the previous sections to derive the geometric transition of entanglement branes.   
Section \ref{resolve} provides a resolution of the entanglement brane boundary state as a sum over stringy objects labelled by a Young tableaux.  These tableaux are associated special configurations of D branes which can backreact to product bubbling Calabi Yau geometries.  We will   introduce closed string states prepared with the insertion of these branes and compute their gravitational entropy.

Finally, 
Section \ref{sec:opensubregion} defines the subregion open string algebra via the combinatorial 
quantization of Chern-Simons theory and develops a diagrammatics for open string  modular flow.  Combining all of the tool developed in this work, we compute anyonic entanglement entropy 
and prove the equality~\eqref{openclosedentropy} for a general class of backgrounds and states.  

\section{ The closed string  TQFT and its gravitational entropy} \label{sec:closedstrings}

The worldsheet theory for the A-model topological string can be consistently defined on six-dimensional Kähler target spaces.  
A key simplification arises because the A model localizes onto holomorphic maps, so that string amplitudes receive contributions only from worldsheet instantons wrapping minimal volume two-cycles in the target space.  For example, when there is a single two-cycle with Kahler volume $t$ , the (single string) free energy takes the form
\begin{align}
    F= \sum_{g,n} g_{s}^{2g-2 } N_{g,n} e^{-nt }
\end{align} 
where $n$ labels  the winding number of the worldsheet, and $N_{g,n}$ are the Gromov- Witten invariants, which give a  virtual counting of holomorphic maps. 

In order to define a bulk string theory  Hilbert space and its factorization, we need a consistent framework for cutting and gluing \emph{multi-string} amplitudes, which generalizes the spacetime path integral logic for defining states and Hilbert space operations in QFT.   For the A model, such a framework is provided by a topological field theory, which captures the full string amplitudes on a restricted class of backgrounds\cite{Aganagic:2004js,Bryan:2004iq}. These backgrounds are total spaces of rank-two complex bundles over a two-dimensional surface $\Sigma$
\begin{align} \label{bundle}
    M = \mathcal{O}(k_{1}) \oplus \mathcal{O}(k_{2}) \longrightarrow \Sigma 
\end{align}
The bundle structure is completely specified by two integers $k_{1},k_{2}$, which determine the Euler Class of the two line bundles: in particular, $\mathcal{O}(k)$ maybe viewed as the complex line bundle associated to a magnetic monopole of charge $k$ \footnote{ $k_{1} $ is also the intersection number of the zero section of $\mathcal{O}(k_{2})$ and $\Sigma$, and similarly for $k_{2}$}.

On these backgrounds,  the full closed string amplitude localizes to worldsheet instantons that wrap the surface $\Sigma$ with arbitrary degree and winding number.  These amplitudes  are equal to the partition functions of a TQFT which lives on the two-manifold $\Sigma$ and  depends on the bundle data $k_{1},k_{2}$. Thus, we can represent  the full string theory amplitudes on the spacetime $M$ in terms of  two-dimensional surfaces decorated by their Euler classes:
\[
\includegraphics[width=0.4\linewidth]{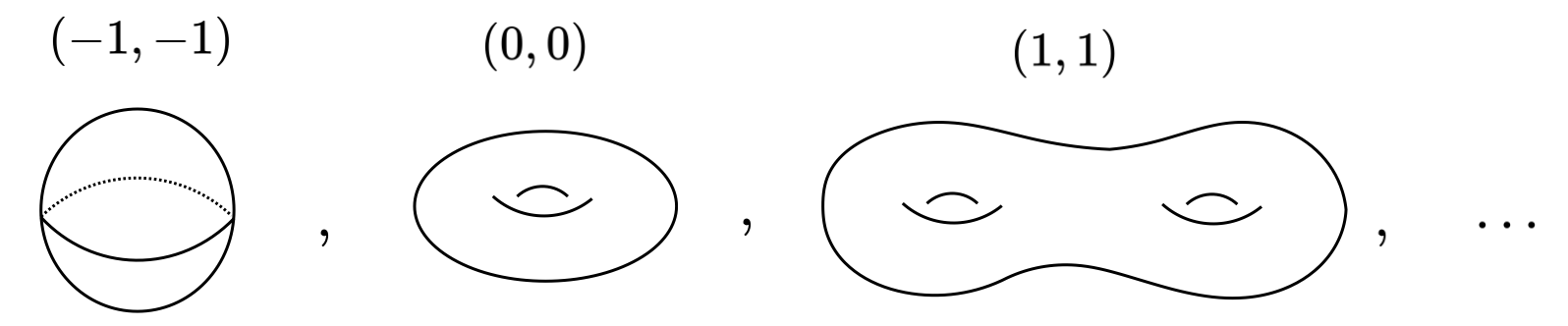}
\]
We emphasize these are diagrams for the target space, and not the worldsheet.
The reduction to a 2D TQFT occurs because in the presence of the non-compact fibers, finite energy worldsheet  configurations localize to branched coverings of $\Sigma$.  A general amplitude can be factorized into basic building blocks by pinching $\Sigma$ along a circle: this breaks up the target space into chunks connnected by a singular a co-dimension 2 fiber over  $p\in\Sigma$ at the pinch\footnote{The fiber over the pinch point is a co-dimension 2 divisor on which the ramification data is specified in relative Gromov Witten theory \cite{IonelParker2003}.  In this context the ramificaiton data is referred to as the tangency condition at the divisor }.
\begin{align}\label{pinch}
    \includegraphics[scale=.1]{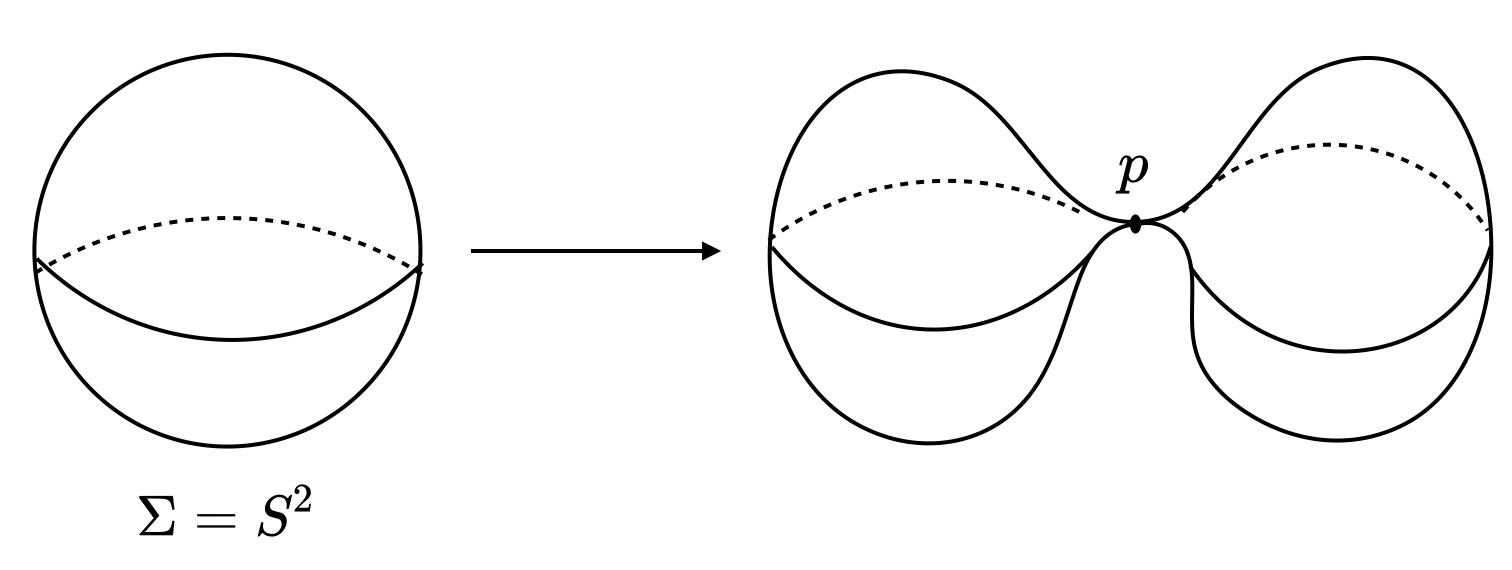}
\end{align}

The amplitude on each building block can then be specified by the ramification data that gives the winding pattern of the closed string worldsheet around the branch point $p$.   This factorizes the string amplitude on the total space time to ampitudes on subregions of the spacetime, where the ramification data is matched and glued.   For example, a connected, spherical worldsheet wrapping the target space sphere  in \eqref{pinch} with degree 3 would be cut into the following worldsheets configurations
\begin{align}
    \includegraphics[scale=.15]{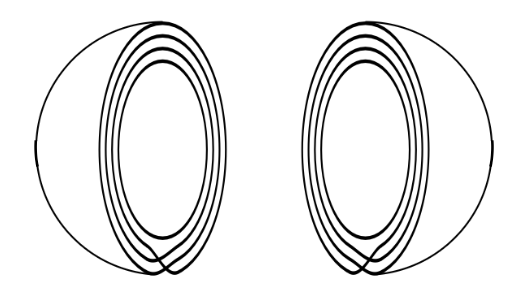}
\end{align}
Remarkably, the gluing rules for these building blocks are equivalent to those of a 2D TQFT, so that the amplitude on the total manifold can be generated by gluing 2D caps and pants that produce $\Sigma$.   The only extra ingredient is the bundle data $k_{1},k_{2}$, which accounts for the effects of worldsheet fluctuations along fibers normal to $\Sigma$.  These Euler classes are additive upon gluing, and leads to a more complicated algebraic structure than that of an ordinary 2D TQFT\footnote{The full TQFT structure is a functor from cobordisms with bundles into the category of vector space.  These are generator by the $(0,0)$ Frobenius algebra, plus the $(0,-1)$ and $(-1,0)$ caps}. In this work,  we will further restrict to target space that satisfy the Calabi Yau condition,  which relates the bundle structure to the Euler character of the base space:
\begin{align}
    k_{1}+k_{2} = - \chi (\Sigma) 
\end{align}
This further reduces the A model TQFT to a theory generated by the following building blocks: 
\begin{align} \label{buildingblocks}
 \mathtikz{\epsilonC{0}{0};
\node at (0,1/2){(-1,0)};} , \quad 
 \mathtikz{\etaC{0}{0};
\node at (0,1/2){(0,-1)};} ,\quad 
\mathtikz{\muC{0}{0};
\node at (0,1/2){(1,0)};} ,\quad  \mathtikz{\deltaC{0}{0};
\node at (0,1/2){(0,1)};},
 \quad
\mathtikz{\idC{0}{0} ;\node at (0,1/2){(0,0)}}
\end{align}
Over these surfaces with boundary, the bundle data $k_{1},k_{2}$ is defined as  \emph{relative} Euler class numbers, which depend on a trivialization on the boundary. They can be expressed in terms of integrals invariants that belong to the relative cohomology of the Riemann surface (See the appendix for a review of relative cohomology).
We have presented these generators as cobordisms with time running vertically from bottom to top: these are  linear maps from the Hilbert space of initial to final circles.  They defined the \emph{closed string} algebra on the Hilbert space, with product given by the pair of pants. This is also the operator algebra for closed strings\footnote{This is a spacetime operator algebra, not the operator algebra on the worldsheet}, since there is a state-operator correspondence mapping the pair of pants to a 3-puncture sphere with operators inserted: 
\begin{align}
\mathtikz{ \deltaC{0}{0}; 
  \node at (-.5,-1.3) {$i$}; 
  \node at (.5,-1.3) {$j$};
  \node at (0,.5) {$k$}; 
  \node at (-1,0) {$(0,1)$};  } =  \vcenter{\hbox{\includegraphics[scale=.2]{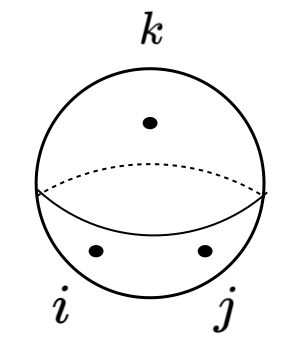}}}
\end{align}

Note that with respect to this choice of product, the unit element is not one of the generators displayed in \eqref{buildingblocks} , but it can be obtained from them by gluing:
\begin{align}
      \mathtikz{\epsilonC{0}{0};
\node at (0,1/2){(0,-1)}; \node at (0,-.5) {\text{unit}} }=\mathtikz{\epsilonC{0}{-1cm} \muC{0}{0} \etaC{-.5cm}{0cm} ;\node at (0,-1.7) {(-1,0)};\node at (1.2, -.3) {(1,0)}; \node at (-.5,.5) {(0,-1)}}
\end{align}
\subsection{Hilbert space of closed strings}
We now  define the closed string Hilbert space assigned to a 
TQFT state cut, which lives on a circle in $\Sigma$.  
The elements of this Hilbert space are boundary conditions for the closed string\footnote{this is similar to how we define states as boundary conditions in a spacetime path integral.}: for a worldsheet corresponding to a degree $n$ map, the boundary condition is given by a set of winding numbers that sums to $n$.   This is determined  by a set of occupation numbers $k_{m}$ specifying how many strings there are with winding number $m$.  These satisfy:
\begin{align}\label{mkn}
    \sum_{m=1}^{\infty}  m k_{m} = n 
\end{align}
The set of all such occupation numbers for arbitrary $n$ can be identified with the occupation numbers for a chiral boson.  We can therefore define the closed string Hilbert space $\mathcal{H}_{\text{closed}}$ as a fock space of states:
\begin{align}
    \ket{\vec{k}} = \prod_{m=1}^{\infty} \alpha_{-m}^{k_{m}} \ket{0} ,\qquad [ \alpha_{m},\alpha_{n}]= m \delta_{n+m}
\end{align}
We will refer to $\ket{\vec{k}}$ as the winding basis.  This basis can be represented by power sums, which span the space of symmetric functions of infinitely many variables. Defining these variables by $x_{i}$, the power sums are 
$p_{k}(x) =\sum_{i=1}^\infty x_{i}^k $
and the wavefunctions of the winding basis are
\begin{align}
    \braket{\vec{x}|\vec{k}} = \prod_{m=1} (p_{m} (\vec{x}))^{k_{m}} 
\end{align}
In this representation, $\alpha_{-m}$ and  $\alpha_{m}$ acts as
\begin{align}
    \alpha_{-m}= p_{m}(x) , \qquad  \alpha_{m} = m \frac{\pd}{\pd p_{m} },\quad m>0
\end{align}
The standard scalar product on symmetric functions given by
\begin{align}\label{hall} \braket{\vec{k}|\vec{k'}} = \delta_{kk'} z_{\vec{k} },
\quad z_{\vec{k}} = \prod_{m=1} m^{k_{m}}(k_{m}!)
\end{align}
We refer to this as the Hall product.  It is
consistent with 
\begin{align}
    \alpha_{-m}^{\dagger} = \alpha_{m},
\end{align}
but note that there is no complex conjugation in the adjoint, because the scalar product is not sesquilinear\footnote{The power sums actually span the space of symmetric functions with rational coefficients, consistent with a real vector space that doesn't require a sesquilinear inner product. }.  
\paragraph{The representation basis}
It will be useful to go to a basis that diagonalizes the pair of pants.   To define this basis, notice that   each $\{k_{m} \}$ satisfying \eqref{mkn}  specifies  a partition of a positive integer $n$, which describes the conjugacy class of a permutation in $S_{n}$.  The permutation is the deck transformation that describe how different branches of the worldsheet are permuted as we go around a pinch point $p$ on the $\Sigma$.  The identification of $\vec{k}$ with a conjugacy class of a permutation allows us to define a basis labelled by a Young Tableaux $R$:
\begin{align} \label{frob}
    \ket{R}= \sum_{\vec{k} \subset S_{n}} \frac{\chi_{R}(\vec{k})}{z_{\vec{k}}}   \ket{\vec{k}}, \qquad \braket{R|R'} =\delta_{RR'}
\end{align}
Here $\chi_{R}(\vec{k})$ is a character of the symmetric group $S_{n}$ , evaluated on the conjugacy class represented by $\vec{k}$.  
The wavefunctions in the representation basis are given by the Schur polynomials
\begin{align}\label{schur}
    s_{R}(x) = \braket{x|R} = \frac{\det x_{j}^{R_{i}-i}}{\det x_{j}^{-i}},\qquad  i,j=1,\cdots \infty 
\end{align}
where $R_{i}$ are the row lengths of the tableaux.  This ratio of Vandemonde determinants arise from the fermionization of the chiral boson. \eqref{unit}
\subsection{TQFT generators and partition functions}
In the representation basis, the generators of the TQFT are given by \cite{Vafa:2004qa,Bryan:2004iq,Donnelly:2020teo}
\begin{align} 
 \mathtikz{\epsilonC{-1}{0};
\node at (0,1/2){(-1,0)};} &= \sum_{R} d_{q}(R) (-1)^{l(R) }q^{-\kappa_{R}/4} e^{-t l(R)} \ket{R}  \label{unit} \\ 
 \mathtikz{\etaC{0}{0};
\node at (0,1/2){(0,-1)};} &= \sum_{R}  d_{q}(R) q^{\kappa_{R}/4} e^{-t l(R)} \bra{R} 
\label{counit} \\
\mathtikz{\muC{0}{0};
\node at (0,1/2){(1,0)};} &=\sum_{R} \frac{
1}{d_{q}(R)} (-1)^{l(R)} q^{\kappa_{R}/4} e^{-t l(R)} \ket{R} \bra{R} \bra{R}
 \label{pants} \\ \mathtikz{\deltaC{0}{0};
\node at (0,1/2){(0,1)};} &=\sum_{R} \frac{1}{d_{q}(R)} q^{-\kappa_{R}/4} e^{-t l(R)} \ket{R} \ket{R} \bra{R}
 \label{cop} \\
\mathtikz{\idC{0}{0} ;\node at (0,1/2){(0,0)}}&= e^{-t \hat{H}} =\sum_{R} e^{-t l(R)}  \ket{R}  \bra{R}   \label{id},
\end{align}
where $l(R)$ is the number of boxes in the tableaux, $\kappa_{R}$ is a quantity  defined in terms of row and columm numbers $i(\Box), j(\Box)$ \footnote{Via Schur Weyl duality $\kappa_{R}$ is also related to the Casimir $C_{2}(R)$ of $U(N)$ via
\begin{align}
    C_{2}(R)= \kappa_{R} + N l(R)
\end{align}}
\begin{align}
    \kappa_{R} = 2 \sum_{\Box \in R} (i(\Box) - j(\Box) ), 
\end{align}
and $d_{q}(R)$ is the quantum dimension of the symmetric group\footnote{In \cite{Aganagic:2004js}, $d_{q}(R)$ is defined without the factor of $i^{l(R)}$}
\begin{align}
   d_{q}(R) &= \prod_{\Box \in R}  \frac{1}{q^{ h(\Box)/2}-q^{- h(\Box)/2}}=\prod_{\Box  \in R} \frac{1}{2 \sinh(\frac{h(\Box) g_s}{2})} ,\quad q=e^{g_{s}}
\end{align}
The ``Calabi-Yau caps" \eqref{unit}, \eqref{counit} can be identified with the topological vertex \cite{Aganagic:2003db}, and the unsual factors of $(-1)^{l(R)}$ come from the vertex gluing rules: we will elaborate on this in the next section.  Finally, note that there is a normalization constant ( independent of the Kahler modulus $t$) that is omitted from these cobordisms, because these are non-local with respect to the TQFT gluing rules.   These can be easily restored by hand,  and will be relevant when we consider open-closed dualities in section \ref{section:transition}

\paragraph{The partition function on a general manifold }
We can obtain the partition function on a general manifold by gluing the TQFT generators.   In this process, it is useful to organize the bundle data in terms of the sum and differences of the Euler classes. The former gives the Euler characteristic, while the latter determines the relative twisting of the bundles.  To change the relative twist systematically, note that gluing the $(-1, 0)$ cap to the $(0,1)$ pair of pants produces the $(-1.1)$ cylinder, which can be used to increase  $k_{2}-k_{1}$ by an arbitrary positive even integer\footnote{In \cite{Donnelly:2020teo}, there was a different choice of generators in which the Calabi Yau caps, had the opposite Euler classes, so they had opposite signs in front of $\kappa_{R}$.  The resulting cobordisms only generated  subclass of CY's because the $(1,-1)$ cylinder was not generated.}. Similarly, there is a $(1,-1)$ cylinder that changes the relative twist in the opposite direction:  
\begin{align}\label{tcyl}
 \mathtikz{\node at (0,.5){(-1,1)};\idC{0cm}{0cm}} = \sum_{R}  (-1)^{l(R)} q^{-\kappa_{R}/2} \ket{R}\bra{R},\quad  \mathtikz{\node at (0,.5){(1,-1)};\idC{0cm}{0cm}} = \sum_{R}   (-1)^{l(R)}q^{\kappa_{R}/2} \ket{R}\bra{R}
\end{align}
On the other hand, to adjust the Euler characteristic $k_{1}+k_{2}$ we can insert a handle adding operator that increases the genus of the manifold:
\begin{align}
    \mathtikz{\muC{0}{0};\deltaC{0}{1cm};
\node at (-1.5cm ,0 cm){(1,1)}} &= \sum_{R} (-1)^{l(R)}\frac{1}{ (d_{q}(R))^2} \ket{R} \bra{R} \nn
&=\sum_{R} \frac{1}{ (\tilde{d}_{q}(R))^2} \ket{R} \bra{R} 
\end{align}
where we defined the modified dimensions
\begin{align}
\tilde{d}_{q}(R)&= \prod_{\Box \in R}  \frac{i}{q^{ h(\Box)/2}-q^{- h(\Box)/2}}=\prod_{\Box  \in R} \frac{1}{2 \sin(\frac{h(\Box) g_s}{2})} ,
\end{align}
to simplify the formulas and  for ease of comparison with \cite{Bryan:2004iq}.
Gluing $g$ of these handles together with the $(0,-1)$ and $(-1,0)$ Calabi Yau caps gives a target space over a genus $g$ surface  with $(k_{1},k_{2})= (g-1,g-1)$ Euler classes.   We can then obtain any even integer value of  $k_{1}-k_{2}$  by  inserting $\frac{k_{1}-k_{2}}{2}$ cylinders of the form \eqref{tcyl}. This gives the partition function
\begin{align}\label{closedamplitude}
    Z(\Sigma_{g}, k_{1}, k_{2}) = \sum_{R}  (-1)^{l(R)\frac{(k_{1}-k_{2})}{2} } \left(\frac{1}{\tilde{d}_{q}(R)}\right)^{2g-2} q^{\frac{(k_{1}-k_{2})}{4}\kappa_{R} } e^{-tl(R)},
\end{align}  
which matches with the partition function in Corollary 7.2 of \cite{Bryan:2004iq}, obtained from (local) Gromov-Witten theory\footnote{This is the partition function for a special case of the equivariant parameters $(t_{1},t_{2})$ in \cite{Bryan:2004iq}, for which $t_{1}=-t_{2}$ } .   Finally, note that for real $q$ the all of the phases in the summand are of the form  
\begin{align}
    e^{ i\pi  (\frac{k_{1}-k_{2} }{2} +g-1) l(R) }
\end{align}
  These phases can be absorbed into an imaginary shift  of $t$ corresponding a shift of the background $B$ field.  Perhaps for this reason, they were exclude in reference \cite{Aganagic:2004js}. However they are present in \cite{Bryan:2004iq} and they do affect the gravitational entropy due to their dependence on the Euler classes of the background.   

\subsection{Gravitational entropy for closed strings }\label{section:replica}
We now define a notion of gravitational entropy for the topological string, generalizing the proposal in \cite{Donnelly:2020teo,Jiang:2020cqo}.  The essential distinction between gravity and QFT is that gravitational entropy is obtained by differentiating an  \emph{on-shell} gravitational path integral evaluated on smooth replica
geometries.   In the A model TQFT, we will use the Calabi-Yau condition as an on-shell condition for the topological string, and define a gravitational replica trick that allows the geometry to back react in order remain a Calabi Yau.   This idea was implemented in \cite{Donnelly:2020teo,Jiang:2020cqo} for the simplest case when $\Sigma=S^2$, and the replica manifold is branched over a single connected region.  In this case, the replica manifold has a $U(1)$ symmetry so that the gravitational replica trick reduces to a Gibbons Hawking entropy.  Below, we generalize the construction to arbitrary $\Sigma$ and subregions.

\subsection*{The gravitational replica trick}

Consider Calabi--Yau threefold M of the form \eqref{bundle}, fibered over a genus-$g$ surface $\Sigma_g$.   We want to define a replica manifold for $\Sigma$, and then extend it to a replica for $M$ by specifying the replica bundle structure.  \cite{Donnelly:2020teo,Jiang:2020cqo} did this for a single  subregion in $\Sigma=S^2$.   Here we generalize this to arbitrary subregions and arbitrary $\Sigma_{g}$.

Recall the replica prescription for a 2D surface $\Sigma_{g}$.  If $\Sigma_g$ admits a time-reflection symmetric slice $S$ with $m$ intervals, the replica surface $\Sigma_{g,n}$ is the $n$-fold branched cover of $\Sigma_g$ over the $2m$ endpoints. 
Its Euler characteristic is given by 
\begin{align} \label{chin}
\chi(\Sigma_{g} (n,m)) = n(2-2g) - 2m(n-1) 
\end{align}

To extend the replica construction to the total space $M$, we must determine the Euler class data $(k_1(n),k_2(n))$ of the bundle over $\Sigma_{g}(n,m)$, given $(k_1,k_2)$ over $\Sigma_{g,1}$. 
The Calabi--Yau condition requires
\begin{align}\label{CYreplica}
k_1(n) + k_2(n) = -\chi(\Sigma_{g}(n,m)) .
\end{align}
This differs from the canonical replica bundle obtained pulling back the $n=1$ spacetime along the projection $\pi:\Sigma_{g}(n,m) \to \Sigma_{g}$, which would give $k_{1}(n)=n k_{1}$ and $k_{2}(n)=n k_{2}$ 
\footnote{$k_{1}$ and $k_{2}$ counts zeroes of a section of the bundle, and these zeros naturally add under replication.}.  We regard \eqref{CYreplica} as the effect of gravitational backreaction, which modifies the bundle data relative to the pull back bundle.

The single constraint \eqref{CYreplica} leaves the difference $k_1(n) - k_2(n)$ undetermined.  Naively, this allows for  different families of replica manifolds parameterized by this function of $n$.   However, the replica construction should not introduce arbitrary new data unrelated to the entanglement structure of a fixed state.  To resolve this issue, we fix the ambiguity by requiring that 
\begin{align} \label{deltak}
    k_{1}(n)-k_{2}(n) =n(k_{1}-k_{2}).
\end{align}
This choice is symmetric between $k_{1}$ and $k_{2}$. Combined with the CY condition, this implies that
\begin{align}
k_{1}(n) &= n k_{1}  +m (n-1)\nn
k_{2}(n) &= n k_{2} + m(n-1)
\end{align}
so that the replication is symmetric in the two line bundles.

We can construct these replica manifolds from TQFT generators.
To see how this works, consider  a Calabi--Yau manifold $M$ with Chern class data $(k_1,k_2)$ fibered over $\Sigma_g$.  We construct the $n$th replica by taking $n$ copies of $\Sigma_g$, and introducing a hole on each copy for each cut, and gluing them cyclically. 

For example, 
 the $n=2$, $m=1$ replica for a  genus 2 surface is takes the form
 \begin{align} 
\vcenter{\hbox{\includegraphics[scale=.1]{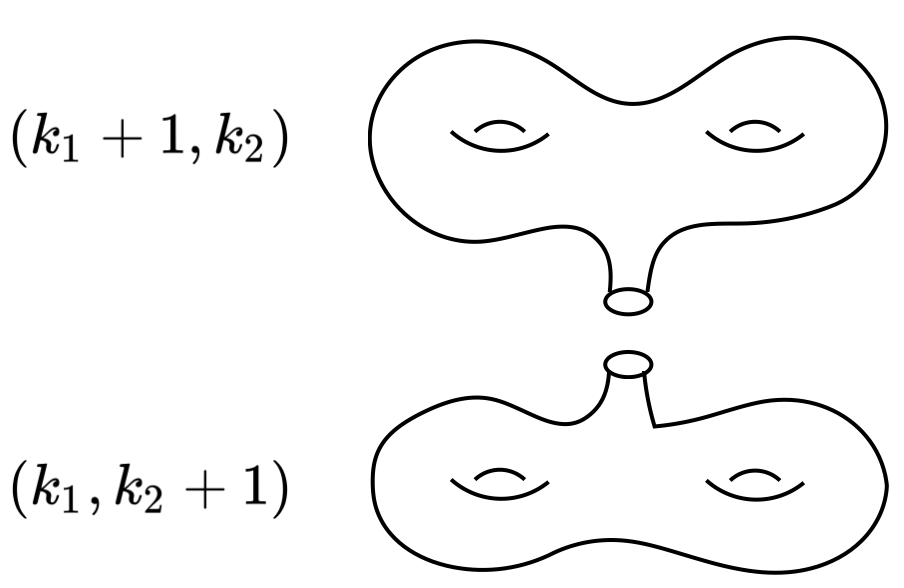} }} = \vcenter{\hbox{\includegraphics[scale=.15]{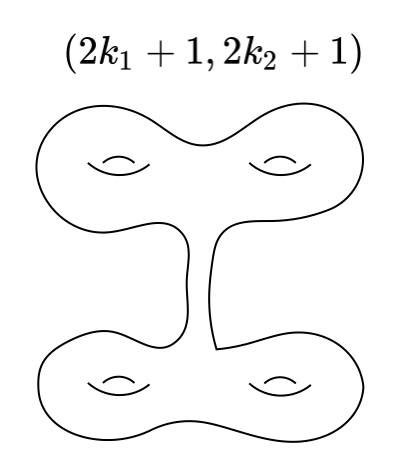}}}
 \end{align} 
Note that on each replica, we could have opened up the branch cut into a hole by removing either a  $(0,-1)$ cap or a $(-1,0)$ from the genus 2 surface.  Either choice is consistent with the CY condition, and we have picked one that ensure that \eqref{deltak} is satisfied.
Similarly,  the $n=3$, and $n=4$ replica are given by:
\begin{align}
\vcenter{\hbox{\includegraphics[scale=.13]{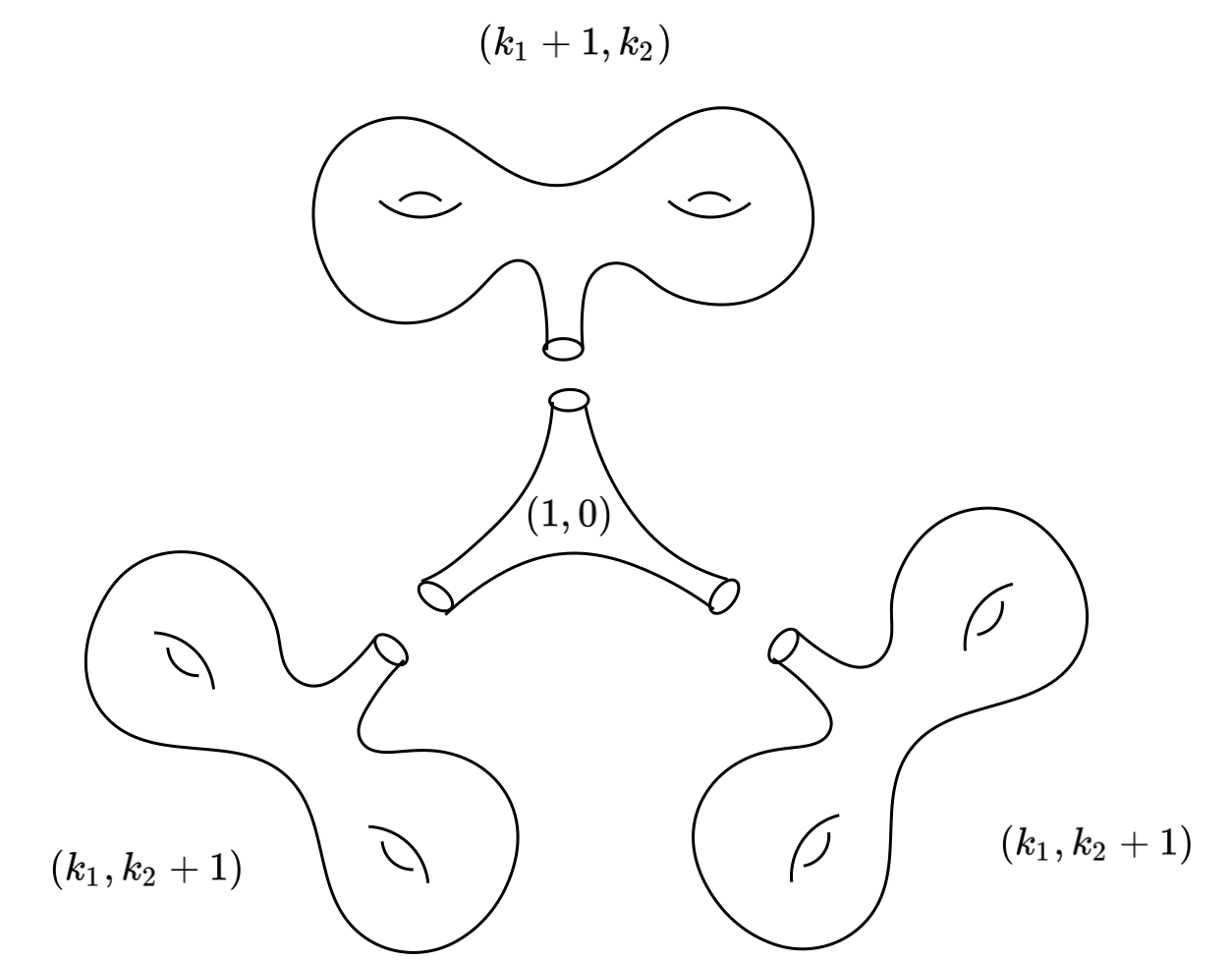} }} ,\qquad   \vcenter{\hbox{\includegraphics[scale=.2]{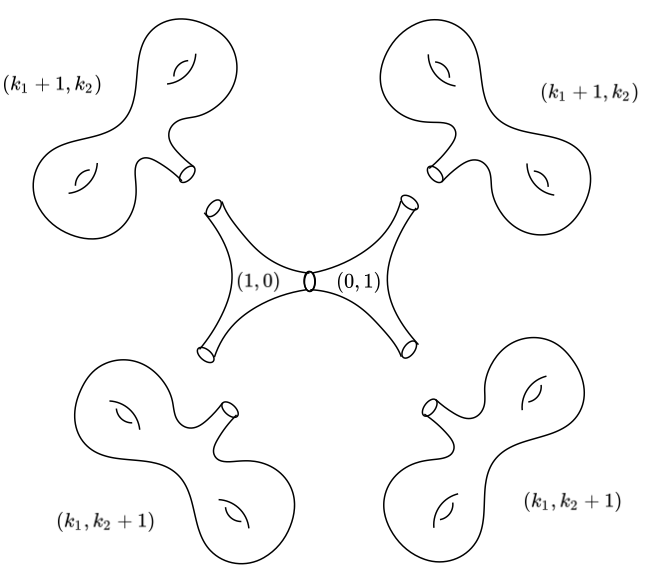}}}
\end{align} 
This leads to the following replica partition function for a single subregion:
\begin{align}
    Z(\Sigma_{g,n}, k_{1}, k_{2} ) = \sum_{R}  (-1)^{l(R)\frac{(k_{1}-k_{2})}{2}} (\tilde{d}_{q}(R))^{2 - 2ng } q^{\frac{n\kappa_{R}}{4} (k_{1}-k_{2})} e^{-nt l(R)}
\end{align}

We define the standard replica entropy by
\begin{align}\label{reptrick}
S_{\text{replica}}
= -\left.\frac{d}{dn}\right|_{n=1}
\frac{Z(\Sigma_{g,n})}{(Z(\Sigma_g))^n}
= \left.(1 - n\partial_n)\right|_{n=1}
\log Z(\Sigma_{g,n}) \, .
\end{align}
For a single connected subregion, this gives:
\begin{align}
    S_{\text{replica}} (m=1)=  -\sum_{R} P(R) 
    \log P(R) +  2\sum_{R} P(R) \log (\tilde{d
    }_{q}(R)),
\end{align}
where $P(R)$ is defined from the partition function at $n=1$:
\begin{align}
    P(R) = \frac{ (-1)^{l(R)\frac{(k_{1}-k_{2})}{2} } (\tilde{d}_{q}(R))^{2-2g}q^{\frac{\kappa_R}{4}(k_1- k_2)}
\, e^{-t\, l(R)}}{Z(\Sigma_{g})}
\end{align}
This is directly analogous to the standard form for the entanglement entropy of a non abelian   gauge theory, with the $\log \tilde{d}_{q}(R) $  term giving the edge mode contribution\footnote{Due to the phases,  this doesn't have the same positive properties as in gauge theory because of the non unitary features of the perturbative topological string amplitudes.  In non-unitary theories, generically complex values of entropies are allowed.  However we expect positivity to be restored by non-perturbative corrections }.   
\subsection{Multiple subregions}
We can generalize this  replica construction to $m$ disconnected subregions as follows.  For each of the $m$ branch surfaces, we connect their $n$ replicas in a cyclicly invariant way, while letting the bundle structure backreact symmetrically in $k_{1}(n)$ and $k_{2}(n)$.   For example, we start with the $g=2$ surface with $m=2$ subregions, the $n=3$ replica would be given by :
\begin{align}
   Z(\Sigma_{2}(n=3,m=2), k_{1},k_{2} )=  \vcenter{\hbox{\includegraphics[scale=.15]{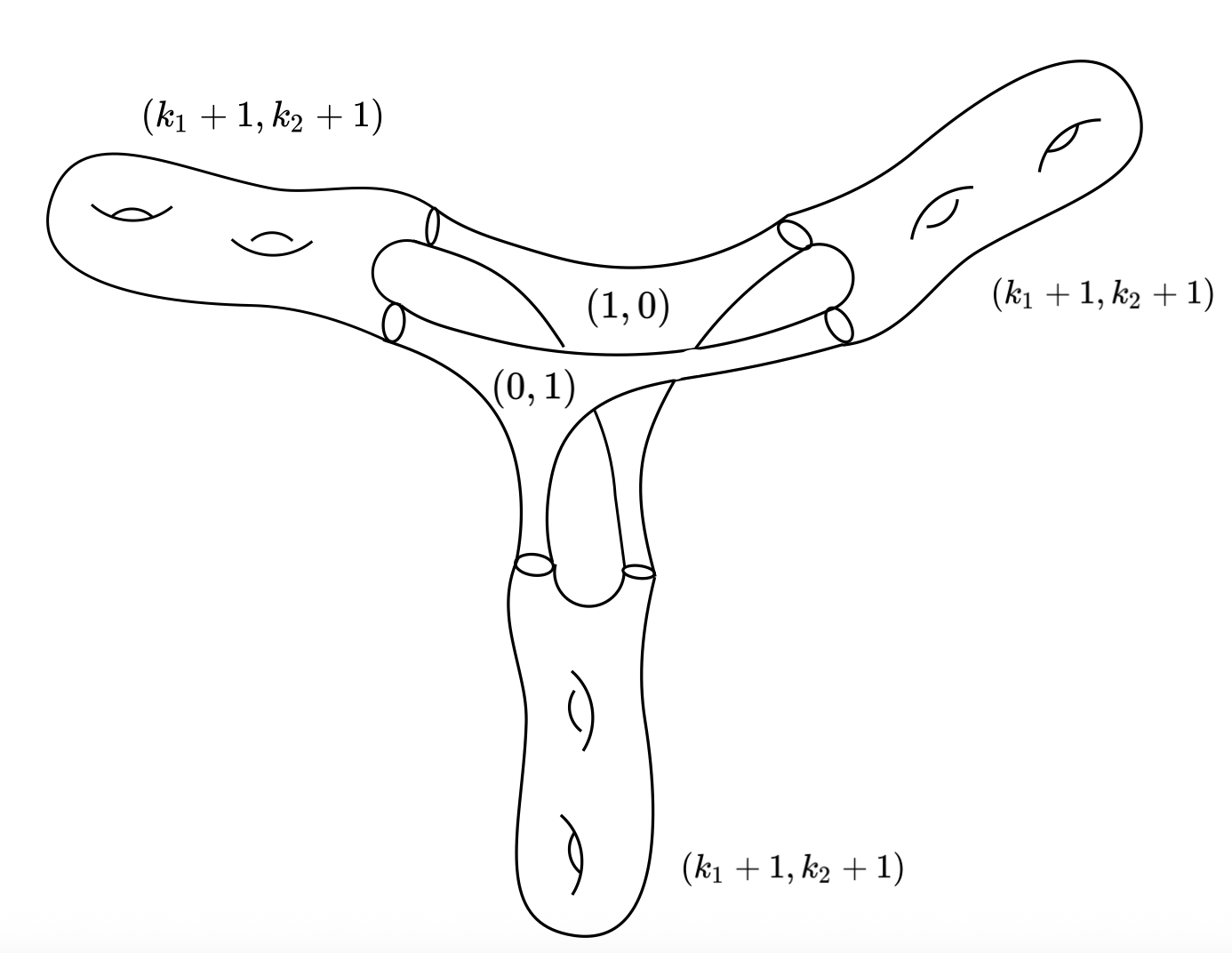}}}
\end{align}
With this prescription, one finds again that
\begin{align}
    k_{1}(n)- k_{2}(n)= n(k_{1}-k_{2} )
\end{align}
The resulting partition function is 
\begin{align}
 Z(\Sigma_{2}(n,m) ,k_{1},k_{2}) =   Z(\Sigma_{g,n}, k_{1}, k_{2} ) = \sum_{R}(-1)^{l(R)\frac{(k_{1}-k_{2})}{2} }  (\tilde{d}_{q}(R))^{n(2-2g) - 2m(n-1)  } q^{\frac{n\kappa_{R}}{4} (k_{1}-k_{2})} e^{-nt l(R)}   
\end{align}
which has replica entropy\footnote{ In a little more detail, using our definition of $P(R)$ , straightforward differentiation gives terms that we can organize as \begin{align}
(1-n\partial_{n} )\log Z(n) = \log Z - \sum_{R}  P(R) (\chi-(2m)) \log d_{q}(R) - P(R) ( \log ( ZP(R)) - \chi \log d_{R})     \end{align}
 There is a cancellation between the two terms with weighted by $\chi$, so that we get the formula as stated in the text. We can express the shannon entropy and edge mode entropy as
 \begin{align}
     S_{\text{shannon} }&=  ( 1-\chi \partial_{\chi } - t \partial_{t} -g_{s} \partial g_{s} )\log Z\nn
     S_{\text{edge} }&=  m \partial_{\chi} \log Z 
 \end{align}}\begin{align}\label{MEE}       S_{\text{replica}} (m)=  -\sum_{R} P(R) 
    \log P(R) + 2m \sum_{R} P(R) \log (\tilde{d}_{q}(R)),
 \end{align}
 where $P(R)$ is defined as before.  The coefficient of the edge mode entropy $\log d_{q}R$ indicates the number of entangling surfaces. 

\section{Local holography and Open-Closed TQFT}\label{sec:extendedTQFT}
To give a statistical interpretation to the gravitational entropies \eqref{MEE}, we introduce a stretched entangling surface on $\Sigma_{g}$ by removing an infinitesmal disk around each entangling point.  If this boundary satisfies the shrinking property,  the resulting target space can be interpreted as a trace of a modular operator on the subregion.   
As we show below, this condition is precisely what promotes the replica entropy to a genuine entanglement entropy.

In  string theory, introducing this boundary is equivalent to local holography: the stretched entangling surface is the support for holographic degrees of freedom, whose backreaction produces a dual target space where the hole is filled in.   The entanglement brane is then defined as the spacetime boundary state that  realizes this filling \cite{Donnelly:2018ppr}.  
\paragraph{Extended TQFT }
In the A model, this heuristic picture has a precise algebraic implementation: introducing shrinkable holes in $\Sigma_g$ is part of an extension of the closed string TQFT to an open-closed TQFT.  To see how this works, consider the resolved conifold.  The total spacetime is a $(-1,-1)$ bundle over a sphere, which is the topological string analogue of euclidean dS$_2$:
\begin{align}
   Z_{res} (t) = \mathtikz{ \epsilonC{0cm}{0cm} \etaC{0cm}{0cm}; \node at (0,1/2){(-1,-1)} } 
\end{align}
$t$ is the area for the sphere.  For the simple case of a single interval subregion, inserting two infinitesmal holes around the entangling points produces an annulus topology. 
The TQFT gluing rule
\begin{align}\label{TQFTglue}
\vcenter{\hbox{\includegraphics[scale=.15]{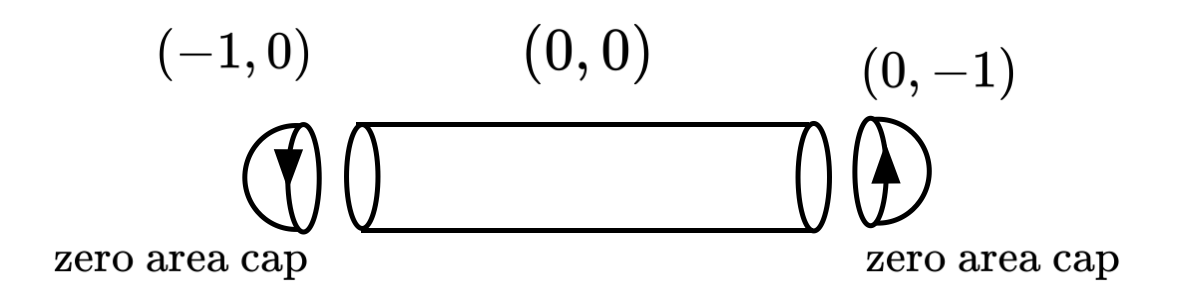}}}=
\mathtikz{ \epsilonC{0cm}{0cm} \etaC{0cm}{0cm};\node at (0,.5) {\footnotesize($-1, -1$)}} 
\end{align} 
imply that the exact boundary state around these holes is given by the zero-area limit of the Calabi–Yau caps\footnote{ For the purposes of reproducing the resolved conifold by gluing, we could have also removed two $(-1,0)$ $(-1,0)$ caps, which would result in an $1.-1$ cylinder. However, we will see that an interpretation of this equality in terms of geometric transitions requires us to remove the $(-1,0)$ and $(0,-1)$ cap. }, which we  denote by $\ket{e}$ and $\bra{\tilde{e}}$ respectively.  They are given by the coherent states \cite{Donnelly:2020teo}:
\begin{align}
  \label{CYcaps}
    \ket{e} &= \exp ( \sum_{n>0} \frac{\alpha_{-n}}{n (q^{n/2} -q^{-n/2})})\ket{0} \nn
     \bra{\tilde{e}} &= \bra{0}\exp ( \sum_{m>0} \frac{-\alpha_{m}}{m (q^{m/2} -q^{-m/2})}) 
\end{align}
Note the shift in background upon integrating out the the Calabi Yau caps: the $(0,0)$ cylinder is a patch of $\mathbb{C}^3$, and the TQFT gluing rules tell us this is dual to the resolved conifold.

A more nontrivial requirement of this boundary state is that it leads to  a trace over the  Hilbert space on an interval subregion
\begin{align}\label{Hopen}
\mathcal{H}_{\text{open}} =\vcenter{\hbox{\includegraphics[scale=.2]{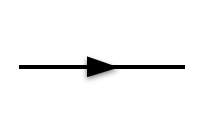}}},
\end{align}
which describes a time slice in the ``open" channel.

In the TQFT framework, the introduction of $\mathcal{H}_{\text{open}}$ is part of the  extension of the closed algebra to an \emph{open algebra }containing new generators:
\begin{align}\label{openfrob}
    \mathtikz{\muA{0}{0}}, \quad \mathtikz{\deltaA{0}{0}}, \quad \mathtikz{ \etaA{0}{0} }, \quad \mathtikz{ \epsilonA{0}{0} },
\end{align}
with the open string fusion defining the product.  Moreover, there are homomorphisms between the closed and open algebras defined by
\begin{align}
    \mathtikz{\zipper{0}{0}}, \quad \mathtikz{\cozipper{0}{0}} 
\end{align}
which allow for interactions between the open and closed algebra that must satisfy a set of compatibility conditions called the Moore-Segal axioms\cite{Moore:2006dw}.  These provide a set of ``re-write" rules that we can use to simplify the diagrammatics of the open-closed algebra. In addition to these relations,  \cite{Donnelly_2019} introduced the entanglement brane axiom:
\begin{align}
\mathtikz{ \epsilonA{0cm}{-1cm} \zipper{0cm}{0cm} }=\mathtikz{\epsilonC{0}{0}  }.
\end{align} 
This axiom
ensures that any holes traced out by the entanglement boundary can be closed. Since our closed TQFT is decorated by Euler classes, we need to decide which cap appears on the RHS of this equation. We will see that a consistent choice is given by \footnote{Note that this cap state is  \emph{not} the unit of the closed algebra with respect to the $(0,1)$ pair of pants} :
 \begin{align}
\mathtikz{ \epsilonA{0cm}{-1cm} \zipper{0cm}{0cm} }=\mathtikz{\epsilonC{0}{0}  ; \node at (1,0) {$(-1,0)$}}
\end{align} 

Given such an extension, the requirement that the entanglememt brane boundary states define a trace becomes a spacetime version of the Cardy condition 

\begin{align}\label{cardy}
\mathtikz{\HidC{0cm}{0cm};\node at (0.5,-.7) {($0,0$)} }
=\mathtikz{ \pairA{0cm}{0cm} \copairA{0cm}{0cm};\node at (0,-1.2) {($0,0$)}  } ,
\end{align}
where we denoted the boundary states by $\ket{\tilde{e}}$ and $\bra{e}$ respectively.  
The RHS is a trace of a propagator $\rho =e^{-tH_{\text{open}}}$ on $\mathcal{H}_{\text{open}}$ now interpreted as a reduced density matrix.

 \paragraph{Extensions with Branes} 
 From a purely mathematical point of view, the most elegant extension of the A model TQFT would involve the q-deformed symmetric group algebra , since closed string states are labelled by permutations and the symmetric group quantum dimensions appear prominently n the partition functions.  However, we will take an alternative approach in order to connect to the physics of D branes.  Mathematically, this corresponds to formulating the extension using algebras of functions on U(N), with $N\to \infty$. The connection to symmetric group then arises via a q-deformed version Schur-Weyl duality, which relates q-deformed $U(N)$ and symmetric group representations. 

Physically, this corresponds to introducing a large N stack of non-compact Lagrangian branes at the stretched entangling surface (the endpoints of the interval in \eqref{Hopen}), whose worldvolume theory is  U(N) Chern-Simons theory.
This is a natural extension of the proposal in \cite{Aganagic:2004js}, which uses the same branes to define codimension-one state cuts in the TQFT.  However, what distiguishes the branes inserted by the entanglement boundary states $\ket{e}$ is that their worldvolume holonomy has to be tuned to a particular value, given by a diagonal matrix $U_{0}$ with matrix elements :
\begin{align}
   (U_{0})_{ij} = \delta_{ij} q^{-i+1/2} \qquad i,j=1,\cdots \infty
\end{align}We will refer non-compact Lagrangian branes with this holonomy as entanglement branes.  These are the stringy degrees of freedom that  underlies the shrinkable boundary condition.  They elevate the Cardy condition in \eqref{cardy} to spacetime duality between an open and closed string background:
\begin{align}
    \label{eshrink}
 \mathtikz{  \epsilonC{0cm}{0cm} \etaC{0cm}{0cm};\node at (0,-.7) {($-1, -1$)}  }=
    \mathtikz{ \pairA{0cm}{0cm} \copairA{0cm}{0cm};\node at (0,-1.2) {($0,0$)}  } 
\end{align}
In the string theory intepretation, the  annulus diagram is the total amplitude for (multiple) strings stretched between two stacks of branes in $\mathbb{C}^3$, while the  sphere diagram is the closed string amplitude on the resolved conifold. 
Their equality follows from a geometric transition in which the entanglement branes dissolve into closed string flux.   These transitions are closely related to the ones studied in \cite{Gomis:2007kz,Okuda:2007ai}: in that framework, the spacetime Cardy condition is the statement that the resolved conifold is a bubbling Calabi Yau arising from a particular configuration of  branes on $\mathbb{C}^3$
\paragraph{Generalized Cardy condition and statistical interpretation of gravitational entropy} 
For a general Calabi Yau fibered over a genus $g$ Riemann surface, the analogue of our spacetime Cardy condition becomes substantially more complex. For example,  consider a genus $2$ manifold, with two entanglememt brane boundaries (blue for branes, and red for negative branes).
\begin{align}
\vcenter{\hbox{\includegraphics[scale=.2]{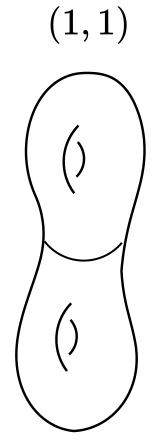}}} = \vcenter{\hbox{\includegraphics[scale=.12]{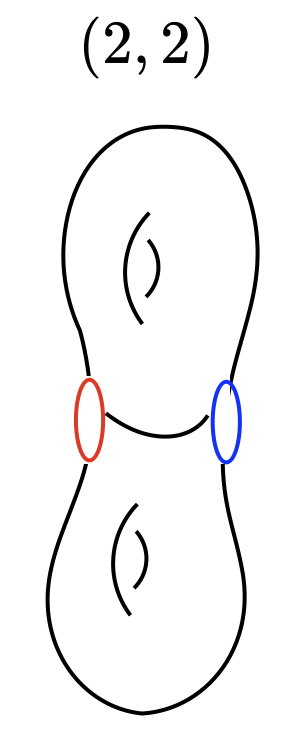}}} ,
\end{align}

The generalized Cardy condition requires that this is the trace of a reduced density matrix $\rho$ on $\mathcal{H}_{\text{open}}$, which now acts as a modular operator that implements a topologically nontrivial evolution. 
We can see this more explcitly, by  creating the two shrinkable holes using the zippers and the open product/co-product:
\begin{align}
    \mathtikz { \vthird{
      \muC{0}{5.4cm}
       \copairC{0}{5.5cm}
      \zipper{0}{4.4cm}  } \vhalf{ \deltaA{0}{2cm}} 
    \vhalf{ \muA{0}{1cm}} \vthird{
      \pairC{0}{-2cm}
      \deltaC{0}{-1cm}
      \cozipper{0}{0} };\node at (1cm,.65cm) {\footnotesize $V$} } 
\end{align}
This diagram makes manifest that the modular flow evolving $V$ back to itself involves a splitting of the interval and then an interaction with an operator produced by the torus with a single boundary:
\begin{align}
    \mathtikz{ \pairC{0}{-2cm}
      \deltaC{0}{-1cm}
      \cozipper{0}{0} }
\end{align}
  
To obtain the modular operator  as an element of the open algebra  $\mathcal{H}_{\text{open}}$ we need to write this as a purely open cobordism.  Fortunately, it was shown in \cite{Donnelly_2019}  that such a re-write is always possible given an extension of the closed TQFT that satisfies the Moore-Segal and entanglement brane axiom.   Such topologically nontrivial modular flows also arise when we introducing multiple subregion on the resolved conifold:
\begin{align}
\vcenter{\hbox{\includegraphics[scale=.15]{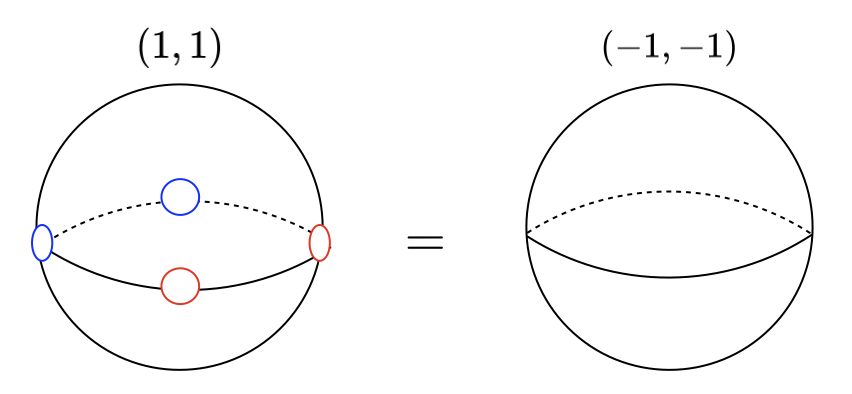}}}  
\end{align}
The modular evolution of the two disjoint interval now involves a cutting and regluing of the interval \cite{Donnelly_2019}. We will explain this in more detail in section \ref{sec:opensubregion} 

Once we have obtained the modular operator $\rho$ on $\mathcal{H}_{\text{open}}$,  we can provide a statistical interpretation of the gravitational entropy by provided we can show that the replica partition function satisfies 
\begin{align} \label{Zntrace}
    Z(n) =  \widetilde{\tr} \rho^n,
\end{align}
where the tilde on the trace accommodates the scenario where fails to be a conventional ``Type I " trace of an ordinary quantum mechanical system.  Indeed, this is expected given shrinkable boundary at the endpoints of the interval. 

\eqref{Zntrace} is not entirely obvious from the definition of the replica trick, because of the higher dimensional nature of the target space and the backreaction that modifies the topology of the total replica Calabi Yau.  However, at the level of 2D topology, this works because gluing $\rho$ $n$ times and then shrinking the $m$ holes around the entangling surface reproduces the 2D replica topology that we defined in section \ref{section:replica}.  This takes some effort to visualize, but one can see easily that the Euler characteristic works out:  by first gluing $\rho$ $n$ times and then filling in the hole, we avoid the naive over-counting that comes with replicating the entangling surface, and that is exactly what the Hurwitz formula \eqref{chin} is designed to account for.
The upshot is that any open extension of the A model TQFT leads the statistical formula
\begin{align}\label{Sent}
S_{\text{replica}} = -\widetilde{\tr} \rho \log \rho 
\end{align}
 
\section{ Closed string Wavefunctions as D brane amplitudes } \label{sec:Dbranewfn}
 In this section, we review the interpretation of $\mathcal{H}_{\text{closed}}$ as wavefunctions on the worldvolume of non-compact Lagrangian branes \cite{Aganagic:2004js}.  We explain how the gluing of these wavefunctions is implemented via brane -negative branes annhilation, and how the relative framing of these wavefunctions gives rise to the relative twist $k_{1}-k_{2}$ in the closed string amplitudes \eqref{closedamplitude}.

\subsection{Large N Chern Simons theory on non-compact branes }
As shown in \cite{Aganagic:2004js},  closed string states of the A model TQFT can be reformulated as D brane amplitudes.
On each connected component of a TQFT state cut, represented diagrammaticaly by a circle in the base manifold $\Sigma$, we insert a large N stack of  non compact Lagrangian branes. These are 3-dimensional branes that wrap a Lagrangian manifold\footnote{This is a half dimensional submanifold where the symplectic form vanishes} in the Calabi Yau.  
They have the topology of a solid, non-compact torus given by $\mathbb{C} \times S^1$, where the non contractible cycle is  identified with the TQFT $S^1$:   
\begin{align}
    \includegraphics[scale=.15]{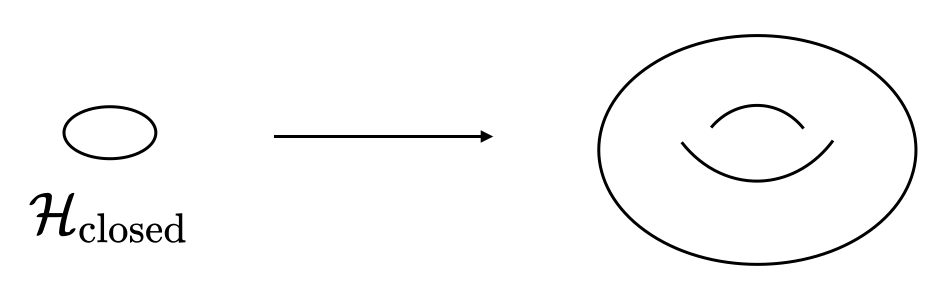}
\end{align}

The worldvolume theory on these branes is $U(N)_{k}$ Chern Simons theory in the large $N$ limit, with the level\footnote{Following \cite{Aganagic:2004js} we do not require $k$ to be an integer. Usually, this is a requirement for the Chern Simons Lagrangian to be gauge invariant.   \cite{Aganagic:2004js}  proposed an analytic continuation to non integer k. }  k related to the  string coupling
\begin{align}\label{q}
      \frac{2 \pi i}{k+N}=g_{s} 
\end{align}
Following \cite{Aganagic:2004js} we identify the chiral boson Hilbert space $\mathcal{H}_{\text{close}}$ with the large N limit of these  D brane amplitudes. Since these non-compact Lagrangians have an asymptotic boundary,  the amplitudes are functions of the Chern Simons boundary condition on the torus boundary.  They can therefore be interpreted as wavefunctions in the large N Chern Simons theory\footnote{The fact that D brane amplitudes can be interpreted as wavefunctions was shown in \cite{Aganagic:2003qj}, where it was found that the amplitudes transform as wavefunctions }. This is shown schematically in the figure below, where an open string worldsheet ending on the branes produce a wavefunction in Chern Simons theory.
\begin{align}
    \includegraphics[scale=.15]{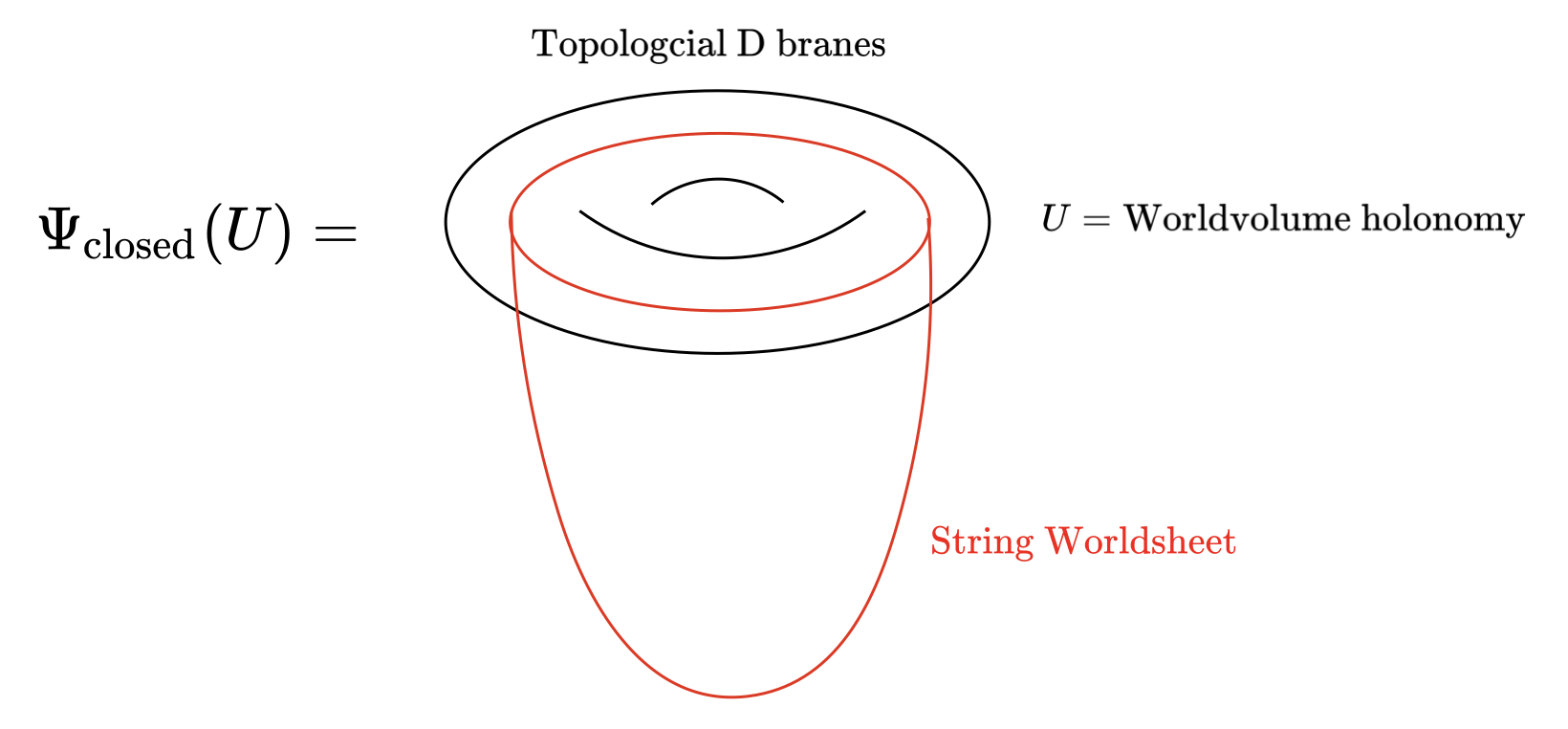}
\end{align}
Thus we identify the closed string Hilbert space and the Chern Simons Hilbert space on the torus
\begin{align}
\mathcal{H}_{\text{closed}}
\;\simeq\;
\lim_{N\to\infty} \mathcal{H}_{T^2}^{\text{CS}}.
\end{align}

The $\simeq$ sign here indicates that there are nontrivial, nonperturbative corrections to this formula.  Omitting these leads to the chiral structure of $\mathcal{H}_{closed}$ and the lack of a sesquilinear inner product. 
\paragraph{Holomorphic factorization at large N}
To understand the origin of these corrections recall the torus  Hilbert space $\mathcal{H}_{T^2}^{\text{CS}}$ at finite N.   This is spanned by a basis of states $\ket{R}_{CS}$ prepared by  a solid torus path integral with a Wilson loop inserted in an integrable representation $R$.  
To match with the topological string amplitudes, it is natural to specify the boundary condition on the torus by the holonomy $U$  around a non-contractible cycle. This choice of this cycle is related to the  choice of framing, which we will discuss at length below \footnote{More generally, we change  this cycle by an $SL(2,\mathbb{Z})$. A general choice of cycle and it's conjugate corresponds to  choice of framing of the Chern Simons amplitude that defines the wavefunction.}.  The corresponding wavefunctions are characters of $U(N)$
\begin{align}
    \braket{U|R}_{CS}= \tr_{R}(U)
\end{align}
They form an orthogonal basis under the inner  product defined by the Haar integral
\begin{align} \label{haar} 
\int dU\, \tr_{R}(U)\tr_{R'}(U^{-1}) = \delta_{RR'}.
\end{align}

To obtain a string theory interpretation of these states, we need to define what
it means to fix a representation in the large-$N$ limit.  We do this by appealing
to the Frobenius relation for the finite-$N$ characters:
\begin{align}\label{Frob}
        \text{tr}_{R}(U)=  \sum_{\vec{k} \subset S_{n}} \frac{\chi_{R}(\vec{k})}{z_{\vec{k}}}    \prod_{i} (\text{tr} U^{i} )^{k_{i}},
\end{align} 
The RHS of this equation is determined by symmetric group data specified by the tableaux for $R$.  Thus, the large N limit of this expression is well defined if the tableaux remains fixed  as $N\to\infty$.  
These representations with finite number of boxes  make up a large-$N$ sector consisting of functions
built from traces $\tr U^k$ with positive powers.  
In the string theory interpretation they arise from the coupling of the worldsheet boundary to the worldvolume gauge field on the D branes, with $k$ specifying how the boundary winds around the torus.  
In the large N limit\footnote{at finite N, these are not orthogonal due to trace relations},they are related to the oscillator states by
\begin{align}
    \tr U^k = \braket{U |\alpha_{-k} |0} = \sum_{i} x_{i}^{k}, \qquad k>0,
\end{align}
where $x_{i}=e^{i\theta_{i}}$ are phases making up the eigenvalues of $U$.  
This identification reproduces \eqref{Frob}, and defines the chiral sector relevant for perturbative
topological string amplitudes.

\medskip
\noindent\textbf{The anti-chiral sector}\quad
On the other hand, an example of a representation whose tableau scales with $N$
is the anti-fundamental, which is a column of $N$ boxes.  These can be captured
by a second sector defined by the same Frobenius formula \eqref{Frob}, but with
the RHS replaced by $\tr (U^{-k})$.  As we explain in more detail in
Section~3.2, these arise from worldsheets that end on negative branes, which
have the opposite chirality from D-branes.  Their power-sum expansion involves
negative modes of $x_{i}$, which we view as functions of an anti-holomorphic
variable $\bar{x}_{i}$:
\begin{align}
    \tr U^{-k} &=\sum_{i} x_{i}^{-k} \nonumber\\
   &\equiv  \sum_{i} \bar{x}_{i}^{k}.
\end{align}
At finite $N$, the unitarity constraint $U^{\dagger}U=1$ implies
$\bar{x}_{i} =x_{i}^*$, so the two sectors are not independent.  However, in
the strict large-$N$ limit, $x_{i}$ and $\bar{x}_{i}$ are treated as
independent variables: the unitarity constraint is relaxed, and holomorphic and
anti-holomorphic degrees of freedom decouple.  The total Hilbert space therefore
holomorphically factorizes:
\begin{align}
\lim_{N\to \infty} \mathcal{H}_{T^2}^{\text{CS}}
\;\hookrightarrow\;
\mathcal{H}_{\text{closed}}^{+}
\otimes
\mathcal{H}_{\text{closed}}^{-}.
\end{align}
Since the perturbative topological string amplitudes are holomorphic, they can
be captured by projecting to the $\mathcal{H}_{\text{closed}}^{+}$ sector:
\begin{align}\label{proj}
    \mathcal{H}_{\text{closed}}^{+}
\otimes
\mathcal{H}_{\text{closed}}^{-}&\to \mathcal{H}_{\text{closed}}^{+} \nn
 \lim_{N\to \infty } \ket{R}_{\text{CS}} &\to \ket{R} 
\end{align}
This gives the precise mapping between the closed string basis $\ket{R}$ and the Chern Simons basis states.  We will see that this projection preserves the local tensor product structure for the Chern--Simons
theory.   However, unitarity is lost because the sesquilinear property of the
inner product  \eqref{haar} is not preserved, in view of the fact that $U^{-1}$ is no longer
interpreted as $U^{\dagger}$. In the chiral theory, the same formula holds as a real bilinear pairing between closed strings states and their dual, which defines the Hall product \eqref{hall}.

\subsection{Negative branes, TQFT gluing, and relative framing}
The two sectors $\mathcal{H}_{\text{closed}}^{\pm}$ have a natural physical
interpretation: they correspond to states produced by worldsheets ending on branes and negative
branes respectively. Negative branes are defined by flipping the overall sign of $D$ brane boundary states \cite{Okuda:2006fb}. They are objects that completely annihilates with ordinary branes, because they have negative tension as well negative RR charges. This differs from brane anti anti-branes annhilation, which leaves behind a tachyon condensate that shifts the background.   This means we can cut closed string amplitudes into open string amplitudes by introducing stacks of branes and negative branes,  and then glue them back together by annhilation without  changing the original background. 

In topological string theory, this canceling of branes and negative branes was proposed as a microscopic mechanism for co-dimension 1 gluing of states in the A model TQFT \cite{Aganagic:2004js}.  In the A model, negative branes can be defined simply by assigning a negative number $-N$ of Chan Paton factors to each worldsheet boundary \cite{Vafa:2001qf}.      In terms of the oscillator basis, this acts as
\begin{align}
    \alpha_{-n} \to - \alpha_{n},
\end{align}
which gives a minus sign for each worldsheet hole. 
For example, the  $(-1,0)$ and $(0,-1)$ Calabi Yau caps are related by this  operation, as shown in \eqref{CYcaps}, indicating that they correspond to  string amplitudes in the presence of branes and negative branes respectively. This implies that the entanglement boundary states $\ket{e}$, $\bra{\tilde{e}}$ inserts stacks of branes and negative branes respectively. 
In the representation basis, the brane to negative brane operation ( for a fixed choice of framing) is implemented by 
\begin{align} \label{Rflip}
\ket{R} \to  (-1)^{l(R)}\bra{R^t},
\end{align} 
where $R^t$ is the transposed tableaux. This
can be inferred from the character identity  
\begin{align}
\chi_{R}(\vec{k}) (-1)^{|\vec{k}|} = (-1)^{l(R)} \chi_{R^t} (\vec{k)},
\end{align}
where $|\vec{k}|\equiv \sum k_{i} $ indicates the number of worldsheet boundaries.  More physically, we can understand the  \eqref{Rflip} as the result of orientation reversal in the worldvolume Chern Simons theory, which flips the sign\footnote{Orientation reversal flips the sign of the action, due to the transformation of epsilon symbol.  This flip the coupling $k$, which is corrected to $k+N$ at 1 loop. } of the quantum corrected version of $N$:
\begin{align}\label{qinv}
 k+N \to  -(k+N) \longleftrightarrow q \to q^{-1} 
\end{align} 
 This is equivalent to the mapping \eqref{Rflip} when applied to the closed string amplitudes \eqref{unit} - \eqref{id} in the representation basis\footnote{In checking this, we must note that $d_{q^{-1}}(R) = (-1)^{l(R)} d_{q}(R)$}. 
Orientation reversal also sends $U\to U^{-1}$, leading to the power sums $\tr U^{-n}$  in the wavefunctions on negative branes.  This is consistent with the fact that TQFT gluing is defined between states and their duals, related by orientation reversal. 
\paragraph{Reflection positivity }

The brane to negative brane mapping \eqref{qinv} gives a physical explanation for the lack of symmetry between the two Calabi Yau caps in \eqref{unit}, \eqref{counit}:  their amplitudes are related by a nontrivial involution sending  $q\to q^{-1}$, rather than the adjoint operation defined by the Hall product that  send $\ket{R}\to\bra{R}$, or equivalently  $U \to U^{-1}$.    
Thus, they correspond to amplitudes on the same background geometry in the presence of a stack of branes and negative branes respectively.  
The TQFT gluing\footnote{In topology,  manifolds are glued with opposite orientations- this is why the gluing rules in \cite{Aganagic:2004js} sends $U\to U^{-1}$.  However, gluing in the cobordism language is done on boundaries with the same orientation.  This is because the ingoing and outgoing boundary is distinguished: we  flip the orientation of outgoing boundaries.  Then glue outgoing to ingoing boundaries with the same orientation gives  a composition of functors,  } then correspond to the annhilation of these branes as originally proposed in \cite{Aganagic:2004js}.  

The branes to negative  branes mapping can be implemented by TQFT cobordisms: for the Calabi Yau caps, this is given by the $(-1,1)$ annulus \begin{align} \label{antibrane}
  \mathtikz{\node at (0,1) {(-1,1)};\copairC {0cm}{0cm}} = \sum_{R} (-1)^{l(R)} q^{-\kappa_{R}/4} e^{-t l(R)} \bra{R}\bra{R}   
\end{align}
In contrast, the $(0,0)$ annulus would implement the adjoint with respect to the Hall inner product.  More generally, a $(k,0)$ state with branes is mapped to a $(0,k)$ state with negative branes by a $(k,-k)$ annulus.   
Gluing a state to its Hall adjoint using this annulus insertion produces TQFT partition functions that are not reflection positive.  This is manifest in the $(-1,-1) $ resolved conifold, obtained by gluing the $(-1,0)$ and $(0,-1)$  Calabi Yau caps.  This implies the partition function is not a norm square, consistent with the holomorphicity ( in $t$ ) of the perturbative topological string amplitude. 

This failure of reflection positivity also shows up when we define a density matrix $\rho$ whose entropy is computed by the gravitational replica trick. For the resolved conifold, this is \cite{Donnelly:2020teo}:
\begin{align}
    \rho =  \mathtikz{ \epsilonC{0cm}{.5cm} \etaC{0cm}{-.5cm}; \node at (0,1){(-1,0)} ;\node at (0,-1){(0,-1)}}
\end{align}  Once again,  the ``ket" and ``bra" do not correspond to the same state. This is a known feature of replica entanglement entropy in non-unitary systems such as quantum group invariant spin chains \cite{Couvreur_2017}, where the path integral replica trick computes the entropy of a density matrix composed of left and right eigenstates, since these are the states selected by path integral evolution.   Interestingly,  the left and right eigenstates in \cite{Couvreur_2017} are also related by sending  $q\to q^{-1}$.   Finally, the same comments   applies to the $(1,0)$ and $(0,1)$ pair of pants. 

  \paragraph{Relative framing}  
Aside from the worldvolume holonomy, the open string amplitude in the presence of non-compact D branes depend on an extra parameter given by the framing of the Lagrangian\footnote{  This is the string theory analogue of the framing anomaly in Chern Simons theory.}.  In the TQFT wavefunctions described above, this framing is given by the difference of the Euler classes.  When we glue together wavefunctions by annhilating branes and negative branes. their relative framing will affect the glued geometry via the difference of the total Euler classes:
\begin{align}\label{relframe}
 \text{relative framing}= k_{1}-k_{2}, 
\end{align}
which appears in the exponent of $q^{\kappa_{R}}$ in the partition function.   Note that this relative framing is zero when we glue together wavefunctions that are  simply related by $q\to q^{-1}$.   This means that this mapping picks the framing of the branes and negative branes to be opposite. However, in general, we have the freedom to choose framings that do not cancel between the branes and negative branes.   This mismatch is allows us to probe the full range of nontrivial intersection numbers on the Calabi Yau, which depends on $k_{1}-k_{2}$ as well as $k_{1}+k_{2}$.  In the example of the resolved conifold, we can change the framing on the negative branes so that annhilation corresponds to  gluing together two $(-1,0)$ Calab Yau caps, which leads to 
\begin{align}
    Z_{res}(t, k_{1}=-2,k_{2}=0) = \sum_{R} (d_{q}(R) )^{2} q^{-\frac{\kappa_{R}}{2}} e^{-tl(R)}
\end{align}

\section{Geometric transition of entanglement branes}\label{section:transition}
In this section, we explain the geometric transition of the entanglement branes (E-brane) from multiple perspectives, expanding upon the discussion in  \cite{Donnelly:2020teo}.  

We begin by recalling that in the Chern Simons langauge, the E-brane boundary state is an eigenstate of the  worldvolume holonomy $D$.    In general, a state of definite holonomy $U$ is a coherent state of the form\cite{Marino:2005sj}
\begin{align}\label{holU}
    \ket{U}
    =
    \exp\left(
        \sum_{n\geq 1}
        \frac{\tr U^n}{n}\alpha_{-n}
    \right)\ket{0},
\end{align}
which satisfies\footnote{It also satisfies $   \braket{U|\vec{k}} = \prod_{m} (\tr U^m   )^{k_{m}}$} 
\begin{align}
   \braket{U|R} &=\tr_{R}(U)
\end{align}

Comparing \eqref{CYcaps} with \eqref{holU}, we see that the  entanglement brane holonomy $U_{0}$ must satisfy:
\begin{align}
    \tr U_{0}^n
    =
    \frac{1}{q^{n/2}-q^{-n/2}}.
\end{align}
In the large \(N\) limit,  this is solved by the diagonal matrix
\begin{align}\label{Dtilde}
   (U_{0})_{ij}
    =
    \delta_{ij}q^{-j+\frac12},
    \qquad
    i,j=1,\ldots,\infty,
\end{align}
provided that we continue $q$ to a real number\footnote{We actually only need to continue $g_{s}$ so it has an imaginary part}. We can write this real holonomy as
\begin{align}
   U_{0} 
    =
    \mathrm{Diag}\{e^{\widetilde x_1},e^{\widetilde x_2},\ldots,
    e^{\widetilde x_N}\},
    \qquad
    \widetilde x_i
    =
    g_s\left(-i+\frac12\right).
\end{align}
and the entanglement brane boundary state is
\begin{align}
    \ket{e} = \ket{U_{0} }   
\end{align}
Similarly, the negative entanglement brane is \cite{Donnelly:2020teo}
\begin{align}
\bra{\tilde{e}} = \bra{U_{0}^{-1}} 
\end{align} 

In the rest of this section, we provide three interpretation of the distinguished worldvolume holonomy $U_{0}$ :

\begin{enumerate}

\item As a modulus for the entanglement branes specifying their spacetime position along the edge of a toric diagram representing the Calabi Yau.
 We give an explicit derivation of the transition of  a single stack of entanglement branes in in $\mathbb{C}^3$, which shows how degenerate strings ( not wrapping any cycles)  stretched between these D brane positions are needed to give an exact open closed  duality.
\item As the large N, double scaled limit of  the balancing element of the quantum group $U(N)_{q}$.  This is the element that defines the quantum trace which appears in our spacetime Cardy condition. 
\item As the shrinkable holonomy around a trivial Wilson line in worldvolume Chern Simons theory.  In skein theory, this is known as the Omega loop, which projects lines linking with it onto the  vacuum .

\end{enumerate}

\subsection{Toric diagrams and brane moduli}
To explain the interpretation of $U_{0}$ as position moduli of D branes, we begin with a basic review of toric diagrams, which describe topological string amplitudes on toric Calabi Yau's.  They are particularly useful for specifying the locations of non-compact Lagrangian branes.  Even though the geometries described by the A model TQFT  are no longer toric when the base manifold has genus bigger than 1, the toric diagrams remain a useful local tool. This is because we can describe the local neigbhorhood around a stack of branes in the TQFT geometries in terms of a local patch that is toric.  
We will derive the geometric transition of entanglement branes in a local $\mathbb{C}^3$ patch and use this as a fundamental
building block of local holography.
\paragraph{Toric diagrams }
Toric Calabi-Yau manifolds can be built by gluing $\mathbb{C}^3$ patches.
Each $\mathbb{C}^3$ patch admits a $T^2\times\mathbb{R}$ fibration over
$\mathbb{R}^3$, which we represent as a trivalent vertex in $\mathbb{R}^2$,
where each edge denotes the degeneracy locus where one of the $S^1$ cycles of
the $T^2$ shrinks.  The toric diagram for the total Calabi-Yau encodes the
gluing of these patches as a cubic Feynman diagram: each trivalent vertex
represents a $\mathbb{C}^3$ patch, each edge is labelled by a K\"{a}hler
parameter, and each edge direction is specified by a primitive vector
$v \in \mathbb{Z}^2$ indicating which $S^1$ cycle degenerates along it.  The
Calabi-Yau condition requires that the three edge vectors at every vertex
satisfy $\sum_i v_i = 0$.

Non-compact Lagrangian branes are located at points on the edges: they wrap
an internal Lagrangian submanifold with topology $\mathbb{C}\times S^1$, where
the $S^1$ is the cycle of the $T^2$ that remains non-degenerate along that
edge.  The framing of such a brane is specified by a vector $f\in\mathbb{Z}^2$
satisfying $f\wedge v = 1$, meaning $f$ and $v$ are independent cycles on the
torus, with $f$ specifying the cycle along which the worldvolume holonomy $U$
is defined.  This equation is invariant under $f\to f+kv$ for integer $k$.   When we glue branes to negative branes, the relative choice of $k$ becomes idetntified with the relative framing parameter in \eqref{relframe}.

The toric diagram for $\mathbb{C}^3$ with 3 stacks of branes is: 
\begin{align}
 \mathtikz{\CthreeVertex{0}{0}}   
\end{align}
The corresponding string amplitude is the topological vertex \cite{Aganagic:2003db}, given by 
\begin{align}
Z_{\text{vertex}}(U_{i}, t_{i}) = C_{R_{1}R_{2}R_{3}} \prod_{i} e^{t_{i}l(R_{i})}\tr_{R_{i}}(U_{i}) 
\end{align}
where $C_{R_{1}R_{2}R_{3}}$ are the vertex coefficients, and $U_{i}$ are the worldvolume holonomies.  
\paragraph{Brane moduli}
Since every point on an edge is a valid Lagrangian, the position 
along an edge is a brane modulus. 
This brane position is measured directly by non-degenerate open-string instantons,  which must wrap relative homology cycles with boundaries on the brane. For a single Lagrangian $L$ in $\mathbb{C}^3$, these relative cycles are disks $D$ in spacetime whose boundary is along the noncontractible cycle of $L$.   They are represented in red in the following diagram:
\begin{align}
\vcenter{\hbox{\includegraphics[scale=.25]{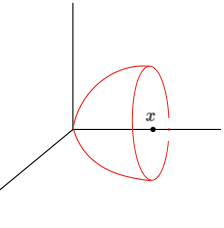} } }= \sum_{R} C_{00R} (q) e^{-xl(R)} \tr_{R} (U)
\end{align}

The Nambu-goto action of these open string instantons produces the symplectic area of this disk:
\[
x= \int_D \omega
\]
Intuitively, this measures the real position of the brane along the edge, since moving the brane changes
the area of the disk that ends on it.

However, the symplectic area does not give the total position modulus for the branes. This is because the area does not by itself define a relative period, which should be invariant under deforming the boundary $\partial D$ within the Lagrangian $L$.  To obtain such an invariant, we need to include a boundary term.  This boundary term arises from the coupling of the worldsheet boundary to the brane, which produces the integral of a 1-form field\footnote{ $phi$ parametrizes the normal deformations of the disk.  If \(v\) is the corresponding normal vector field, then the change
of the disk area is
\begin{align}
    \delta\int_D\omega
    =
    \int_D\mathcal{L}_v\omega
    =
    \int_D d\,\iota_v\omega
    =
    \oint_{\partial D}\iota_v\omega .
\end{align}
Thus \(\phi\) may be identified with \(\iota_v\omega\), and its boundary period
measures the displacement of the brane along the toric edge.  The transformation
\(\phi\to \phi+df\) is a Hamiltonian deformation of \(L\), hence a gauge
redundancy of the open modulus.} $\phi$ on the worldvolume. The total position modulus is then 
\begin{align}
x=\int_{D } \omega +\int_{\partial D} \phi
\end{align} 
The 1 form $\phi$ can be viewed as the imaginary part of a worldvolume gauge field and parametrizes the traverse deformations of the brane.  Indeed, such a complexification of the gauge field is needed for consistency with the complexification of the Kähler form $\mathcal{J}$ via the inclusion of the spacetime B field:
\begin{align}
    \mathcal{J} &= B+i\omega
\end{align}
Consistency with higher form gauge invariance of these 2 form fields requires the introduction of a complexified gauge field
\begin{align}
    \mathcal{A} &= A+i\phi 
\end{align} 
where \(A\) is the Hermitian Chern-Simons gauge field on the brane.  The complex combination
\begin{align}\label{openmodulus}
    z
    =
    i\left(
    \oint_{\partial D}\mathcal{A}
    +
    \int_D \mathcal{J} \right)
\end{align} is then invariant under the higher form gauge transformations.

For multiple D branes in a stack,  $\mathcal{A}$ is matrix valued, so the shift by the integral of the Kahler form in $z$ should be understood to be proportional to the identity  $N\times N$ matrix.  The open instanton then couples to the complexified holonomy
\begin{align}\label{complexholonomy}
    P\exp - i\left(
    \oint_{\partial D}\mathcal{A}
    +
    \int_D \mathcal{J}
    \right) =U e^{-t} ,
\end{align}
where $U=  P\exp i\left(
    \oint_{\partial D}\mathcal{A}\right) $.
Turning on a diagonal, real $U$ corresponds to separating the D branes around the stack position $t$, with the $i$ th D brane located at $x_{i}$\footnote{ It may seem surprising that the worldvolume \emph{holonomy} encodes the
\emph{transverse position} of the branes, since in the physical string these
are independent degrees of freedom.  This ultimately arises from the holomorphic nature of the topological string worldsheet, as explained in \cite{Aspinwall:2004jr,Gomis:2006mv}. See the appendix for more details.}, where 
\begin{align}
e^{-x_{i}}=e^{-t} U_{ii} 
\end{align}

The fact that the Kahler form can be incorporated into the complex holonomy via $U\to Ue^{-t}$ is consistent with fact that the non degenerate instantons ending on the branes give:
 \cite{Aganagic:2003db}:
\begin{align}\label{vertex}
    Z_{\text{vertex}}(\mathbb{C}^3,\, Ue^{-t})&= \sum_R (-1)^{l(R)}d_q(R) q^{-\kappa_R/4}\, e^{-tl(R)}\tr_R(U)\nn
    &= \sum_R (-1)^{l(R)} d_q(R) q^{-\kappa_R/4}\, \tr_R(Ue^{-t}) 
\end{align}
Here we used the identity $\tr_{R}(Ue^{-t}) = e^{-tl(R)} \tr_{R}(U)$, which shows that the instanton action can be absorbed into the complex holonomy.   
\paragraph{Branes along the fiber}
In the TQFT language, the vertex with one stack of branes is just the Calabi Yau cap:
\begin{align}
  \mathtikz{ \etaC{0cm}{0cm} ;\node at (0,-1/2){(-1,0)}}    = \vcenter{\hbox{\includegraphics[scale=.1]{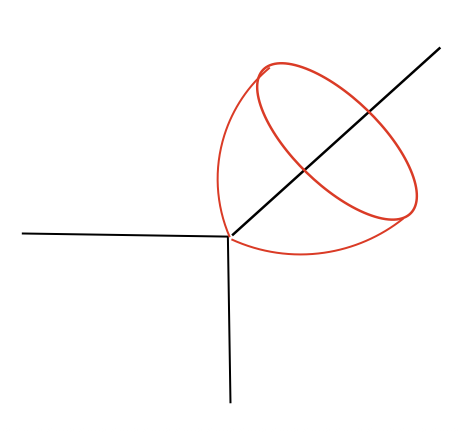} } } 
\end{align}
This describes the local neighborhood of any point $P$ on the base manifold $\Sigma$.  
As explained in \cite{Aganagic:2004js}, adding another stack of branes to the vertex then corresponds to inserting a puncture at the point $P$, with the branes wrapping the fiber directions. For $n$ punctures with branes in the fiber, the topological string amplitude is 
\begin{align}
    Z(\Sigma_{g,n})  = \sum_{R, Q_{1},\cdots Q_{n}} \frac{W_{RQ_{1}}\cdots W_{RQ_{n}}}{W_{R,0}^{2g-2+n}}q^{\frac{(k_{1}-k_{2})\kappa_{R}}{4}} e^{-tl(R)} \tr_{Q_{1}} (U_{1}) \cdots \tr_{Q_{n}}(U_{n}),
\end{align}
where
\begin{align}
W_{RQ}(q) =\lim\limits_{N\to\infty} q^{\frac{N(l(R)+l(Q))}{2}} \frac{S_{RQ}(q,N)}{S_{00}(q,N)},
\end{align} 
and $U_{i}$ are the worldvolume holonomies.  

Both the gluing of topological vertices and the A model TQFT can be interpreted in terms of cancelling branes and negative branes.  Thus the two formalisms are intimately connected, and they give useful complementary perspective on the target space geometry. 

\subsection{The local geometric transition}
Here we give explicit derivation for the transition of a single stack of entanglement branes in $\mathbb{C}^3$ and show how their backreaction produces the resolved conifold partiton function.

There are two effects of introducing entanglement branes. First, the contribution from non-degenerate worldsheets is modified by setting $U=U_{0}$ in the vertex.  This is obtained by computing $\tr_{R}(U_{0})$  using the properties of the Schur function \eqref{schur},  which satisfy \cite{Marino:2005sj} 
\begin{align}
\tr_{R}(U_{0})= s_{R}( x_{i}=q^{-i+1/2})  = 
d_{q}(R)q^{\kappa_{R}/4}.
\end{align}. Plugging this into the vertex gives
\begin{align}
    Z_{\text{vertex}}(\mathbb{C}^3,g_{s},U_{0}) &= \sum_{R} (-1)^{l(R)} (d_{q}(R)) q^{-\kappa_{R}/4} e^{-t l(R)} \tr_{R}(U_{0})\nn
    &=\sum_{R}(-1)^{l(R)} (d_{q}(R))^{2}  e^{-t l(R)}
\end{align} 
This is exactly the non-degenerate worldsheet contribution to the resolved conifold partition function, as expected from the TQFT logica.    However, there are also contributions from degenerate worldsheets which are omited from the local description of the  TQFT.  
These arise because 
the discrete holonomy $U=U_{0}$ produces small separations of the branes on the stack by a distance $g_{s}$.  This produces massive strings stretched between them, which do not wrap any relative homology cycles.  They generate zero-area worldsheets  whose 1-loop partition functions contribute  a factor of the form \cite{Gomis:2007kz}
\begin{align}\label{annuli}
   \lim_{N\to \infty} \xi(q)^N \exp (- \sum_{n=1} \frac{1}{n} \sum_{1\leq i <j \leq N} q^{-n(i-j)}),
\end{align}
where each  prefactor 
\begin{align}
    \xi (q)  \equiv \prod_{m=1}^{\infty} (1-q^m )^{-1}
\end{align}
is the partition function of a chiral boson associated to each brane, corresponding to degneerate worldsheets ending on the same brane, while the  exponential factor comes from a  fermion determinant that arises when we integrate out the open strings between different branes.  To account for the full ampllitude, the contribution \eqref{annuli} has to be introduced in addition to setting $U=U_{0}$ in \eqref{vertex}.    In the appendix, we show that 
\begin{align}
 \lim_{N\to \infty} \xi(q)^N \exp (- \sum_{n=1} \frac{1}{n} \sum_{1\leq i <j \leq N} q^{-n(i-j)}) = \prod_{k=1} \frac{1}{(1-q^{k})^{k}},
\end{align} 
where $M(q)= \prod_{k} \frac{1}{(1-q^k)^{k}}$
is the MacMahon function that gives the contribution of degenerate worldsheets to the resolved conifold partition function !  Thus, incorporating the worldsheets connecting the separated branes produces an exact transition between $\mathbb{C}^3$ and the resolved conifold.  In terms of toric diagrams, this transition is given by:
\begin{align}
\includegraphics[scale=.3]{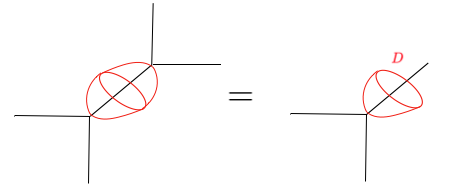}
\end{align}

Finally,  to obtain a transition to  a $(-k,-2+k)$ resolved confiold , we can modify the entanglement brane boundary state by inserting the $(1,-1)$ annulus, which alters the framing of the branes.   This produces the boundary state
\begin{align}
    \ket{e,k}= \sum_{R} d_{q}(R) q^{\frac{(2-2k)\kappa_{R}}{4}} e^{-tl(R)}\ket{R},
\end{align}
which gives the desired transition.

\paragraph{Transition with multiple stacks}  
The local nature of the transition described abve means that it is straightforward to generalize to multiple stack of branes located at different circles on the base manifold of the TQFT.   Moreover, for even number of stacks we can introduce an equal number of branes and negative branes.  This has the advantage that the degenerate contributions 
cancel up to a phase, since each stack of negative branes gives:
\begin{align}
    M(q^{-1} ) =  \frac{(\prod_{n=1} q^{k^{2}} )}{M(q)}\to \frac{ e^{\pi i /12 } }{M(q)},
\end{align}
where we used the brane to negative brane map and zeta function regularization.  So in these cases the TQFT description of the transition is exact. 

As a simple check of this more general scenario, we can consider the transition of $\mathbb{C}^3$ with a stack of branes and negative branes. The entanglement boundary state for negative branes is
\begin{align}
    \bra{\tilde{e}} = \bra{0}\exp ( \sum_{m>0} \frac{-\alpha_{m}}{m (q^{m/2} -q^{-m/2})}) .
\end{align}
The oscillator algebra implies that
\begin{align}
   \braket{\tilde{e}|e^{-t l(R)}|e} &= \exp \sum_{n=1}^{\infty} \frac{-e^{-nt}}{n ([n]_{q})^{2}}\nn
   &=Z_{res}(t),
\end{align}  
 This is exactly the transition described in \eqref{cardy}.   In terms of toric diagrams, this is depicted as :
\begin{align}
\vcenter{\hbox{\includegraphics[scale=.1]{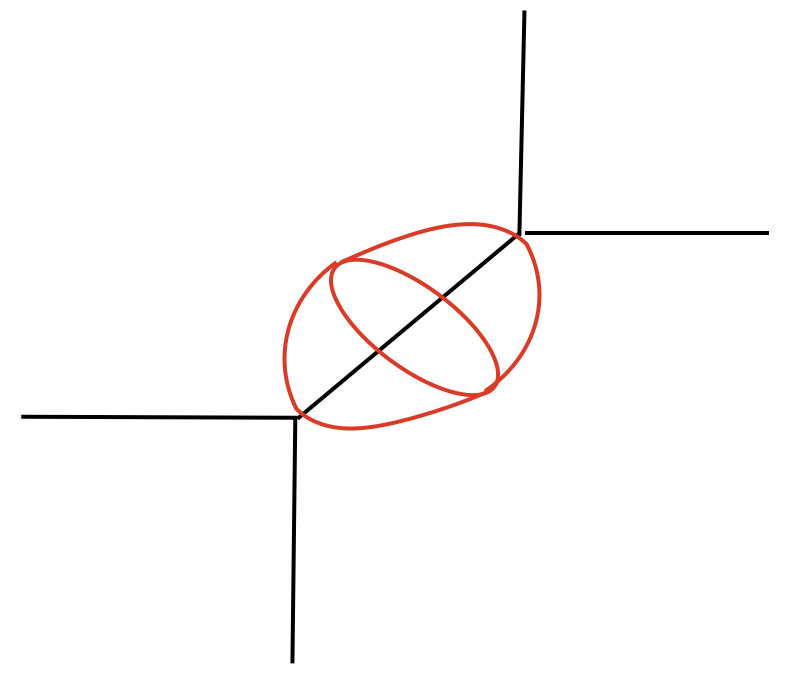}}} = \vcenter{\hbox{\includegraphics[scale=.2]{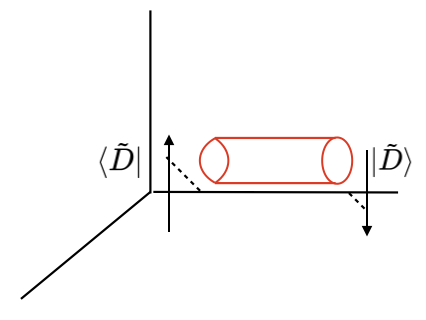}}}
\end{align}

\subsection{Worldsheet description of the transition }
The local transition described above has a simple target space description, in which a relative homology cycle ending on branes closes up into an absolute homology cycle when these branes dissolve.  As explained in \cite{Gomis:2006mv}, this can be viewed as a type of surgery on spacetime: A tubular neighborhood of  topology $(\text{3-ball)} \times \mathbb{C}^2 \times S^1$ is removed around the branes, and $S^2 \times \mathbb{C}^2 \times (\text{disk})$ is glued in, with the disk combining with the relative cycle to create a 2 -cycle.

In this section, we discuss the worldsheet description of the transition \cite{Gomis:2006mv}. 
We begin by introducing t'Hooft coupling in to the Chern Simons wavefunction on the entannglement branes.   We do so by rescaling the worldvolume holonomy:\footnote{Here constant means independent of the Chan Paton factors  } 
\begin{align} \label{D}
  D&=U_{0}e^{\frac{g_{s} N}{2}}\nn
    &=  \text{Diag}  \{e^{x_{1}}, e^{x_{2}} \cdots  e^{ x_{N}} \} ,\quad 
     x_{i}   = g_{s} (- i +\frac{1}{2} + \frac{N}{2}). 
\end{align}
This introduces the  t'hooft coupling
\begin{align}
t'= g_{s}N/2
\end{align}
into the game.
The precise factor of  $\frac{1}{2}$ was chosen because  resulting group element $D$ has a special significance from the point of view of representation theory:   It can be expressed in terms of the Weyl vector $\rho$ and the Cartan generater $H$ as
\begin{align}
D= q^{\braket{\rho,H}}
\end{align}
Moreover, when $q$ is a phase,
$D$ is a $U(N)$ matrix that defines its quantum dimesions via
\begin{align}\label{qdim}
    \dim_{q}(R)= \tr_{R}(D) .
\end{align} As we explain in more detail in \ref{section:subregion}, introducing $D$ as the worldvolume holonomy leads to an interpretation of the transition as a large N quantum trace on an open string  subregion algebra.  

Here, we focus on the Chern Simons worldvolume interpretation.  For this purpose, the crucial  observation  is that
$D$ is exactly the quantum corrected holonomy around a trivial Wilson loop in Chern Simons theory \cite{ELITZUR1989108}.  
\begin{align}
   \braket{U|0}=\vcenter{\hbox{ \includegraphics[scale=.2]{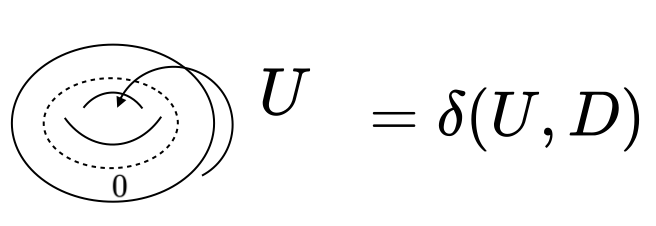}}}
\end{align}
This means that the boundary state $\ket{D}$ glues in a solid torus with no Wilson loop insertions.  As explained in \cite{Wong:2025asd},  this defines a shrinkable boundary condition  for Chern Simons theory\cite{Wong:2025asd}, which is shifted from the identity due to quantum corrections. 
\paragraph{Closing up worldsheet boundaries} 
We now take the large N limit  with the t'Hooft parameter $t'$ fixed, and show how this produces closed strings on the resolved conifold. In particular, we want to show how the boundary of the open string instantons on the branes close up in the large $N$ limit.  
 
To see how this works in a specific example,  consider open strings on $\mathbb{C}^3$ with a finite $N$ stack of branes and negative-branes, separated by distance $t_{op}$ in the target space.   
The total string amplitude comes from non-degenerate instantons corresponding to open strings stretched between the two stacks. 
\begin{align}\label{C3openstrings}
Z_{\text{open}}(\mathbb{C}^3,D) =  \vcenter{\hbox{\includegraphics[scale=.23]{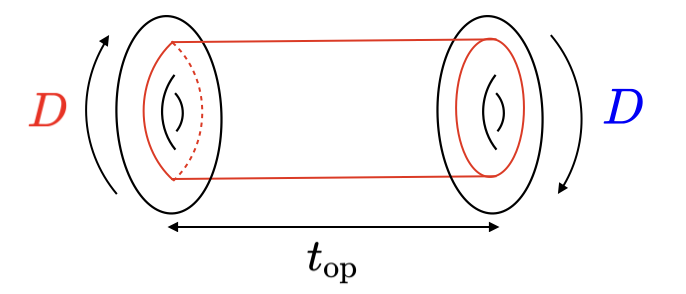}}} 
\end{align}

In the Chern Simons theory, integrating out open strings stretched between the branes inserts the  Ooguri-Vafa operator \cite{Ooguri:1999bv}
\begin{align} \label{ooguri}
  \sum_{R}   \ket{R^{t}} (-1)^{l(R)} e^{-t_{\text{op}}l(R)} \bra{R}.
\end{align}

Since the holonomy $D$ is imposed  along the non-contractible cycle of the Lagrangian branes, it glues an S-dual solid torus with no Wilson loops.  This compactifies the Lagrangian branes so that the open strings are stretched between two $S^3$:
\begin{align} \label{GVfiniteN}
\vcenter{\hbox{\includegraphics[scale=.15]{figures/Stretchedstrings.png}}}&= \vcenter{\hbox{\includegraphics[scale=.15]{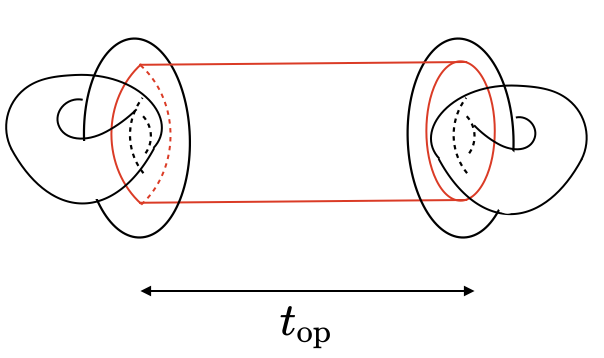}}}\nn
\sum_{R} \tr_{R^{t}}(D) (-1)^{l(R)}e^{-t_{op} l(R)}tr_{R} (D)&= \sum_{R}\dim_{q^{-1}}(R) e^{-t_{op} l(R)} \dim_{q}(R) 
\end{align}

On the RHS, we evaluated the traces using \eqref{qdim}, used brane to anti-brane map $q \to q^{-1}$.
We can interpret the quantum dimensions 
\begin{align}
\dim_{q}(R)= \braket{\tr_{R}(U)}_{S^3}
\end{align} 
as the Chern Simons expectation value of Wilson loops on $S^3$.  These Wilson loops are identified with the instanton boundary.  In the t'Hooft limit, the perturbative $1/N$ expansion produces a sum over all ribbon Feynman diagrams on $S^3$ that end on the Wilson loops: this represents an interaction between the stretched open string boundaries and the degrees of freedom on the branes.  In fact, this is where all the nontrivial $g_{s}$ dependence enters the open string partition function.  In the worldsheet description, summing over the Chern Simons ribbon graphs closes up the worldsheet boundary into smooth disk:
\begin{align}
    \includegraphics[scale=.1]{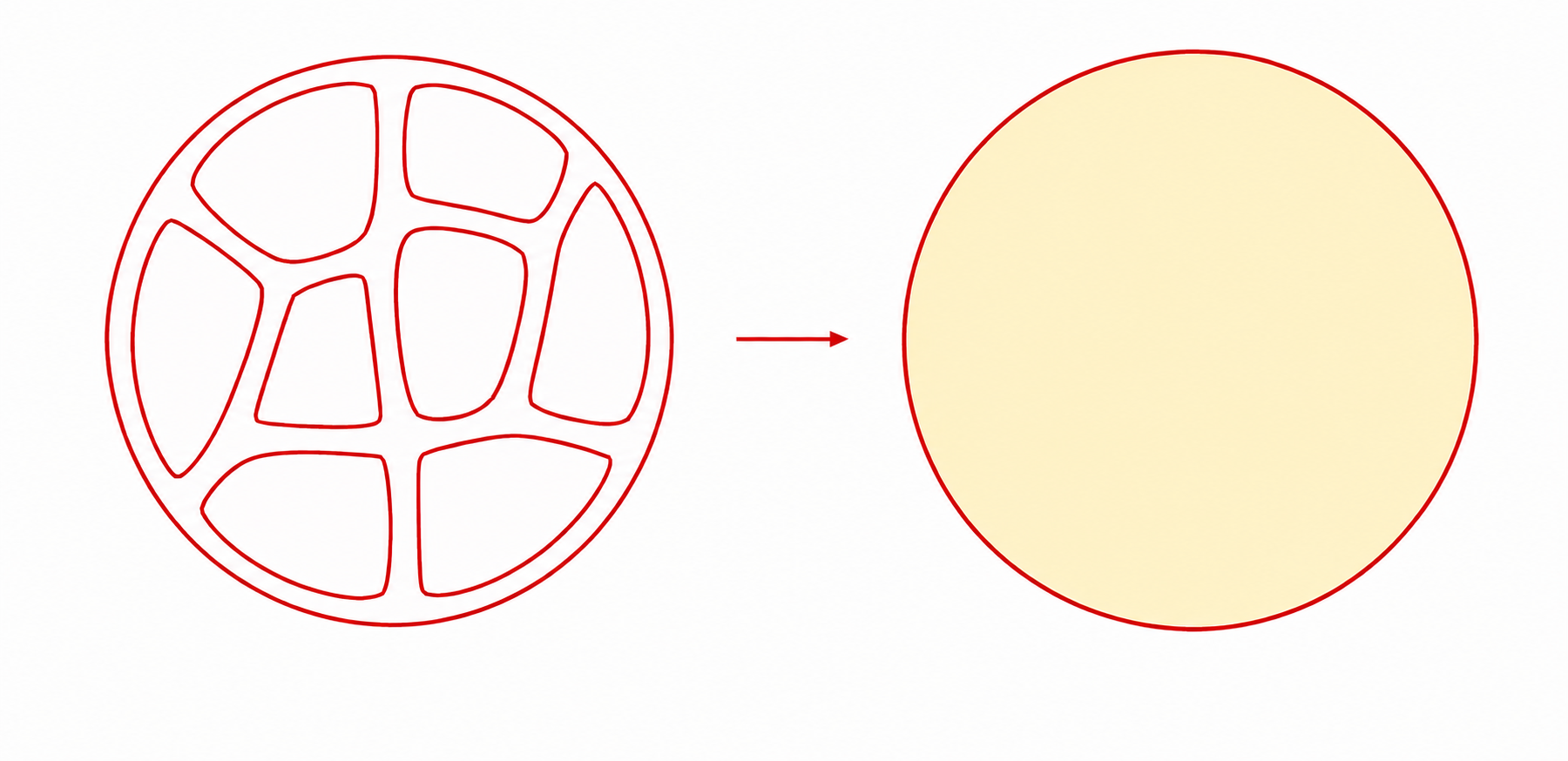}
\end{align}

Remarkably, this also closes up the hole in spacetime, since it produces a geometric transition 
where the relative cycle ending on the branes close up into an $S^2$ with Kahler parameter given by $t_{op}$.  The precise transition requires a double scaled limit that also sends $t' \to \infty$\cite{Donnelly:2020teo}. This can be seen explicitly by performing a large $N$ t'Hooft expansion, combined with  Laurent series expansion in $e^{t'l(R)}$
\begin{align}\label{dscale}
    \dim_{q}R &= q^{\frac{Nl(R)}{2}} d_{q}(R) q^{\kappa_{R}/4} + \cdots \nn
            &= e^{t' l(R) }d_{q}(R) q^{\kappa_{R}/4} + O (1)
\end{align} 
Similarly if we send q to $q^{-1}$, the same limit extracts the lowest term of the Laurent expansion proportional to $q^{-Nl(R)/2}$
\begin{align}
 \dim_{^{-1}q}R &= q^{-\frac{Nl(R)}{2}} d_{q^{-1}}(R) q^{-\kappa_{R}/4} + \cdots \nn
            &= e^{-t' l(R) } d_{q^{-1}}(R) q^{-\kappa_{R}/4}+ O (1)
\end{align}
Finally, applying these identities to \eqref{GVfiniteN} and using the fact that 
\begin{align} 
d_{q^{-1} } (R)  = (-1)^{l(R)} d_{q}(R)  
\end{align}
gives 
\begin{align}
Z_{\text{open}}(\mathbb{C}^3,D)= \sum_{R} d_{q}(R)^{2} (-1)^{l(R)} e^{-t_{\text{op}}},
\end{align}
which
reproduces the resolved conifold partition function. 
Finally, note that the large $t'$ limit is needed to extract the symmetric group quantum dimemsion $d_{q}(R)$ from the $U(N)$ quantum dimensions.  
\section{Resolving the entanglement brane} \label{resolve}
In the previous sections, we considered the geometric transition of a 
distinguished configuration of equally spaced D-branes and identified 
the corresponding fixed holonomy boundary state.  It is a superposition over all representations:
\begin{align} \label{branesum}
    \ket{D} &=\sum_{R} Z_{R} \ket{R} 
\end{align}
where $Z_{R}$ is a regularized, large N limit of the $U(N)_{q}$ quantum dimension $\dim_{q}R$.   This equation says $R$ is the conjugate variable to the holonomies \footnote{We will see that there is a discrete set of such holonomies} arising in the non-abelian Fourier transform on a (quantum) group.  What type of physical objects do these conjugate variables represent?  One answer is that $R$  labells superpositions \eqref{frob}  of winding strings. However, as shown in \cite{Okuda:2007ai, Gomis:2007kz}, for a single row or column of $O(N)$ boxes, these strings transition into branes and negative branes respectively.  Furthermore, when the tableaux have $O(N)$ rows as well as columns, these branes backreact on the geometry to produce ``Bubbling Calabi Yau's".  More specifically if we write the open string partition function between E branes in terms of Chern Simons expectation values of Wilson loops $W_{R}(q)$ on $S^3$:
\begin{align}
    Z_{\text{open}}(\mathbb{C}^3,D) =  \sum_{R} \braket{W_{R}(q)}_{S^3}e^{-t l(R)} \braket{W_{R}(q^{-1})}_{S^3},
\end{align}
 then the Fourier coefficients 
\begin{align}
    Z_{R}= \lim_{N\to \infty} \braket{W_{R}}_{S^3}
\end{align}
are the partition functions of the bulk stringy objects described above.  This is completely analogous to the holographic description of Wilson loops in AdS/CFT.  This allows us to interpret   $ Z_{\text{open}}(\mathbb{C}^3,D) $ as a sum over superselection sectors of stringy objects, which contribute independently to the geometric transition of the total entanglement brane configuration.

As an application of this gravitational interpretation, we will  incorporate the insertion of these R-type objects into the 
state preparation of TQFT closed string states and compute the resulting 
gravitational entropy. In the TQFT language, these states involve the 
insertion of punctures or defect lines labelled by a tableaux $R$. We will be especially interested in the case when $R$ is a rectangular tableaux with $O(N^2)$ boxes.  In the large $N$ limit, inserting these branes allow us to compute the entropy of Bubbling Calabi Yau geometries.

\subsection{Discrete holonomies and $R$ type branes }
It was shown in \cite{Okuda:2007ai} that a state $\ket{R}$ where $R$ is a rectangular tableaux of $O(N^2)$ boxes corresponds to a configuration of D branes on $\mathbb{C}^3$ with discrete worldvolume holonomy
\begin{align} \label{discrete}
    U_{R}=\text{Diag}  \{e^{x_{1}}, e^{x_{2}} \cdots  e^{ x_{N}} \} ,\quad 
     x_{i}   = g_{s} (R_{i}- i +\frac{N+1}{2} )  ,
\end{align}
where $R_{i}$ labels the ith row length of the tableaux.  In the spacetime interpretation, these are non-uniformly spaced D branes at position $x_{i}$.  Such configurations were studied in \cite{Okuda:2007ai} an example of a bion in topological string theory.

Note that in Chern Simons theory on a torus,  $U_{R}$ is the holonomy around Wilson loops in the representation $R$ \cite{ELITZUR1989108}.  
In contrast with the entanglement brane $\ket{D}$, the fixed holonomy state $\ket{U_{R}}$ glues in an S-dual solid torus with a Wilson loop inserted. The discreteness of these holonomies is consistent with the interpretation of \eqref{branesum}, as a Fourier transform between conjugate variables. 
\paragraph{Branes on a general toric Calabi Yau}
We would like to consider the insertion of $R$ type branes into geometries other than $\mathbb{C}^3$, which may have a nontrivial Kahler class. This  will allow us to prepare new states in the closed string TQFT.  On these more general toric geometries,
\cite{Gomis:2007kz} showed that such R-type branes have holonomies given by 
 \begin{align}
   U_{R} = \text{Diag}  \{e^{-x_{1}}, e^{-x_{2}} \cdots  e^{ -x_{N}} \} ,\quad 
     x_{i}   = g_{s} (R_{i}- i +\frac{1}{2} + N)  ,
\end{align}
where $x_{i}$ once again denotes the position of the branes on an edge of a toric diagram.  Note the shift of $g_{s}N/2$ relative to branes in $\mathbb{C}^3$.  
It was shown in \cite{Gomis:2006mv} that these branes transition into closed string geometries with additional fluxes given by 
\begin{align}
     t_{i} = g_{s} l_{i} 
 \end{align}
where $l_{i}$ are the Maya coordinates of the tableaux.  These D branes can be inserted either as a puncture or a line defect on the base manifold of a TQFT geometry.   

\subsection{Entanglement entropy of bubbling Calabi Yau's}
For simplicity, we consider the insertion of such branes on the $(-1,$ -1)  conifold geometry. For the branes that are extended , the total open-closed partition function  was computed in \cite{Gomis:2006mv}:
\begin{align}    Z_{res}(t,R)=\vcenter{\hbox{\includegraphics[scale=.1]{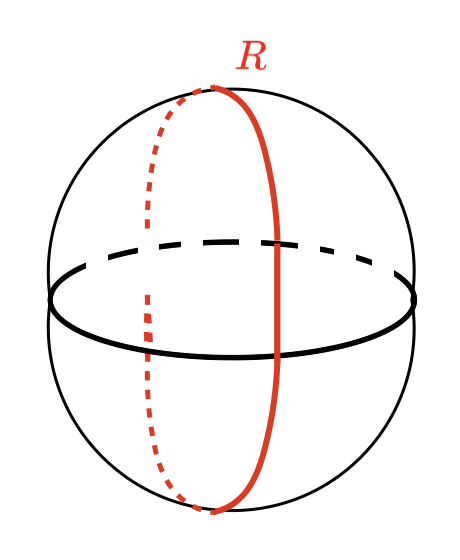}}} = M(q)e^{-\sum_{n} \frac{e^{-nt}}{n[n]_{q}^2}} \prod_{i<j} (1- e^{(x_{i}-x_{j})}) \prod_{i} \exp (\sum_{n=1}\frac{e^{-nx_{i}} + e^{-n(t-x_{i})}}{n[n]_{q}})
\end{align} 
It matches with the large $N$ limit of the corresponding Wilson loop in Chern Simons theory. The  term $M(q)e^{-\sum_{n} \frac{e^{-nt}}{n[n]_{q}^2}} $ comes from the closed worldsheets and reproduces the partition function without branes, while  $ \prod_{i<j} (1- e^{(x_{i}-x_{j})})$ comes from degenerate worldsheets connected the separated branes. The contribution from non-degenerate instantons ending on branes come entirely from the last term, where $x_{i}$ and $t-x_{i}$ describes the relative periods of disks attaching to the the brane from two sides: 
\begin{align}
\includegraphics[scale=.1]{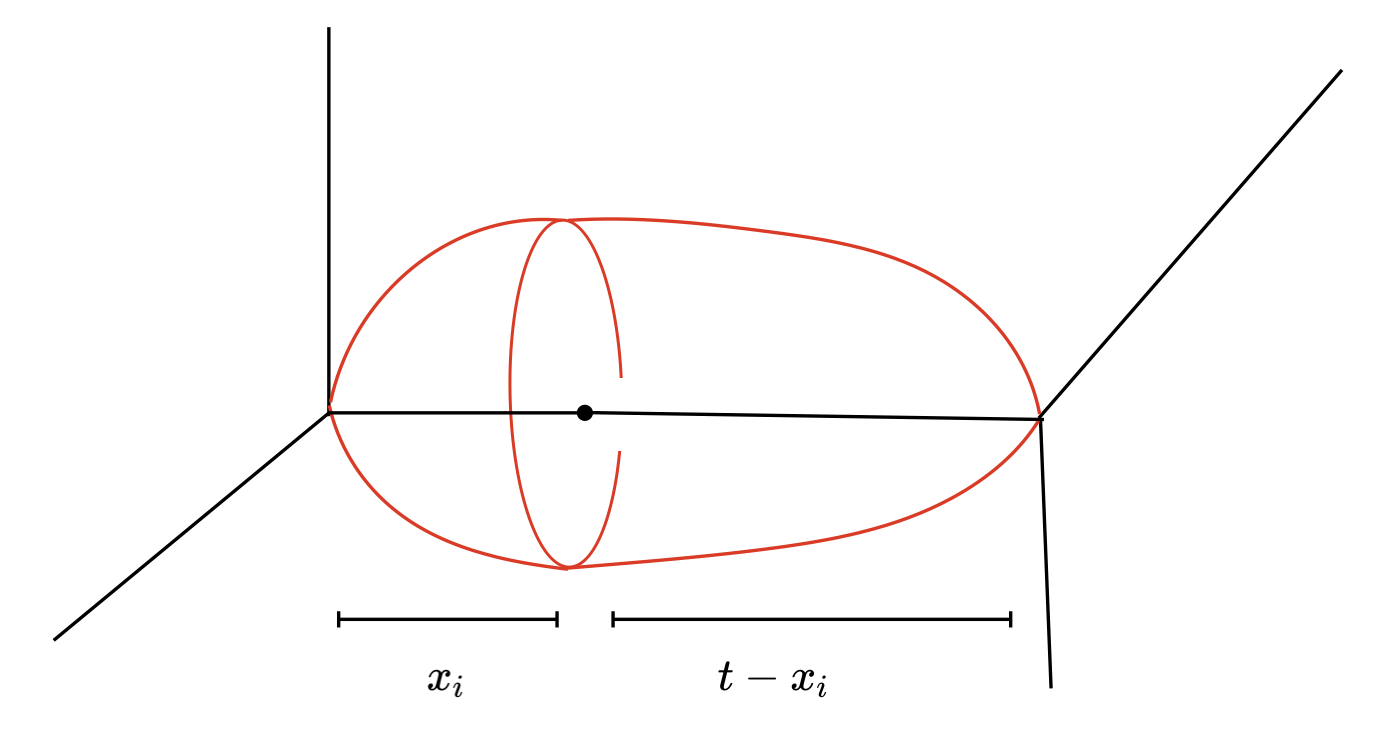}
\end{align}

The symmetry of the sphere implies we can compute the gravitational entropy as before using the Gibbons Hawking prescription \cite{Donnelly:2020teo}:
\begin{align}\label{GH}
    S= (1-t\partial_{t}) \log Z_{res}(t,R)
\end{align}
This gives the gravitational entropy of the resolved conifold without branes, plus an additional contribution from the branes given by
\begin{align}
    \Delta S(R) =  (1- t \sum_{n=1} 
    \frac{e^{-n(t-x_{i})}}{[n]_{q}})
\end{align}
We expect that this computes the entanglement entropy of a state given by
\begin{align}
\ket{\psi_{R}}=\vcenter{\hbox{\includegraphics[scale=.2]{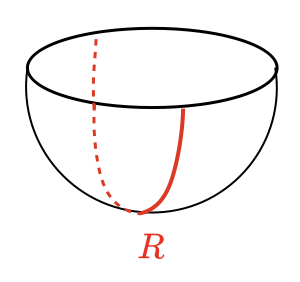}}}
\end{align}
We will give a proof of this in \cite{Wong:2026}, using the connection of topological string theory to q-deformed two dimensional Yang Mills.

Similarly, we can consider branes inserted on punctures. For two punctures on  the base sphere of the resolved conifold, we have the partition function
\begin{align}
    Z_{res}(t,Q,\bar{Q}) &= \vcenter{\hbox{\includegraphics[scale=.1]{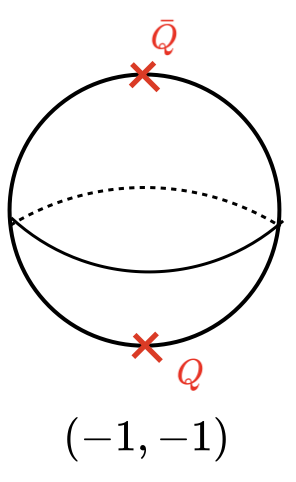}}} \nn 
    &= \sum_{R} \tilde{d}_{q}(R)^2 W_{RQ}W_{R\bar{Q}}   \tr_{R}(U_{Q})\tr_{R} (U_{\bar{Q}}) e^{-t l(R)}
    \end{align}
where $U_{Q}$, $U_{\bar{Q}}$ are holonmies on the branes.    The gravitational entropy is again given by the Gibbons Hawking formula \eqref{GH}.  It takes the form of \eqref{MEE}, with the probability factor replaced by
\begin{align}
P(R) = \frac{\tilde{d}_{q}(R)^2 W_{RQ}W_{R\bar{Q}}   \tr_{R}(U_{Q})\tr_{R} (U_{Q}) e^{-t l(R)}}{Z_{res}(t,Q,\bar{Q})}
\end{align}
This computes the entropy of a state prepared by a hemisphere with a stack of D branes insertion in the fiber above the puncture.


\section{Subregion algebras, quantum trace, and modular flow}\label{sec:opensubregion}
We have seen that the geometric transition of entanglement branes provides an open string dual to the closed string backgrounds in the A model TQFT. We now provide canonical description of these  open strings by defining their operator algebra and the corresponding GNS Hilbert space.  This is not the conventional Hilbert space of open strings:  due to the  interactions of the string worldsheet with the entanglement branes, the effective open string degrees of freedom  are anyons transforming in the quantum group $U(\infty)_{q}$. This can be interpreted as an effect of averaging over background gauge fields.   We will describe the modular flow of these subregion open strings and compute their anyonic entanglement entropy.

The anyonic Chan Paton factors of these open strings define a set of subregion edge modes.  Via a q-deformed Schur Weyl duality, we can superpose the subregion open strings so that the Chan Paton factors organize into superselection sectors labelled by a tableaxu $R$.  These $R$ sector edge modess provide an internal Hilbert space for the R-type stringy objects described in section \ref{resolve}.   The picture is suggestive of a type of defect holography, where a bulk spacetime is reconstructed from summing over all defects.   Moroever, the edge mode entanglement can be interpreted as the result of the pair creation of these stringy objects from the vacuum.  Finally,  given a suitable factorization map from closed to open strings, we show that the anyonic entanglement entropy of open strings reproduces the gravitational entropy of closed strings.  

\subsection{Closed string algebra and combinatorial quantization } \label{section:subregion}
In the previous section, we identified the closed string Hilbert space $\mathcal{H}_{\text{closed}} $ with the Hilbert space of $U(N)$ Chern Simons theory  at large $N$.  This suggests we can arrive at a string theory subregion as a large N limit of a subregion in Chern Simons theory.  However, a naive application of this idea runs into an immediate problem.  In the standard quantization of Chern Simons theory, the subregion Hilbert spaces consists of infinite dimensional representations of the $U(N)$ Kacs-Moody algebra.  The associated factorization map on the torus produces a UV infinite entanglememt entropy, which does not match the  UV finite replica entropy computed in the section \ref{sec:closedstrings}.  This mismatch can be attributed to the non-topological nature of the usual gapless  entanglement boundary condition, which breaks diffeomorphism invariance along the entangling surface. 

On the other hand, as explained in \cite{Wong:2025asd},  the quantum group quantization of Chern Simons theory \cite{Mertens:2025ydx, Alekseev:1994au,Alekseev:1994pa}  naturally gives rise to a topological subregion that preserves diff invariance.   We will refer to these methods generically as  combinatorial quantization.  This quantization scheme avoids UV infinites by bypassing the local gauge field, and directly quantizing the matrix elements $U_{ij}$ of the holonomies, which are treated as the fundamental variables\footnote{The Poisson brackets of classical holonomies was worked out in \cite{Fock:1998nu}}.  
Upon quantization, they satisfy the quantum group algebra of $U(N)_{q}$, the $q$ deformed unitary group, with the $q$ deformation parameter given in \eqref{q}.   This means that the matrix elements $U_{ij}$ are promoted to operators that do not commute, but rather satisfy commutation relations specified by a quantum R matrix.  

A Hilbert space can be constructed from this operator algebra via the GNS construction \cite{Wong:2025asd}, which identifies algebra elements with states of the GNS Hilbert space.  In the case of the torus, the Hilbert space $\mathcal{H}_{T^2}$ of gauge invariant states is  spanned by quantum characters of $U(N)_{q}$.
These are defined via the quantum trace acting on an element $U \in U(N)_{q}$ \footnote{We choose to insert $D^{-1}$ instead of $D$ in the defintion of $\widetilde{tr}$ in order to align with choices we make in the open string algebra }: 
\begin{align}\label{qchar}
    \widetilde{tr}_{R}(U) &\equiv\tr_{R}(D^{-1} U) \nn
    &= \sum_{i} (D^{-1}_{R})_{ij} R(U)_{ji},
\end{align}
where $D$ is the ``balancing element" of the quantum group defined in  \eqref{D},  and $D_{R}$ is the same element in the representation $R$. 
The matrix $D$ is defined in terms of the R-matrix and double twist, which are braiding data associated to representation category of $U(N)_{q}$. Its insertion defines a \emph{categorical} trace
$\widetilde{\tr}_{R}$ on an irrep of $U(N)_{q}$. This is a trace on the Hilbert space of an anyon labelled by $R$, whose worldline traces out the Wilson loop.  It satisfies a twisted version of cyclicity and is used to define anyonic entanglement entropy:see \cite{Wong:2025asd} for a detailed review.   

The quantum characters \eqref{qchar} span the space of quantum group invariant functions\footnote{These are invariant under the quantum group version of conjugation.} on $U(N)_{q}$.  
They are orthonormal with respect to an inner product defined by a q-analogue of the Haar measure: 
\begin{align}
\int dU \widetilde{\tr}_{R}(U) \widetilde{\tr}_{R'} (U) &= \delta_{RR'}
\end{align}

There is an obvious isomorphism between the Hilbert space spanned by ordinary characters and the one spanned by quantum characters:
\begin{align}
   \tr_{R} (U) \rightarrow \widetilde{\text{tr}}_{R} (U).
\end{align}
Therefore, we can define the Hilbert space of the  closed A model TQFT in terms of quantum characters.  This means that we now interpret the wavefunctions for closed strings states $\ket{R}$ as :
\begin{align} \braket{U|R}=\widetilde{\tr}_{R}(U)
\end{align}

\paragraph{q-deformed boundary states }   Since combinatorial quantization elevates the matrix elements $U_{ij}$ to non-commuting operators, the state defined in \eqref{qchar} is a  superposition of operators rather than a function of a definite holonomy.   While this feature becomes important when we factorized closed strings into open strings, it is not relevant to computations of the closed A model TQFT that does not require cutting open the trace\footnote{\cite{deHaro:2006uvl} noted a similar phenomenon in q-deformed gauge theory, where the quantum trace is needed only when cutting open Wilson loops}.  Therefore, in formulating the closed theory, we will restrict to the maximal torus of diagonal holonomies, where the formulas for the quantum characters simplify.
Importantly, this allows us to make sense of the closed string boundary states corresponding to the $R$ type branes defined in \ref{resolve} , which involve discrete holonomies  $U$ belonging to the maximal torus of $U(N)$. In particular, this includes the shrinkable holonomy corresponding to the entanglement branes.  To avoid confusion, we will use the symbol $\chi_{R}(U)$ to denote restriction of the trace to the maximal torus (with $\chi(U)$ denoting the character in the fundamental rep), and write our closed string wavefunction as
\begin{align}
\braket{U|R}=\widetilde{\chi}_{R}(U),\qquad  \braket{U|\vec{k}} =
 \prod_{n=1}^{\infty} (\widetilde{\chi} (U^{n}))^{k_{n}}
\end{align}

To explain this in more detail, recall that the  maximal torus for $U(N)$ consists of diagonal $N\times N$ matrices of phases.  Writing $H_{i}$ for  a matrix with nonzero entry on the ith diagaonal, an element of the maximal torus is
\begin{align}\label{Uexpa}
    U= e^{\vec{a}\cdot \vec{H} }
\end{align}
The classical characters are then given by
\begin{align}
    \chi_{R}(e^{\vec{a} \cdot \vec{H}}) = \sum_{\mu \in W_{R}} e^{\vec{a} \cdot \vec{\mu}}
\end{align}
where $\vec{\mu}$ is a weight of the representation $R$.  For a quantum group, we still have a Cartan algebra defined by $N$ commuting generators $H_{i}$ from which we can obtain generators of the maximal torus:
\begin{align}
    K_{i}= q^{H_{i}}
\end{align}
As in \eqref{Uexpa}, a diagonal quantum group element can be expressed as
\begin{align}
    U=\prod_{i} K_{i}^{a_{i}}
\end{align}
and evaluating the quantum character on this gives
\begin{align}
    \widetilde{\chi}_{R}(U)= \chi_{R}(q^{\rho }\prod_{i} q^{a_{i}} ) = \chi_{R}(q^{\rho + a})
\end{align}
Thus, along the maximal torus, the quantum characters are just the ordinary characters shifted by the Weyl vector.
This means that  the Weyl vector part of the discrete holonomies in \eqref{discrete}  can be absorbed into the quantum trace.  In this formulation of the A model TQFT, the holonomy state $\ket{U}$ which implements the geometric transition of entanglement branes is the identity $\ket{U}=\ket{1}$.

\subsection{The topological subregion algebra }
We now define a topological  subregion in Chern Simons theory, which we identify as a subregion of the closed string theory.  First, let us recall that Chern Simons theory in combinatorial quantization is equivalent to a q-deformed lattice gauge theory \cite{Alekseev:1994au,Alekseev:1994pa}.   The quantization scheme proceeds as follows. On each spatial manifold $\Sigma$, an extended state space is first defined by triangulating $\Sigma$, and then assigning the quantum group algebra  $L^{2}(U(N)_{q})$ to each link.    This is the space of functions on quantum group elements  $U$ living on a link which defines an algebra under pointwise multiplication.  Via the GNS construction this defines an extended state space on each link, from which the gauge invariant Hilbert space $\mathcal{H}_{\Sigma}$ is constructed  by imposing gauge constraints at the vertices, and applying a quantum traces around each loop.   The quantum trace renders the physical states invariant under quantum group gauge transformations. 

Following the standard approach to Hilbert space factorization  in lattice gauge theory, we can define a subregion Hilbert space by  lifting the gauge constraints that cross the entangling surface.  On the torus, we choose a triangulation that split the spatial cycle into two links, corresponding to two cylindrical regions $V$  and $\bar{V}$ as shown below: 
\begin{align}\label{Tsplit}
\includegraphics[scale=.2]{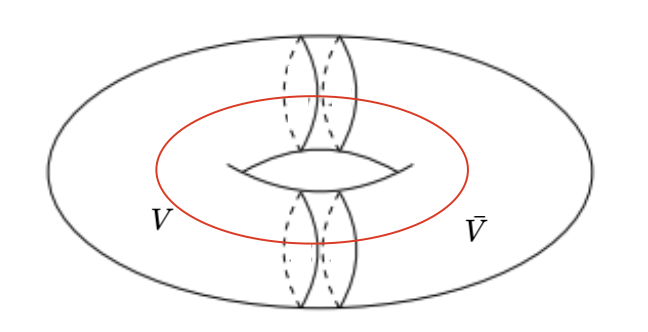}
\end{align}
Then the subregion Hilbert space $\mathcal{H}_{V}$ is just the  GNS Hilbert space for the algebra on a single link, given by $L^{2}(U(N)_{q})$.   Note that  the subregion algebra is bigger than the gauge invariant algebra on the torus, because gauge constraints have been lifted. This leads to anyonic edge modes on the boundary of each link.

We interpret  $L^{2}(U(N)_{q}$ as an effective Hilbert space for open strings stretched between entanglement branes.  A single open string is represented by the wavefunction $U_{ij}$, a matrix element in the fundamental (co)representation of  $U(N)_{q}$.    A general n-string state lablelled by Chan Paton factors $I,J$  is given by the product 
\begin{align} \label{ostrings}
        \braket{U|IJ} &= U_{i_{1}j_{1}} \cdots U_{i_{n}j_{n}},
\end{align}
and  satisfy anyonic statistics specified by the R matrix.
 
As in the case of the closed string,  there are  superpositions of open string states that are labelled by a tableaux $R$.  These arise from a q-analogue of the Schur Weyl duality, which implies we can obtain irreps of   $U(N)_{q}$ via a braided symmetrization/anti symmetrization\footnote{The action of the permutation group in the usual Schur Weyl duality is replaced with its q-deformation, which produces the Hecke algebra. See \cite{deHaro:2006uvl} for more details}   of the indices $I,J$.   We denote these representation basis elements by $R_{ab}(U)$ , with $a,b,=1,\cdots \dim R$ labelling the symmetrized  Chan Paton factors.  They form an orthogonal basis with respect to the inner product defined by the q-haar functional $h: L^{2}(U(N)_{q}) \to \mathbb{R} $
\begin{align}
    h(R_{ab}^*(U) R_{cd}(U)) =\frac{\delta_{ac}\delta_{bd} D_{bb}^R }{\dim_{q}R}
\end{align}
The Hilbert space thus  admits the decomposition 
\begin{align}
    L^{2}( U(N)_{q}) &= \bigoplus_{R} V^{*}_{R} \otimes  V_{R}\nn
    &= \text{span} \{R_{ab}(U) \cdots a,b=1 \cdots \dim R \},
\end{align}
which represents a  q-analogue of the Peter Weyl theoem.

Each R sectors can be interpreted as the entangled  Hilbert space of the $R$ type stringy objects described in section \ref{resolve}, with $V_{R}$ and $V_{R}*$ representing the internal Hilbert space of the object and its ``negative" counterpart.  

\subsection{Open string modular flow and ribbon ZX calculus} 
The subregion GNS Hilbert space we have defined is based on the algebra of functions on a compact  quantum group with pointwise multiplication as the product.   However there is also a Frobenius algebra on this function space defined by the convolution product.   These two algebraic structures  produce a  ``ZX calculus "\footnote{While the ZX diagrammatic calculus was developed for finite dimensional Hilbert space, we recently generalized it to the infinite dimensional setting involving $L^{2}(G)$ for a compact group \cite{Wong:2026igo}, and quantum groups} that provides an efficient description of quantum information processes on  the subregion \cite{Wong:2026igo}.  
In particular, the Frobenius algebra operations produce the basic building blocks that generate subregion modular flow. These building blocks can be represented by ribbon diagrams as we now review. 

To begin with, we represent the cylinder Hilbert space $L^{2}( U(N)_{q})$ subregion as an interval
\begin{align} \label{cylinder}
 \vcenter{\hbox{\includegraphics[scale=.2]{figures/Hopen.png}}}=   \vcenter{\hbox{{
\includegraphics[scale=.1]{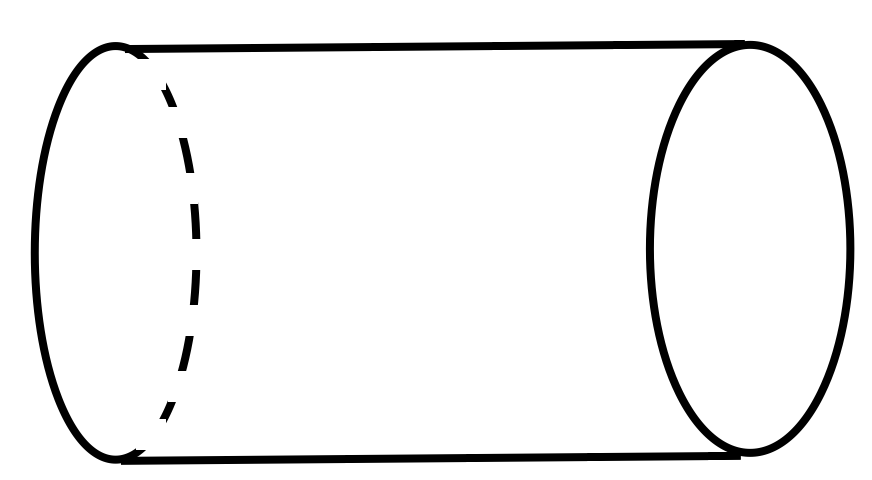} }}},
\end{align}
so that its  modular time evolution traces out a ribbon. 
The Peter Weyl theorem implies this ribbon can also be understood in terms of entangled anyons:
\begin{align} \label{openprop}
\mathtikz{\idA{0}{0}}=\bigoplus_{R} \mathtikz{\stripRR{0cm}{0cm}} = \bigoplus_{R}  1_{R^*\otimes R} e^{-t l(R)}
\end{align} 
This defines a propagator with a Boltzmann factor $e^{-t l(R)}$ which gives the dependence on the ``relative" Kahler parameter and plays the role of a convergence factor.   
The ribbon can split and join, producing the open algebra generators in \eqref{openfrob}.   In addition, there is a ``copy" operation on the quantum group, described by the stacking of the worldsheets
\begin{align}
    \mathtikz{\nablaAflip{0}{0}},
\end{align}
which corresponds to sums of products of Clebsch Gordon coefficients for the quantum group. See \cite{Wong:2025asd,Wong:2026igo} for more details.

The Peter Weyl theorem implies these ribbon graphs can be interpreted in terms of anyon worldlines. For example, the product is a fusion of intervals given by
\begin{align}
 \mu&= \mathtikz{\deltaA{0}{0}}  = \sum_{R} \frac{1}{\sqrt{\dim_{q}R}} \vcenter{\hbox{
\includegraphics[scale=.1]{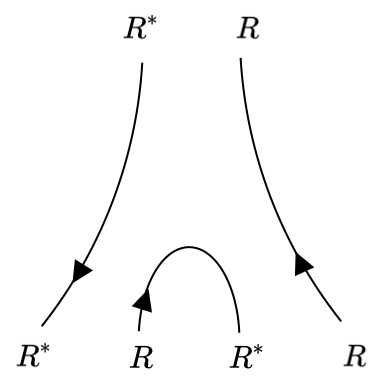}}}
\nn
\mu&: \ket{R, a,b } \otimes \ket{R',c,d}  \longrightarrow \frac{1}{\sqrt{\dim_{q}R}}  \delta_{RR'} D^{R}_{bc} \ket{R,a,d}
\end{align}

While the co-product describes the factorization of one interval into two: 
\begin{align}
  \Delta&=  \mathtikz{\muA{0}{0}}  =\sum_{R} \frac{1}{\sqrt{\dim_{q}R}}\vcenter{\hbox{\includegraphics[scale=.1]{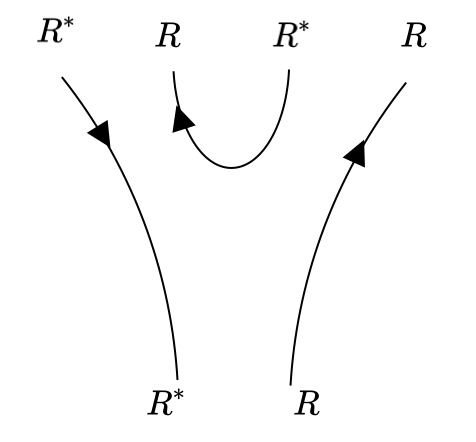}}} \nn 
  \Delta&:   \ket{R a,b}   \to   \frac{1}{\sqrt{\dim_{q}R}} \sum_{c}  \ket{R ,a,c} \otimes \ket{R,c,b}.
\end{align}
This  correspond to the (co) representation property 
$R(U)_{ab}  \to  \sum_{c} R(U)_{ac} \otimes R(U)_{cb}$, and represents an entangling operation between two subregions.

Notice that the product and co-product involves the pair annhilation and creation of anyons,  with the pair creation responsible for the edge mode entanglement between two  subregions.   Similarly, the unit and co-unit are given by
\begin{align}
\mathtikz{\epsilonA{0}{0}}   &=\sum_{R} \vcenter{\hbox{
\includegraphics[scale=.1]{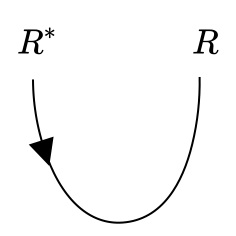}}}=\sum_{R,a}  \sqrt{\dim_{q}R} (D_{aa}^{R})^{-1} \ket{R, a,a}\nn
\mathtikz{\etaA{0}{0}}   &=\sum_{R} \vcenter{\hbox{
\includegraphics[scale=.1]{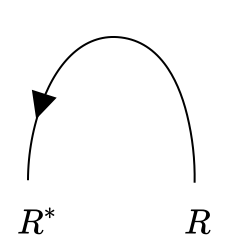}}} : \ket{R,a,b} \to \sqrt{\dim_{q}R} \delta_{ab},
\end{align}
corresponding to summing over all pair creations and pair annhilations respectively.   Thus, these subregion operations cut open the anyon loops dual to the vacuum diagrams of bulk stringy objects, and exposes their internal quantum number labelled by $a,b$.

The q-deformation introduces a nontrivial braiding operation:
\begin{align}
\mathtikz{\rightbraidA{0}{0}}
\end{align}
so that the non-commutativity of point wise multiplication can be expressed as 
\begin{align}
    \mathtikz{\nablaAflip{0}{0}  \rightbraidA{0 cm}{0cm} ;\node at (-.5cm ,-1.3 cm){A} ;\node at (.5cm ,-1.3 cm){B} ;\node at (0cm ,1.3 cm){C}} =\bigoplus_{A,B,C} R_{AB}^{C} \mathtikz{\nablaAflip{0}{0} ;\node at (-.5cm ,-.3 cm){A} ;\node at (.5cm ,-.3 cm){B} ;\node at (0cm ,1.3 cm){C} },
\end{align}
where $R_{AB}^{C}$ is an R-matrix for the regular representation of $U(N)_{q}$.

\paragraph{The Dehn twist} We can apply a Dehn twist to the cylinder \eqref{cylinder},  along a circle that is hidden in the ribbon notation.  This is diagonal in the  representation basis: 
\begin{align}
T \ket{R,a,b} = q^{ \frac{C_{2}(R)}{2}}  \ket{R,a,b} 
\end{align}
where 
\begin{align}
    C_{2}(R) =\kappa_{R} + N l(R)
\end{align}
is the quadratic Casimir.  
This operation is essential in producing the relative framing needed to match with the closed string partition functions.  We will represent it as the insertion of the  $T$ operator along  the interval: 
\begin{align}
\mathtikz{\pairAT{0}{0} \copairAT{0}{0}}
\end{align}
\paragraph{The quantum partial trace}
Gluing the co-unit to the product or the unit to the co-product leads respectivelt to a pairing of states (cup) and a bulk thermofield double  (cap) connecting two subregions:
\begin{align}\label{capcup}
\mathtikz{\copairA{0}{0}} &: \ket{R,a,b} \otimes \ket{R,c,d} \to  \delta_{ad} \delta_{bc} D_{bc}^{R}\nn
\mathtikz{\pairA{0}{0}}&= \sum_{R,a,b} (D_{aa}^{R})^{-1} \ket{R,a,b} \ket{R,b,a}
 \end{align}
Note the different placement of the $D$ and $D^{-1}$.  These operations defined adjoint operations that map between states and their duals.    
There are two ways of gluing the ``cup" and the ``cap". One produces the zigzag id:
\begin{align}
\mathtikz{\copairA{0}{0} \pairA{1cm}{0cm} }= \mathtikz{ \idA{0}{0}}
\end{align}
and the other produces a \textbf{quantum trace} on the effective open string Hilbert space: 
\begin{align}
\mathtikz{\pairA{0}{0} \copairA{0}{0}} &=  \sum_{R,a,b} (D_{aa}^{R})^{-1} D_{bb}^{R}e^{-t l(R)} \nn
&= \sum_{R} (\dim_{q}R)^{2} e^{-t l(R)}\nn
&= \widetilde{\tr}_{L^{2}(U(N)_{q}} (e^{-t l(R)})
\end{align}
Notice that while the balancing elements cancelled in \eqref{capcup}, they combine to make $(\dim_{q}R)^{2} $ in the quantum trace.   These properties are captured more elegantly in the anyon notation, where they amount to a choice of compatible orientation of the anyon lines, and the rule that an anyon loop produces a factor of $\dim_{q}R$.
For example, the ribbon cup and caps are equivalent to the anyon diagrams:
\begin{align}
    \includegraphics[scale=.1]{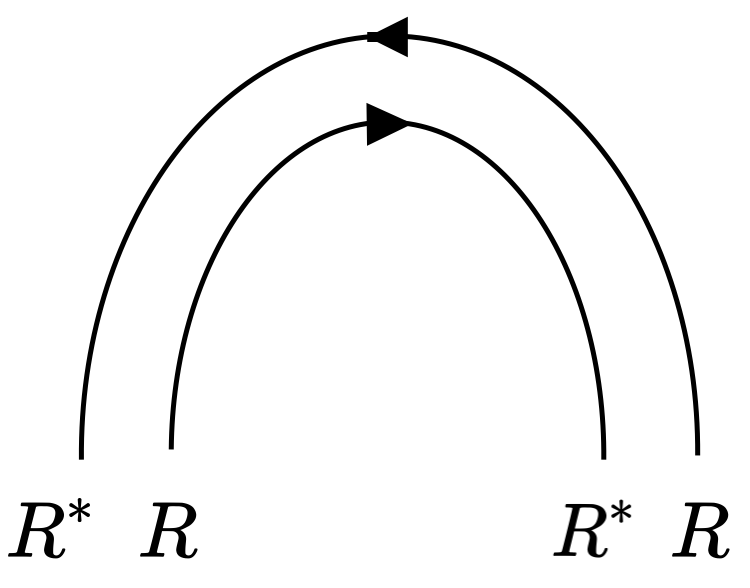} ,\qquad    \includegraphics[scale=.1]{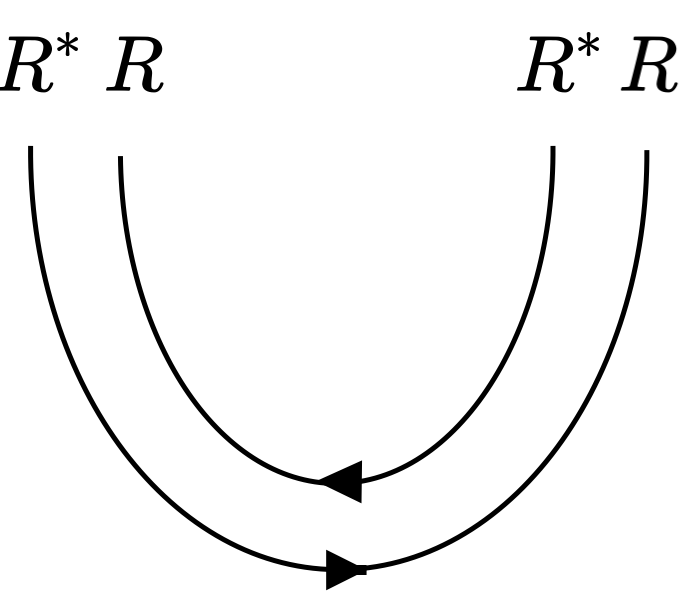} 
\end{align}
Both ways of gluing the ribbon caps and cups are compatible with these  orientations, and the cancellations of the balancing element corresopnds to the zigzag identity for each anyon line. See
\cite{Wong2025} for a more detailed review of how the balancing element enters in the Hilbert space description of anyons, as well as an interpretation of the quantum trace as a renormalized trace.

Finally, applying these adjoint operations  to a general density matrix on a subregion defines a left and right \textbf{quantum partial trace} :
\begin{align}\label{partial trace}
      \mathtikz{ \idA{-3.5cm}{0cm}; \pairA{-3cm}{-1cm}; \copairA{-3cm}{0cm}; 
\idA{-1.5cm}{1cm};
\idA{-1.5cm}{-1cm};
\draw[thick] (-3cm,-1cm) rectangle  (-1cm,0cm);} ,\quad \mathtikz{ \idA{-.5cm}{0cm}; \pairA{-1cm}{-1cm}; \copairA{-1cm}{0cm}; 
\idA{-2.5cm}{1cm};
\idA{-2.5cm}{-1cm};
\draw[thick] (-3cm,-1cm) rectangle  (-1cm,0cm);} .
\end{align} 
This is equivalent to the conventional anyonic partial trace, which is used to define anyonic reduced density matrix and its associated entanglement entropy and negativity \cite{Bonderson:2017osr, Shapourian:2020tmc}.   
It is important to keep in mind that left and right partial tracing are distinguished due to the presence of nontrivial braiding.   In particular, given a bipartition of anyon lines into $A$ and $B$ :
\begin{align}
   \rho =  \mathtikz{ \idA{-1.5cm}{1cm} ; \idA{-1.5cm}{-1cm};
\idA{-2.5cm}{1cm};
\idA{-2.5cm}{-1cm};
\draw[thick] (-3cm,-1cm) rectangle  (-1cm,0cm);\node at (-1.5,1.5) {$b$};\node at (-2.5,1.5) {$a$} } ,
\end{align}
we define the reduced density matrix on $a$ by right partial tracing:
\begin{align}
\rho_{a} = \widetilde{\tr}_{b,R}  \,\, \rho
   =  \mathtikz{ \idA{-.5cm}{0cm}; \pairA{-1cm}{-1cm}; \copairA{-1cm}{0cm}; 
\idA{-2.5cm}{1cm};
\idA{-2.5cm}{-1cm};
\draw[thick] (-3cm,-1cm) rectangle  (-1cm,0cm);} .
\end{align}
But we define the anyonic entanglement entropy using the left partial trace on $a$ \cite{Couvreur_2017}:
\begin{align}
    S_{a} \equiv - \widetilde{\tr}_{a,L} \rho_{a} \log \rho_{a} 
\end{align}
As shown in \cite{Kato:2013ava} and recently reviewed in \cite{Wong:2025asd}, anyonic entanglement entropy has an operational interpretation in terms of distilling of anyon bell pairs.

\subsection{The large N open-closed algebra and anyonic entanglement entropy }
To obtain a match between the anyonic entanglement entropy of open strings and the gravitational entropy of closed strings, we must  apply the double scaled limit \eqref{dscale} directly to the ribbon diagrams.   In other words, we take the large N limit of the subregion algebra, associated to the category Rep$(U(N)_{q}$, rather than the limit of the anyonic entanglement entropy as computed in Chern Simons theory.   This has the effect of producing the quantum dimensions $d_{q}(R)$ for each anyonic loop, which implements the geometric transition of the entanglement branes.  
This produces a compatibility between the subregion algebra and the closed string algebra presented in section \ref{sec:closedstrings}, as we now show. 

Consider the double scaling limit on the ribbon diagrams:
\begin{align}
 \mathtikz{\deltaA{0}{0}}  &: \ket{R, a,b } \otimes \ket{R',c,d}  \longrightarrow \frac{1}{\sqrt{d_{q}(R)q^{\kappa_{R}/4}}}  \delta_{RR'} D^{R}_{bc} \ket{R,a,d}\nn
 \mathtikz{\muA{0}{0}}  &:   \ket{R a,b}   \to   \frac{1}{\sqrt{d_{q}(R) q^{\kappa_{R}/4}}} \sum_{c}  \ket{R ,a,c} \otimes \ket{R,c,b}\nn
 \mathtikz{\epsilonA{0}{0}}   &=\sum_{R,a}  \sqrt{d_{q}(R) q^{\kappa_{R}/4}} (D_{aa}^{R})^{-1} \ket{R, a,a}\nn
\mathtikz{\etaA{0}{0}}   & : \ket{R,a,b} \to \sqrt{d_{q}(R) q^{\kappa_{R}/4}}\delta_{ab},
\end{align}
In these expressions, it is implicit that when we glue the diagrams together to produce a trace of the balancing element $D^{R}$, we apply the chiral double scaled limit so that:
\begin{align} \label{clockwise}
    \sum_{a} D^{R}_{aa} =\sum_{a} q^{-a +\frac{N+1}{2}}  \longrightarrow d_{q}(R) q^{\kappa_{R}/4 } q^{Nl(R)/2}
\end{align}
and for the inverse:
\begin{align}\label{cclockwise}
\sum_{a} (D^{R}_{aa})^{-1} \to  (-1)^{l(R)} d_{q}(R) q^{-\kappa_{R}/4} q^{Nl(R)/2}
\end{align}
Given these limits, any holes produced by the splitting and joining of the ribbons can be closed.  This expresses an important of the open string algebra.   For example, 
the isometric property of the co-product is encoded in the relation:
\begin{align}
     \mathtikz{\muA{0}{-1cm} \deltaA{0}{0}} = \mathtikz{\idA{0}{0}} \longleftrightarrow \sum_{R} \frac{1}{d_{q}(R) q^{\kappa_{R}/4}}  \vcenter{\hbox{\includegraphics[scale=.1]{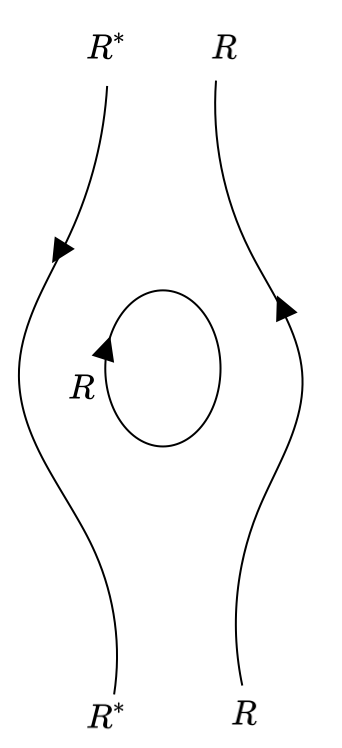}}} = \sum_{R} \mathtikz{\stripRR{0cm}{0cm}}
\end{align}
In the anyon langauge, this follows because every loop can be replaced by a factor of the quantum dimension, in accordance with diagrammatic rules of a modular tensor category\footnote{In the double scaled large N limit, each loop in sector $R$ is replaced by a factor of $d_{q}(R)q^{\kappa_{R}/4}$ , which cancels against the amplitude for the process in each R sector }. 

Thus, filling in a hole with a disk is implemented by summing over all vacuum loops of stringy objects around it.  This leads to an intepretation of local holography as a direct sum of defect holography: the latter relates an anyon loop to a single vacuum diagram of a bulk stringy object, where as the former relates the entanglement brane superposition of these anyon loops to a patch of bulk spacetime.

Finally, we note that in the anyon language, the two different limits in \eqref{clockwise}, \eqref{cclockwise} are associated to different orientations of anyon loops:
The ``interior" anyon loops in the clockwise orientation gives  \eqref{clockwise}, whereas ``exterior" anyon loops in the counterclockwise orientation gives \eqref{cclockwise}.  
The sign factor of $(-1)^{l(R)}$ for loops in the counterclockwise orientation is an important feature of the large N limit, and is  needed to produce the transition  between  open string and closed string backgrounds.  For example, this rule produces the transition between $\mathbb{C}^3$ and the resolved conifold:
\begin{align}
\mathtikz{ \pairA{0cm}{0cm} \copairA{0cm}{0cm};\node at (0,-1.2) {($0,0$)}  } =
 \mathtikz{  \epsilonC{0cm}{0cm} \etaC{0cm}{0cm};\node at (0,-.7) {($-1, -1$)}  }= \sum_{R} (-1)^{l(R)} d_{q}(R)^2  e^{-tl(R)}.
\end{align}
While we previously derived this transition using the Ooguri Vafa operator \eqref{ooguri},  our formulation of the subregion algebra provides an interpretation of this partition function as a large N quantum trace. 
\paragraph{Relating the open and closed sectors}
To give a systematic relation between the open and closed algebra, we define homomorphisms between them in the large N limit.   From the perspective of large $N$ Chern Simons theory, the closed Hilbert space corresponds to the center algebra $\mathcal{C}(L^{2}(U(\infty)_{q})$ with respect to the convolution product.   The center is spanned by the quantum characters, which are naturally embedded  into into the subregion algebra $L^{2}(U(\infty)_{q}$  by the ``zipper":
\begin{align}
    \mathtikz{\cozipper {0}{0} } : \ket{R} \to \sum_{a}    \frac{(D_{aa}^{R})^{-1} }{\sqrt{d_{q}(R) q^{\kappa_{R}/4}} }  \ket{R a, a}
\end{align}
This cuts the closed strings into an open strings.   The fact that it is a homomorphism means that
\begin{align}
\tikz[baseline=0cm] { \cozipper{0cm}{1cm} \deltaC{0cm}{0cm} ;\node at (-1,0) {$(0,1)$}} \quad = \quad 
\tikz[baseline=0cm] { \deltaA{0cm}{1cm} \cozipper{-0.5cm}{0cm} \cozipper{0.5cm}{0cm} }
\end{align}
and that it maps the closed unit to the open unit
\begin{align}\label{untitounit}
    \mathtikz{\epsilonC{0}{-1 cm} \cozipper {0}{0}; \node at (-1,-1) {$(0,-1)$} } = \mathtikz{\epsilonA{0}{0}},
\end{align}
This can be readily verified with our formulas for the open closed cobordisms.  Notice that this gives another check that with our conventions, the closed unit is not the generator appearing in \eqref{unit}, but is related to it by applying a relative twist.  

The inverse operation is given by the co-zipper is
\begin{align}
    \mathtikz{\zipper{0}{0}} : \ket{R,a,b} \to \frac{1}{\sqrt{d_{q}(R) q^{\kappa_{R}/4}}}\delta_{ab} \ket{R},
\end{align}
which glues open strings back into closed strings. 
These two operations cuts a closed strings into open strings and then glues open strings back into a closed strings.  The entanglement brane axiom holds for these ``windows":
\begin{align}
     \mathtikz{\zipper{0}{0}\cozipper {0}{-1cm} } = \mathtikz{\idC{0}{0}}
\end{align}

Combining this with the co-product produces the factorization map from the torus to the cylinder: 
    \begin{align}
\mathtikz{\cozipper{0}{0} \muA{0cm}{1cm}}: \mathcal{H}_{ \text{closed}} &\to \mathcal{H}_{V} \otimes \mathcal{H}_{\bar{V}}\nn
\widetilde{\text{Tr}}_{R} (U) &\longrightarrow \sum_{a,b=1}^{\dim R}  D^{R}_{aa}R_{ab} (U_{1}) R_{ba}(U_{2}).
\end{align} This maps open an arbitrary closed string state into an entangled state of open strings.  Note that in this context, where we cut open the closed string trace, it is important to treat $U$ as a general quantum group element, rather than an element of the maximal torus.
\paragraph{Anyonic entanglement entropy}
We now have all the ingredients needed to compute the anyonic entanglement entropy for a subregion of a general closed string state.    Using the factorization map, we can cut open any closed string state and obtain its reduced density matrix diagrammatically.  The simplest example is the state corresponding to the Calabi cap, which can be factorized to produce the entangled thermofield double state of open strings:
\begin{align}
 \mathtikz{\epsilonC{0}{-1cm} \cozipper{0}{0} \muA{0cm}{1cm}} = \mathtikz{\muA{0cm}{1cm} \epsilonA{0}{0}} = \sum_{R,a,b} D_{aa}^R \ket{R,a,b} \ket{R,b,a}
\end{align}

The factorized anyonic density matrix is    
\begin{align}
  \rho =\mathtikz{\pairA{0}{1 cm} \copairA{0}{-1cm}}
\end{align}
Applying a quantum partial trace and restoring the dependence on the Kahler parameter $t$ gives the reduced density matrix
\begin{align}
    \widetilde{\tr} \rho  = \mathtikz{\pairA{0cm}{1cm} \copairA{1cm}{1cm} \copairA{0cm}{-1cm} \pairA{1cm}{-1cm} \idA{1.5cm}{0cm}\idA{1.5cm}{1cm}} =\mathtikz{\idA{0cm}{0cm}\idA{0cm}{1cm}}=\sum_R e^{-t l(R)} \mathbf{1}_{R}=\sum_{R,a,b}  (\tilde{d}_{q}(R))^{2} e^{-t l(R)}  \widehat{\mathbf{1}_{R}},
\end{align}
where 
\begin{align}
     \widehat{\mathbf{1}_{R}} = \frac{\mathbf{1}_{R} }{\tilde{d}_{q}(R))^{2}  }
\end{align}
is the density matrices within each $R$ block   normalized with  respect to the quantum trace.  
The anyonic entanglement entropy is defined as the Von Neumann entropy of the anyonic density matrix $\hat{\rho}$  \emph{normalized with  respect to the quantum trace}\cite{Bonderson:2017osr}.  So defining 
\begin{align}
    \hat{\rho}=  \sum_{R} P(R)  \widehat{\mathbf{1}_{R}},\qquad   P(R) = \frac{ (\tilde{d
    }_{q}(R))^2 e^{-tl(R)}}{Z}, \qquad Z =  \sum_{R} (\tilde{d}_{q}R)^2 e^{-tl(R) }
\end{align} 
the anyonic entanglement entropy is given by
\begin{align}\label{AEE}
    S &= -\widetilde{\tr} \hat{\rho} \log \hat{\rho}
    = -\sum_{R} P(R) \log P(R)  + 2 P(R) \log \tilde{d}_{q}(R),
\end{align}
This agrees with the gravitational entropy of the resolved conifold, as demanded by the entanglement brane axiom. 
We can generalize this computation in several directions.
First, we can introduce a nontrivial relative framing in to the closed string background.   For example, consider the $(0,-2) $ sphere.
\begin{align}
    \mathtikz{ \epsilonC{0}{0} \etaC{0}{0} \node at (1,0) {$(0,-2$)} }  = \sum_{R} d_{q}(R)^2 q^{\frac{\kappa_{R}}{2}} e^{-tl(R) }
\end{align}
It is produced by an open string trace with the insertion of a twist: 
\begin{align}
\mathtikz{\pairAT{0}{0} \copairAT{0}{0}} =\sum_{R} d_{q}(R)^2 q^{\frac{\kappa_{R}}{2}} e^{-tl(R) },
\end{align}
where the the large N limit of $q^{C_{2}(R)}$ produces the $q^{\frac{\kappa_{R}}{2}}$ factor, plus a term $q^{Nl(R)/2}$ that is absorbed into the Kahler modulus.  
We can compute the anyonic entanglement entropy of this state from its reduced density matrix, which is now given by:
\begin{align}
    \widetilde{\tr} \rho  = \mathtikz{\pairAT{0cm}{1cm} \copairAT{1cm}{1cm} \copairA{0cm}{-1cm} \pairA{1cm}{-1cm} \idA{1.5cm}{0cm}\idA{1.5cm}{1cm}} =\mathtikz{\idAT{0cm}{0cm}}=\sum_R q^{\kappa_{R}/2}e^{-t l(R)} \mathbf{1}_{R}=\sum_{R,a,b}  q^{\kappa_{R}/2}e^{-t l(R)}  |R,a,b\rangle \bra{R,a,b}.
\end{align}
The anyonic entanglement entropy proceeds as before and takes the form \eqref{AEE}, with the probability factor $P(R)$ modified to
\begin{align}
P(R)=   \frac{ (d_{q}(R))^2q^{\kappa_{R}/2} e^{-tl(R)}}{Z}, \qquad Z =  \sum_{R} (d_{q}R)^2 q^{\kappa_{R}/2} e^{-tl(R) }
\end{align}

Second, we can also consider multiple disconnected  subregions.   Starting with the same pure state given by the Calabi Yau cap, we can factorized the closed string Hilbert space into the tensor product of 4 open string Hilbert spaces: 
\begin{equation}
\mathtikz{ \epsilonC{0cm}{-2cm} \muA{-.72cm}{1cm} \muA{.72cm}{1cm} \semiwidemuA{0cm}{-1cm} \cozipper{0cm}{-1cm} ; \node at (-2,-1) {$(0,-1)$} }
\end{equation} 
Using the identity \eqref{untitounit},  the associated density matrix
\begin{align}
\includegraphics[scale=.2]{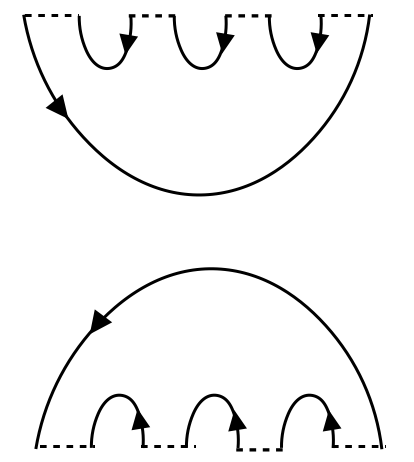}
\end{align}

Applying the quantum partial trace to two of the four intervals produces the anyonic reduced density matrix 
\begin{align}\label{two} 
\rho = \vcenter{\hbox{\includegraphics[scale=.4]{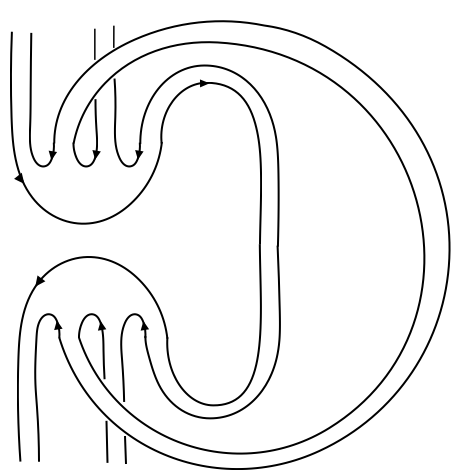} }}  &= \mathtikz{
\idA{1.5cm}{0cm} \muA{0cm}{0cm}
\idA{-.5cm}{1cm} \leftbraidA{1cm}{1cm}
\deltaA{0cm}{2cm} \idA{1.5cm}{2cm}
\deltaA{0cm}{-1cm} \idA{1.5cm}{-1cm}
\idA{-0.5cm}{-2cm} \rightbraidA{1cm}{-2cm}
 \muA{0cm}{-3cm} \idA{1.5cm}{-3cm}
} ,
\end{align}
which can be exhibited as the linear map:
\begin{align} 
\rho : \ket{R ,a,b} \ket{R,c,d} \to \frac{D_{bb}^{R}D_{cc}^{R}D_{dd}^{R}(D_{aa}^{R})^{-1} e^{-tl(R)}}{(d_{q}(R))^{2} q^{\kappa_{R}/4}} \ket{R ,a,b} \ket{R,c,d} 
\end{align}
In this case, we see qualitative differences with the modular flow for a single region:  the flow involves a nontrivial  topological interaction of the two intervals, which split and reconnect.  These correspond to pair creation of anyons which then fuse back with a different anyon partner into the vacuum, occuring with relative probablity $d_{q}(R)) $. This occurs once in the upper half of the diagram, corresponding to the half modular flow $\rho^{1/2}$, and then again in the lowerhalf, given the total normalization factor of  $\frac{1}{(d_{q}(R))^{2} q^{\kappa_{R}/4}}$. 

The density matrix in each sector, normalized with respect to the left partial trace \footnote{The left partial trace produces four anyon loops, two clockwise and two counter-clockwise which leads to the $\tilde{d}_{q}(R)^4$ factor }, is given by
\begin{align}
 \widehat{\mathbf{1}_{R}} \otimes  \widehat{\mathbf{1}_{R}} = \frac{\mathbf{1}_{R} \otimes\mathbf{1}_{R} }{\tilde{d}_{q}(R))^{4}  },
\end{align} 
which we can substitute into the total normalized density matrix:
\begin{align}
    \hat{\rho}=  \sum_{R} P(R)  \widehat{\mathbf{1}_{R}}  \otimes \widehat{\mathbf{1}_{R}},\qquad   P(R) = \frac{ (\tilde{d
    }_{q}(R))^2 e^{-tl(R)}}{Z}, \qquad Z =  \sum_{R} (\tilde{d}_{q}R)^2 e^{-tl(R) }
\end{align} 
The anyonic entanglement entropy is given by
\begin{align} \label{anyonEE}
    S &= -\widetilde{\tr} \hat{\rho} \log \hat{\rho}
    = -\sum_{R} P(R) \log P(R)  + 4 P(R) \log \tilde{d}_{q}(R),
\end{align}
which nontrivially matches with the gravitational entropy. 
Finally, we also add arbitrary handles to the state preparation, which repeats the same story with a new value for $P(R)$.

\subsection{Averaging and the quantum trace}
Finally, we would like to make a brief remark on the stringy origin of the quantum trace, which has played a crucial role in the  open string computations of this section. 

In the conventional quantization of the topological string, there is a Hilbert space of ordinary Chan Paton factors with an ordinary trace that gives rise to factors of $Ng_{s}$ for each boundary of the worldsheet.   However, applying this trace to the multi open string  Hilbert space does not produce the geometric transition to a closed string background.  Indeed, consider the partition function obtained by integrating out open strings stretched between two stacks of ordinary branes and negative branes with trivial holonomy\footnote{
Here we used the fact that $\tr_{R}(1)=\tr_{R^t}(1)$ in the large N limit. } :
\begin{align}
     Z_{\text{open}}(U=1,V=1)=\sum_{R} \tr_{R\otimes R^*}( (-1)^{l(R)}e^{-t_{op} l(R)})
\end{align}
Here we applied the Ooguri-vafa operator \eqref{ooguri} with $U=V=1$, corresponding to stacks of  branes with zero separation.  This partition function produces an ordinary (fermionic) trace over the open string Hilbert space $L^{2}(U(N))$. But it does not produce a closed string partition function since it fails to produce q -deformed dimensions.

On the other hand, entanglement branes support a  special value of the world volume holonomy which produces a coarse graining.  As we have seen, strings stretched between these branes give: 
\begin{align}
Z_{\text{open}}(U=D,V=D)=\sum_{R} \braket{W_{R}(q^{-1})} e^{-t_{op} l(R)}\braket{W_{R}(q)},
\end{align}
where $\braket{W_{R}(q)}$ is the expectation value of a Wilson loop in $S^3$.  This involves a path integration over the worldvolume Chern Simons gauge field, which is interpreted as an averaging over the background from the worldsheet point of view.
This coarse graining effect is captured by the q-deformation of the subregion algebra, and changes the ordinary trace into a quantum trace.

\section{Conclusion}

In this work, we gave a precise realization of local holography \cite{Wong:2025kpz} in the A model topological string, and showed how it is applied systematically in canonical computations of  gravitational entropy.   Mathematically, local holography is realized by the extension of the A model TQFT, which introduces a subsystem open string algebra given by $L^{2}(U(\infty)_{q})$.  We explained how this algebra arises from  open strings stretched between entanglement branes, and showed how the large N quantum trace implements the transition of these branes.  Furthuremore, we showed that these transitions are related to defect holography of stringy objects labelled by representations of $U(\infty)_{q}$, and illustrated how the pair creation and annhilation of these objects arise in the modular flow of subsystem open strings.  

The consistency between our open string and closed string algebra depended delicately on a chiral, double scaled limit that relates the quantum dimensions of $U(N)_{q}$ and the q-deformed symmetric group.  Perhaps a simpler formulation of the open string algebra can be obtained  by introducing the q deformed quantum group algebra  directly.  This maybe related to recent ideas in open-closed Gromov-Witten theory as formulated in  \cite{Troost:2026pmx}.   In this context, we might expect that open string modular flow can be via the angular quantization of closed string worldsheets as discussed in  \cite{Jafferis:2021ywg}.
Further development of these ideas include generalizing to the B  model, the refined topological string, and the statistical mechanical description of the A model in terms of Calabi Yau crystals.

The perturbative topological amplitudes discussed in this work are chiral and non-unitary.  This leads to some peculiar features of the associated entanglement entropy.  We plan to investigate the restoration of unitarity via  non-perturbative corrections defined by q-deformed two dimensional Yang Mills \cite{Aganagic:2004js}.  This produces  non- chiral topological string amplitudes that are in principle compatible with a positive, sesquilinear inner product.   We will also apply known techniques in the computation of Wilson loops in q-deformed Yang Mills to give a statistical interpretation of the gravitational entropy of bubbling Calabi Yau's.

Ultimately, we are interested understand local holography and the formulation of quantum information measures in physical string theories.   For this purpose, the tensionless string in AdS3 provides a particular useful toy model that is closely related to the A model \cite{Knighton:2024ybs}.  It would be interesting to see if some of the ideas in our work can be used in that setting.

\section*{Acknowledgements}
We thank Dan Jafferis and Jan Troost, for discussions on topological string theory, and Thomas Mertens, Joan Simon, and Qifeng Wu for discussion on quantum group algebras and topological subregions, and William Donnelly for use of his tikz macros.    We thank the Simons Center for Geometry and Physics, and ICTP Trieste for hospitality during the completion of this work.  
GW is supported by STFC grant
ST/X000761/1 and the Oxford Mathematical Institute.

\appendix

\section{Relative cohomology} \label{relative}
Consider a closed from $F$ satisfying $dF=0$ over a sphere. We can interpret it as the field strength of a magnetic monople.  If we never cross any magnetic charges, the magnetic flux out of a sphere is a topological invariant, which defines a cohomology class
\begin{align}
    \int_{S^2} F  \in H^{2}(S^{2},\mathbb{R})
\end{align}
By Gauss law it just counts the total magnetic charge inside.   On a disk $D$, the flux of $F$ is no longer invariant under  general deformation that moves the boundary\footnote{Relatedly, the integral is no longer gauge invariant under a higher form gauge transformation $F \to F + d \eta$, where $\eta$ is a one form.}.   If we write $F=d A$ in annular neighborhood of the boundary $\pd D$, we see that deforming $\pd D$ changes the flux by
\begin{align}
    \int_{\pd D} A
\end{align}.   However we can obtain a topological invariant by adding boundary term that subtracts these change: 
\begin{align}
\int_{D} F - \int_{\pd D} A \in H^{2}(D,\pd D, \mathbb{R}) 
\end{align}
This is a topological invariant.    It is also gauge invariant under 1 form gauge transformations. 
\subsection{Euler classes}
When $F$ is the curvature of a real bundle over $S^2$, the flux can be interpreted as an Euler class.   
Let $L \to X$ be a complex line bundle over an oriented manifold $X$ with a smooth boundary $\partial X $. The Euler class of the underlying real bundle is the same as the Chern class but for an $SO(N)$  bundle
\begin{align}
e(L_{\mathbb{R}}) = c_1(L).
\end{align} 

If $\nabla$ is a connection with curvature $F$, then on a closed manifold one has
\begin{align}
 e(L)= \frac{i}{2\pi} \int_{X} F.
\end{align}
When $X$ has a boundary, this integral is not invariant by itself.   If we think of $F$ as the field strength,  then $e(L)$ is the magnetic flux out of $X$.  When $X$ is closed, this is topological invariant by Gauss law, since the integral counts the total charge inside.  But if $X$ has a boundary, and a general deformation that moves the boundary will change the flux.

To obtain a well-defined quantity, one must choose a nowhere-vanishing section $s$ of $L$ along $\partial X$. This defines a relative class
\begin{align}
e(L,s) \in H^2(X,\partial X;\mathbb{Z}),
\end{align}
The section $s$ defines a local gauge field $A$ via
\begin{align}
    \nabla s = i A s
\end{align}
and we can express the relative Euler class as
\begin{align}
e(L,s)=
\frac{i}{2\pi}\int_X F
-
\frac{1}{2\pi}\int_{\partial X} A,
\end{align}
which is invariant under changes of trivialization up to integers.

Geometrically, this integer is the obstruction to extending $s$ to a nowhere-zero section over all of $X$. Equivalently, it is the sum of the indices of the zeros of a generic extension of $s$ into the interior.

\paragraph{Example: Euler Characteristic of a disk}
Consider a complex line bundle  $L \to D^2$ over a disk.  This is a $U(1)$ bundle that we can identify with the tangent bundle of the disk, which is  the $SO(2)$ frame bundle with spin connection $\omega$. The identification arises from the relationship between $\omega$ and the Vielbeins:

Given this identification, we have a local relation
\begin{align}
   F&= d \omega \nn
   &=\sqrt{g} R
\end{align}
On a closed manifold $\Sigma$ the Euler class of $L$ is just the Euler characteristic:
\begin{align}
  e(L )= \int_{\Sigma} F= \int _{\Sigma}\sqrt{g} R =\chi (\Sigma)
\end{align}
From the perspective of the line bunde, this integral is non-zero because we cannot set 
 $F=d\omega $ globally. Instead this relation only holds patch by patch, glued together by a gauge transformation of $\omega$.

An open manifold like a disk can be covered by a single patch. And since the disk is flat, we might expect that  
\begin{align}\label{naive}
\int_{D^2} F = \oint_{\pd D^2}  \omega = 0 
\end{align}
However, there is a constraint on the tangent bundle of the \emph{smooth} disk with no conical singularities, which is that the total cone angle is $2\pi$.  This angle\footnote{Recall that $K_{ab}= \nabla_{a}n_{b}$ measures the derivative of the unit normal. So the Gibbons Hawking term tells us how this normal vector rotates as well traverse the boundary circle, which measures the cone angle of the disk.  } is measured by the Gibbons-Hawking boundary term, which equals the integral  of the spin connection
\begin{align}
 \int_{\pd D^2}  \sqrt{\gamma} K =\oint_{\pd D^2} \omega = 2 \pi 
\end{align}  
From the point of view of the complex bundle, this constraint picks out a now-where vanishing section on the boundary.  This choice is not compatible with \eqref{naive}.     Due to this  incompatiblity, we must once again define the gauge field $\omega $ in local patches, and glue together the patches with a nontrivial transition function. 

The global, gauge invariant quantity which defines a topological invariant in this situation is
\begin{align}
e(L,s)&=
\frac{1}{2\pi}\int_X F
-
\frac{1}{2\pi}\int_{\partial X} \omega
\end{align}
This is just the Euler Characteristic of the disk, with the boundary term given by the integral of the extrinsic curvature

More generally, one chooses a section $s$ on the boundary of the disk that can wind $n$ times, and the relative cohomology class measures the winding number $n$.  These corresponds to disks with $2\pi n$ cone angle.

\subsection{Boundaries with corners }
For the tangent bundle, Gauss--Bonnet becomes
\begin{align}
\chi(\Sigma) =
\frac{1}{2\pi}\int_\Sigma K, dA
+
\frac{1}{2\pi}\int_{\partial \Sigma_{\mathrm{smooth}}} k_g, ds
+
\sum_{p} \frac{\pi - \alpha_p}{2\pi},
\end{align}
where $\alpha_p$ is the interior angle at a corner $p$.
For the complex line bundle, the interior angles $\alpha$ are replaced by phases that glue together the sections $s_{p}$ on each edge.
\medskip

\section{An identity for the MacMahon function}
We show that
\[
\lim_{P \to \infty} \xi(q)^P \exp\!\left(
- \sum_{n=1}^{\infty} \frac{1}{n} \sum_{1\le i < j \le P} q^{n(j-i)}
\right)
= M(q),
\]
where
\[
\xi(q) = \prod_{m=1}^{\infty} (1 - q^m)^{-1}, 
\qquad
M(q) = \prod_{m=1}^{\infty} (1 - q^m)^{-m}.
\]

This proof was obtained with the help of ChatGPT, by feeding in similar formulas in \cite{Okuda:2007ai,Gomis:2007kz,Gomis:2006mv} and then providing it with the conjecture for this identity.

We begin by rewriting the double sum as follows:

For fixed separation $r = j-i$,
\[
\sum_{1 \le i < j \le P} q^{n(j-i)}
= \sum_{r=1}^{P-1} (P-r)\, q^{nr}.
\]

Thus the exponent becomes
\[
- \sum_{n=1}^{\infty} \frac{1}{n} \sum_{r=1}^{P-1} (P-r) q^{nr}.
\]

Next, we take the logarithm. 

Letting
\[
S_P(q) = \xi(q)^P \exp\!\left(
- \sum_{n=1}^{\infty} \frac{1}{n} \sum_{r=1}^{P-1} (P-r) q^{nr}
\right),
\]

we have
\begin{align*}
\log S_P(q)
&= -P \sum_{m=1}^{\infty} \log(1 - q^m)
- \sum_{r=1}^{P-1} (P-r) \sum_{n=1}^{\infty} \frac{q^{nr}}{n}.
\end{align*}.

Using
\[
- \sum_{n=1}^{\infty} \frac{q^{nr}}{n} = \log(1 - q^r),
\]
we obtain
\[
\log S_P(q)
= -P \sum_{m=1}^{\infty} \log(1 - q^m)
+ \sum_{r=1}^{P-1} (P-r) \log(1 - q^r).
\]

Exponentiating gives: 

\[
S_P(q)
= \prod_{m=1}^{\infty} (1 - q^m)^{-P}
\prod_{r=1}^{P-1} (1 - q^r)^{P-r}
=
\left(\prod_{r=1}^{P-1} (1 - q^r)^{-r}\right)
\left(\prod_{r=P}^{\infty} (1 - q^r)^{-P}\right).
\]

Finally, we take the limit  as $P \to \infty$:

\begin{itemize}
\item For fixed $r$, the exponent approaches $-r$, giving
\[
\prod_{r=1}^{\infty} (1 - q^r)^{-r} = M(q).
\]

\item The tail satisfies
\[
\log\!\left(\prod_{r=P}^{\infty} (1 - q^r)^{-P}\right)
\sim P \sum_{r=P}^{\infty} q^r \to 0,
\]
for $|q|<1$, so this factor tends to $1$.
\end{itemize}
Thus we obtain the formula

\[
\boxed{
\lim_{P \to \infty} S_P(q) = M(q).
}
\]

\bibliographystyle{utphys}
\bibliography{topstring}
\end{document}